\documentclass[aps,twocolumn,superscriptaddress,longbibliography,amsmath,amssymb,amsfonts,citeautoscript]{revtex4-1}
\usepackage{graphicx}
\usepackage{bm}
\usepackage{color}
\usepackage{epstopdf}
\usepackage{amsmath}
\usepackage{amssymb}
\usepackage{ulem}
\usepackage[urlcolor=blue,colorlinks=true,citecolor=blue,linkcolor=blue,pdfstartview={FitH},bookmarks=false]{hyperref}
\usepackage{ulem}

\newcommand{\expect}[1]{\langle #1 \rangle}

\newcommand{\be}{\begin{equation}}
\newcommand{\ee}{\end{equation}}
\newcommand{\bea}{\begin{eqnarray}}
\newcommand{\eea}{\end{eqnarray}}

\newcommand{\en}{\varepsilon}

\newcommand{\s}{\sigma}
\newcommand{\G}{\Gamma}
\newcommand{\up}{\uparrow}
\newcommand{\down}{\downarrow}

\usepackage{ulem}
\usepackage{verbatim} 

\begin{document}

\title{Hallmarks of Majorana mode leaking into a hybrid double quantum dot}

\author{Piotr Majek}
\email{pmajek@amu.edu.pl}
\affiliation{Institute of Spintronics and Quantum Information, Faculty of Physics, A.~Mickiewicz University, 61-614 Pozna{\'n}, Poland}

\author{Grzegorz G\'{o}rski}
\email{ggorski@ur.edu.pl}
\affiliation{Institute of Physics, College of Natural Sciences, University of Rzesz\'{o}w, 35-310 Rzesz\'{o}w, Poland}

\author{Tadeusz Doma\'{n}ski}
\email{doman@kft.umcs.lublin.pl}
\affiliation{Institute of Physics, M. Curie-Sk\l{}odowska University, 20-031 Lublin, Poland}

\author{Ireneusz Weymann}
\email{weymann@amu.edu.pl}
\affiliation{Institute of Spintronics and Quantum Information, Faculty of Physics, A.~Mickiewicz University, 61-614 Pozna{\'n}, Poland}

\date{\today}

\begin{abstract}
We investigate the spectral and transport properties of a double quantum dot laterally
attached to a topological superconducting nanowire, hosting the Majorana zero-energy modes.
Specifically, we consider a geometry, in which the outer quantum dot is embedded between
the external normal and superconducting leads, forming a circuit.
First, we derive analytical expressions for the bound states in the case of an uncorrelated system
and discuss their signatures in the tunneling spectroscopy.
Then, we explore the case of strongly correlated quantum dots
by performing the numerical renormalization group calculations,
focusing on the interplay and relationship between the leaking Majorana mode
and the Kondo states on both quantum dots.
Finally, we discuss  feasible means to experimentally probe the in-gap quasiparticles
by using the Andreev spectroscopy based on the particle-to-hole scattering mechanism.
\end{abstract}

\maketitle

\section{Introduction}
\label{sec:introduction}

Two quantum dots contacted in various arrangements with external macroscopic reservoirs have been proposed 
as promising building blocks of future nanoelectronic, spintronic and quantum information technologies 
\cite{vanderWiel-2002,Hanson-2007}. For instance, double quantum dot (DQD) configurations provide a versatile platform for the implementation of spin-based quantum information processing systems \cite{Nowack-2007}.
Moreover, a rapid progress in materials science of superconducting
hybrid nanostructures \cite{Rodero-11,Sherman.2017} stimulated vivid interest in constructing quantum bits out of the Andreev bound states
\cite{Rodero-11,Larsen-2015,deLange-2015,Luthi-2018,PitaVidal-2020}.
Further perspectives for the realization of topological superconducting qubits
are related to nanostructures involving DQDs coupled to topological superconducting wires,
hosting the Majorana zero-energy modes (MZM) at their ends, the so-called Majorana wires (MW)
\cite{Silva-2016,Ivanov-2017,Su-2017,Rancic-2019,Cifuentes-2019,Weymann-2020}. 
Such platforms allow for the implementation of fault-tolerant quantum computing protocols,
which are in the center of interest of quantum information research \cite{Kouwenhoven-2020}.

\begin{figure}[b]
\centering
\includegraphics[width=1\columnwidth]{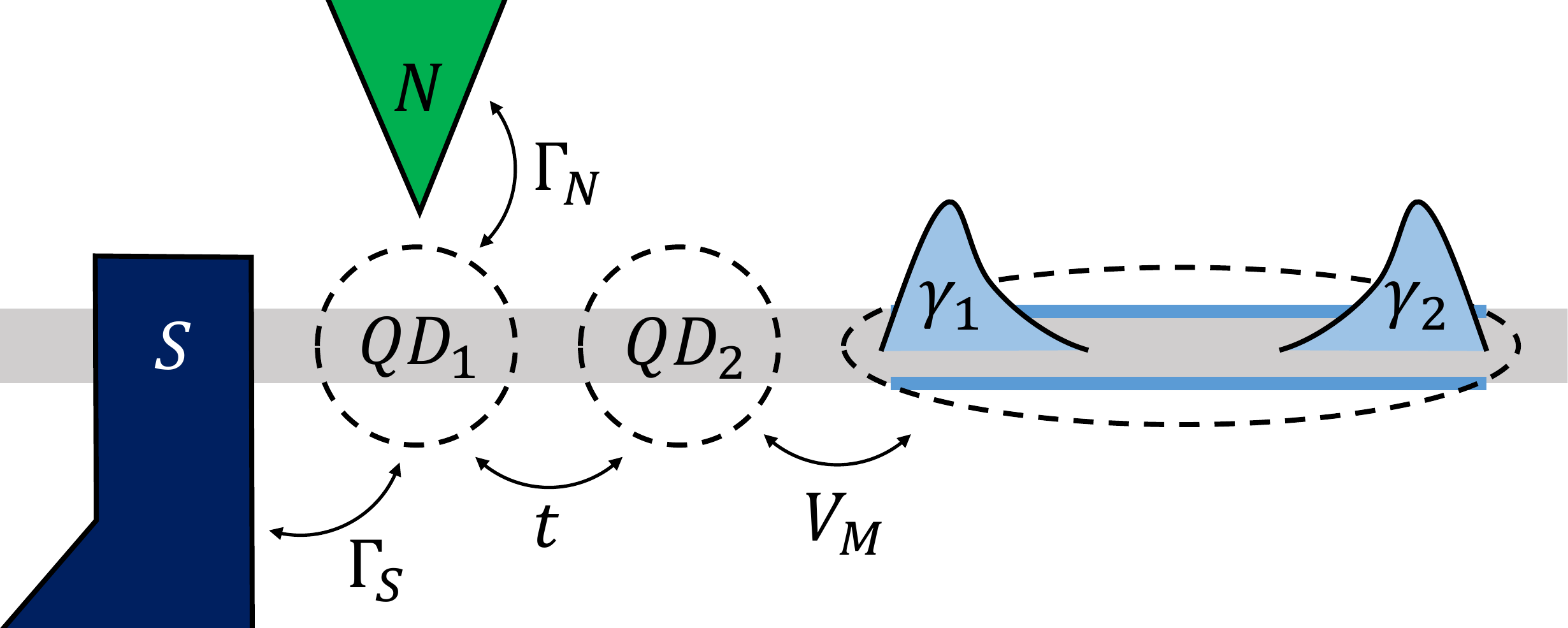}
\caption{Scheme of the considered hybrid structure, comprising the quantum dot (QD$_{1}$) placed between the normal (N) and superconducting (S) electrodes and side-attached to the second quantum dot (QD$_{2}$) bridged with the topological superconducting nanowire, hosting Majorana end-modes
	described by the operators $\gamma_1$ and $\gamma_2$.
An STM tip, attached to QD$_{1}$ with the coupling strength $\G_{N}$,
allows for probing the spectroscopic features of the system.}
\label{scheme}
\end{figure}

The main motivation for studying the hybrid nanostructures
composed of quantum dots and Majorana wires is associated with a tendency
of such end-modes to leak into the neighboring objects \cite{Deng-2016}.
Effectively, this gives rise to fractional values of the differential conductance,
which serve as fingerprints of the exotic character of MZM \cite{Prada_review-2020}. 
Various situations have been investigated so far,
considering mainly single quantum dots and exploring the interplay of correlation effects with the Majorana quasiparticles \cite{Deng-2018,Gorski-2018}. 
In this regard, much less attention has been paid to hybrid systems composed of double quantum dots
hybridized with topological superconducting wires.
Therefore, this work aims to shed some light on the transport properties of DQD-MW hybrid structures,
focusing specifically on the setup displayed in Fig.~\ref{scheme}.

It is important to note that the properties of double quantum dots
proximitized with conventional superconductors have been studied experimentally by the tunneling spectroscopy, using InAs \cite{Sherman.2017,Grove_Rasmussen.2018,Estrada_Saldana.2018,Paaske-2020,Estrada-2020}, InSb \cite{Su.2017}, Ge/Si \cite{Zarassi.2017}, carbon nanotubes \cite{Cleuziou.2006,Pillet.2013}
and by the scanning tunneling microscopy (STM) applied 
to magnetic dimers deposited on superconducting surfaces \cite{Ruby.2018,Franke-2018,Choi.2018,Kezilebieke.2019}.
In-gap bound states of the double quantum dots (dimers) have been thoroughly analyzed by a number groups \cite{Choi-2000,Zhu-2002,
Tanaka.2010,Zitko.2010,Konig.2010,Droste.2012,Grifoni.2013,Sothmann-2014,Yao2014Dec,Meng.2015,Zitko-2015,
Wrzesniewski-2017,Glodzik.2017,Frolov-2018,Scherubl_2019,Zonda.2019,Wang-2019,Leijnse.2019},  
in particular predicting  quantum phase transitions, in which the total spin could vary between
the singlet, doublet and triplet states \cite{Morr-2003,Morr-2006}.
However, the properties of proximitized DQDs, additionally interacting 
with Majorana modes, are much less explored \cite{Brunetti-2013}.

Recently, a fusion of individual bound states into their molecular (hydrogen atom-like)
structure has been realized by a controllable change of the hybridization
between the quantum dots contacted with superconducting reservoirs \cite{Baumgartner-2021}. 
A similar process for topological bound states would be highly desirable, therefore this issue triggers intensive activities  \cite{Flensberg_review-2021}.
Motivated by such achievements and trends, we study the transport behavior of a hybrid double quantum dot setup, where the central quantum dot (QD$_{1}$) is embedded between the superconducting (S) and the normal metallic (N) lead (provided by e.g. an STM tip), forming a circuit. We assume  QD$_{1}$ to be connected through the second quantum dot (QD$_{2}$) to the topological superconducting nanowire, see Fig.~\ref{scheme}.
Practically, such quantum dots can be considered as being a piece of a non-topological segment of the nanowire consisting of two sites,
in  analogy to the experimental hybrid system that provided the first evidence for
leakage of the Majorana quasiparticles \cite{Deng-2016,Deng-2018}.
To obtain the most reliable predictions for the behavior of the considered system when correlations play an important role,
we resort to the density-matrix numerical renormalization group (NRG) calculations \cite{Wilson1975Oct,Andreas_broadening2007,Bulla2008Apr,Toth2008,NRG_code}.
Moreover, to provide the complete picture and deepen the understanding of transport properties of the considered DQD-MW hybrid structure,
we also perform analytical calculations for an uncorrelated case. 

Our study reveals that: 
(i) besides  the conventional in-gap quasiparticle branches
(originating from the hybridization of both quantum dots)
there appear additional structures induced by the Majorana mode in the form of constructive/destructive 
interference pattern imprinted on the spin-down/spin-up sectors of QD$_{1}$;
(ii) in the Kondo regime (when the Coulomb repulsion prevails over the superconducting proximity effect) 
the spin-resolved spectral functions  indicate the detrimental/constructive influence of the Majorana mode 
on $\uparrow$/$\downarrow$ spin sectors of QD$_{1}$. For such strongly correlated system,
we predict  the optimal Andreev conductance near a crossover from
the doublet to the BCS-type singlet configurations of proximitized QD$_{1}$.

The paper is organized as follows. In Sec.\ \ref{sec:model} we introduce the microscopic model of the studied DQD-MW system,
describing our setup. Next, in Sec.\ \ref{sec:uncorrelated} we discuss the properties of the considered hybrid structure,
neglecting the Coulomb repulsion on both quantum dots.
In Sec.\ \ref{sec:correlated} we address the correlation effects in the Kondo regime, using the numerical renormalization group approach.
Finally, we summarize our findings in Sec.~\ref{sec:summary}.

\section{Model}
\label{sec:model}

In what follows, we analyze the spectroscopic and transport properties of N-QD$_{1}$-S
branch (see Fig.~\ref{scheme}), focusing on the subgap energy region. 
The second quantum dot QD$_{2}$ transmits the Majorana mode(s), which in turn 
affects the transport by interferometric effects. We study those effects in 
detail, considering the fully polarized case, $V_{M\downarrow}=V_M$ and 
$V_{M\uparrow}=0$, where 
$V_{M\sigma}$ is the coupling between the second dot and Majorana wire for spin $\sigma$.
Some results for the arbitrary spin-dependent couplings $V_{M\sigma}$
\cite{Klinovaja-2017, Prada-2017,Deng-2018,Schuray-2018} are presented in Appendix \ref{polarization}.

Our hybrid structure (Fig.~\ref{scheme}) can be modeled by the following Hamiltonian
\be
\label{eq:H}
H = H_{\rm N} + T_{\rm N} + H_{\rm DQD} + H_{\rm MW} + H_{\rm SC} ,
\ee
where 
\be
\label{eq:HLeads}
H_{\rm N} = \sum_{\mathbf{k}\sigma}
\en_{N\mathbf{k}} c^\dag_{N\mathbf{k}\sigma} c_{N\mathbf{k}\sigma} 
\ee
describes the metallic lead  with 
the operators $c^\dag_{N\mathbf{k}\sigma}$ creating electrons with spin $\s$,
momentum  $\mathbf{k}$ and energy $\en_{N\mathbf{k}}$. The second term in Eq.~(\ref{eq:H})
describes the tunneling processes between the metallic lead and the first quantum dot
\be
\label{eq:HTun}
T_{\rm N} = \sum_{\mathbf{k}\sigma} V_{N} \left(d^\dag_{1 \s}
c_{N\mathbf{k}\sigma} + c^\dag_{N\mathbf{k}\sigma} d_{1 \s} \right),
\ee
where $V_{N}$ is the momentum independent tunneling matrix element and 
$d^\dag_{1 \s}$ operator creates electrons with spin $\s$ at the central quantum 
dot. When this quantum dot is coupled to the external contacts, it leads to 
the broadening of the dot level described by $\Gamma_N = \pi \rho_N V_N^2$, 
where $V_N$ is assumed to be real, while $\rho_N$ is the density of states of 
the metallic lead. For calculations we  use the bandwidth of metallic lead as 
a convenient energy unit ($D \equiv 1$).

The double quantum dot part is given by
\bea \label{eq:HDDM}
H_{\rm DQD} &=& \sum_{j\s} \en_j d_{j\s}^\dag d_{j\s}
+ \sum_{j} U_j d_{j\uparrow}^\dag d_{j\uparrow} d_{j\downarrow}^\dag 
d_{j\downarrow}
\nonumber\\
&&+ \sum_\s t\left( d_{1\s}^\dag d_{2\s} +  \rm{h.c.} \right) ,
\eea
where $d_{j\s}^\dag$ creates a spin-$\sigma$ electron
on the $j$-th quantum dot with energy $\en_j$.  
The repulsive Coulomb potential $U_j$ between the opposite spin electrons on 
individual quantum dots shall be assumed equal $U_1 = U_2 = U$. 
These quantum dots are interconnected through the hybridization, denoted by $t$.

The low-energy quasiparticles of the topological nanowire can be described by the following term
\be
H_{\rm MW} = \sqrt{2} V_M (d^\dag_{2\downarrow} \gamma_1 + \gamma_1 
d_{2\downarrow}) + i \en_M \gamma_1 \gamma_2.
\ee
The first part couples the spin-$\downarrow$ electrons of the second quantum 
dot with the Majorana mode described by the operator $\gamma_1$ through the tunneling element $V_M$.
The role of the coupling of Majorana quasiparticles
to both spins of QD$_{2}$ is briefly discussed in Appendix.
The Majorana operators can be rewritten in terms of an 
auxiliary fermion operator $f$
as  $\gamma_1 = (f^\dag+f)/\sqrt{2}$ and $\gamma_2 = i(f^\dag-f)/\sqrt{2}$.
In the case of a short nanowire, we should assume the 
overlap $\en_M$ between the wave functions of the Majorana modes.
However, here we focus on the long wire case,
i.e. when the Majorana quasiparticles do not overlap and $\varepsilon_{M}=0$.

The last part of  the Hamiltonian (\ref{eq:H}) refers to superconducting substrate  and its coupling to QD$_{1}$
\begin{eqnarray}
H_{\rm SC} &=& \sum_{\mathbf{k}\sigma}
\en_{S\mathbf{k}} c^\dag_{S\mathbf{k}\sigma} c_{S\mathbf{k}\sigma} 
- \left( \Delta_{SC} \sum_{\mathbf{k}} c^\dag_{S\mathbf{k}\uparrow}
c^\dag_{S-\mathbf{k}\downarrow} + {\rm h.c.} \right)
\nonumber \\
&+& \sum_{\mathbf{k}\sigma} V_{S} \left(d^\dag_{1 \s}
c_{S\mathbf{k}\sigma} + c^\dag_{S\mathbf{k}\sigma} d_{1 \s} \right) .
\label{eq:supercond}
\end{eqnarray}
In the limit of large pairing gap, $\Delta_{SC}\rightarrow \infty$,
these terms give rise to the proximity induced on-dot pairing 
\bea
\label{eq:HSC}
H_{\rm SC} \approx - \Delta_{1} (d_{1 \up}^\dag d_{1 \down}^\dag + d_{1 \down} d_{1 \up}) .
\eea
with the effective pairing potential $\Delta_{1}=\Gamma_{S}$,
and $\Gamma_S$ denoting the coupling strength between QD$_1$
and the superconductor.

\section{The case of uncorrelated system}
\label{sec:uncorrelated}

Let us start our analysis by investigating the spectral and transport properties
in the case of uncorrelated quantum dots, i.e. when $U=0$.
For this purpose, we use the Green’s function method.
The influence of the topological nanowire on the quantum dots
can be captured by the matrix Green's function defined in the particle-hole (Nambu) notation 
 $\hat {\cal{G}}(\omega )=\left< \left< \Psi ;\Psi^\dag \right>\right> _{\omega}$,
with $\Psi  = (d_{1\uparrow},d_{1\uparrow}^\dag,d_{1\downarrow},d_{1\downarrow}^\dag,d_{2\uparrow},d_{2\uparrow}^\dag,d_{2\downarrow},d_{2\downarrow}^\dag,f,f^\dag)$.
For $U=0$, the matrix equation for the Green's function takes the following form
\begin{widetext}
\begin{eqnarray} 
{\cal{G}}^{-1}(\omega) =\omega \hat{I}+
\left( \begin{array}{cccccccccc}  
-\varepsilon_{1}+{i\Gamma_N} &0&0& {\Gamma_{S}}& -t & 0& 0 & 0 & 0 & 0\\
0&\varepsilon_{1}+{i\Gamma_N}&-{\Gamma_{S}}& 0 & 0 & t& 0 & 0& 0& 0\\
0&-{\Gamma_{S}}&-\varepsilon_{1}+{i\Gamma_N}& 0 & 0 & 0& -t & 0&0 & 0\\
{\Gamma_{S}}&0&0&\varepsilon_{1}+{i\Gamma_N} & 0 & 0& 0 & t& 0 & 0\\
-t & 0& 0 & 0&-\varepsilon_{2}&0&0& 0&0 & 0 \\
0 & t& 0 & 0&0&\varepsilon_{2}&0& 0&0 &0 \\
0 & 0& -t & 0&0&0&-\varepsilon_{2}& 0&  -V_{M} & -V_{M} \\ 
0& 0& 0 & t&0&0&0&\varepsilon_{2}& V_{M} & V_{M} \\
0 & 0 &0 & 0& 0 & 0 & -V_{M} & V_{M}& 0 & 0\\
0 & 0 &0 & 0& 0 & 0 &-V_{M} & V_{M}& 0    & 0 
\end{array}\right) ,
\label{Gr66}
\end{eqnarray} 
\end{widetext}
where $\hat{I}$ stands for the identity matrix.

\subsection{Resonant bound states for $\Gamma_{N}=0$}
\label{sec:Resonant_states}

Consider first the conventional bound states of the proximitized double quantum dot in the absence of coupling to the Majorana nanowire \cite{Gorski-2021}.
In the limit of $\Gamma_{N}\rightarrow 0$ such in-gap resonant states are formed at 
\begin{eqnarray}
\varepsilon_{AD1}^{\pm}=\pm \frac{1}{\sqrt{2}} \sqrt{A + \sqrt{A^2-4B}},\nonumber\\
\varepsilon_{AD2}^{\pm}=\pm \frac{1}{\sqrt{2}} \sqrt{A - \sqrt{A^2-4B}} ,
\label{EA0}
\end{eqnarray}
where $A=\varepsilon_1^2+\varepsilon_2^2+\Gamma_{S}^2+2t^2$ and 
$B=(\varepsilon_1\varepsilon_2-t^2)^2+\left(\varepsilon_2\Gamma_{S}\right)^2$. The quasiparticles $\varepsilon_{AD1}^{\pm}$ represent the Andreev bound states of QD$_{1}$, 
$E_{A}^{\pm}=\pm \sqrt{\varepsilon_1^2+\Gamma_{S}^2}$, now slightly modified by the hybridization $t$ to QD$_{2}$. The other quasiparticles $\varepsilon_{AD2}^{\pm}$ originate from the  energy level of  QD$_{2}$ owing to the induced electron pairing (via its coupling to QD$_{1}$).

The attachment of the topological superconductor nanowire to the double quantum dot substantially affects the spectrum of this setup, revealing signatures of the zero-energy Majorana mode. Now, besides the initial quasiparticles $\varepsilon_{AD1}^{\pm}$ and $\varepsilon_{AD2}^{\pm}$ there emerge additional states at
\begin{eqnarray}
\varepsilon_{MD1}^{\pm}=\pm \frac{1}{\sqrt{2}} \sqrt{A_M + \sqrt{A_M^2-4B_M}},\nonumber\\
\varepsilon_{MD2}^{\pm}=\pm \frac{1}{\sqrt{2}} \sqrt{A_M - \sqrt{A_M^2-4B_M}},
\label{EAM}
\end{eqnarray}
where $A_M=A+4 V_{M}^2$ and $B_M=B+4V_{M}^2(\varepsilon_1^2+\Gamma_{S}^2+t^2)$. Let us notice that such new features depend on the coupling strength $V_{M}$. We recognize that the quasiparticle states $\varepsilon_{MD1}^{\pm}$ and $\varepsilon_{MD2}^{\pm}$ are driven by a direct coupling of the QD$_{2}$ and an indirect coupling of the QD$_{1}$ (via QD$_{2}$)
to the Majorana mode. 
The quasiparticle energies $\varepsilon_{ADi}^{\pm}$ and $\varepsilon_{MDi}^{\pm}$ ($i=1,2$) are displayed in Figs.~\ref{figA1_ed20} and \ref{figA1_ed2} by dashed lines.

\subsection{Quasiparticle spectrum for $\Gamma_{N}\neq 0$}
\label{sec:Spectral}

\begin{figure} 
\centering
\includegraphics[width=1\linewidth]{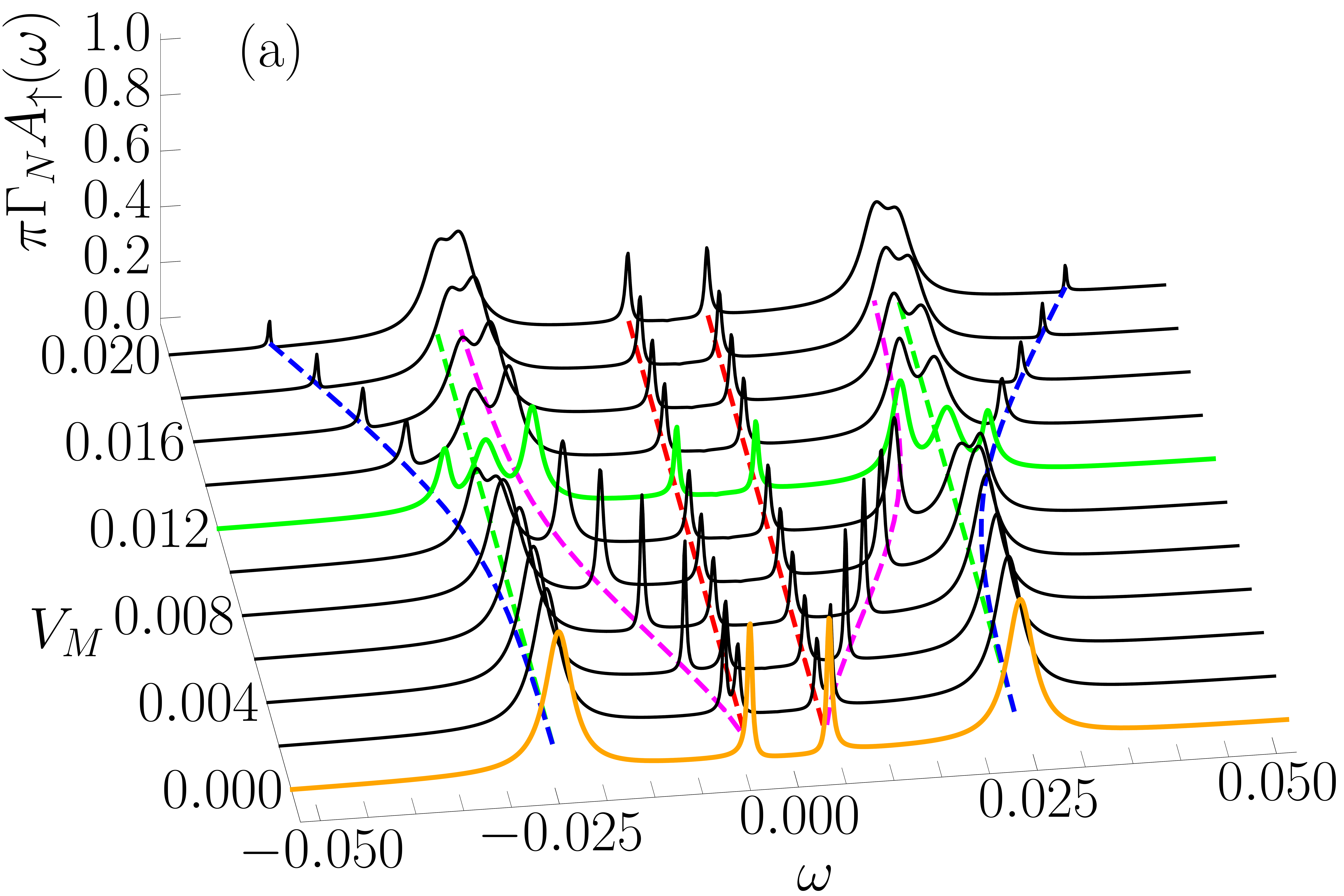}
\includegraphics[width=1\linewidth]{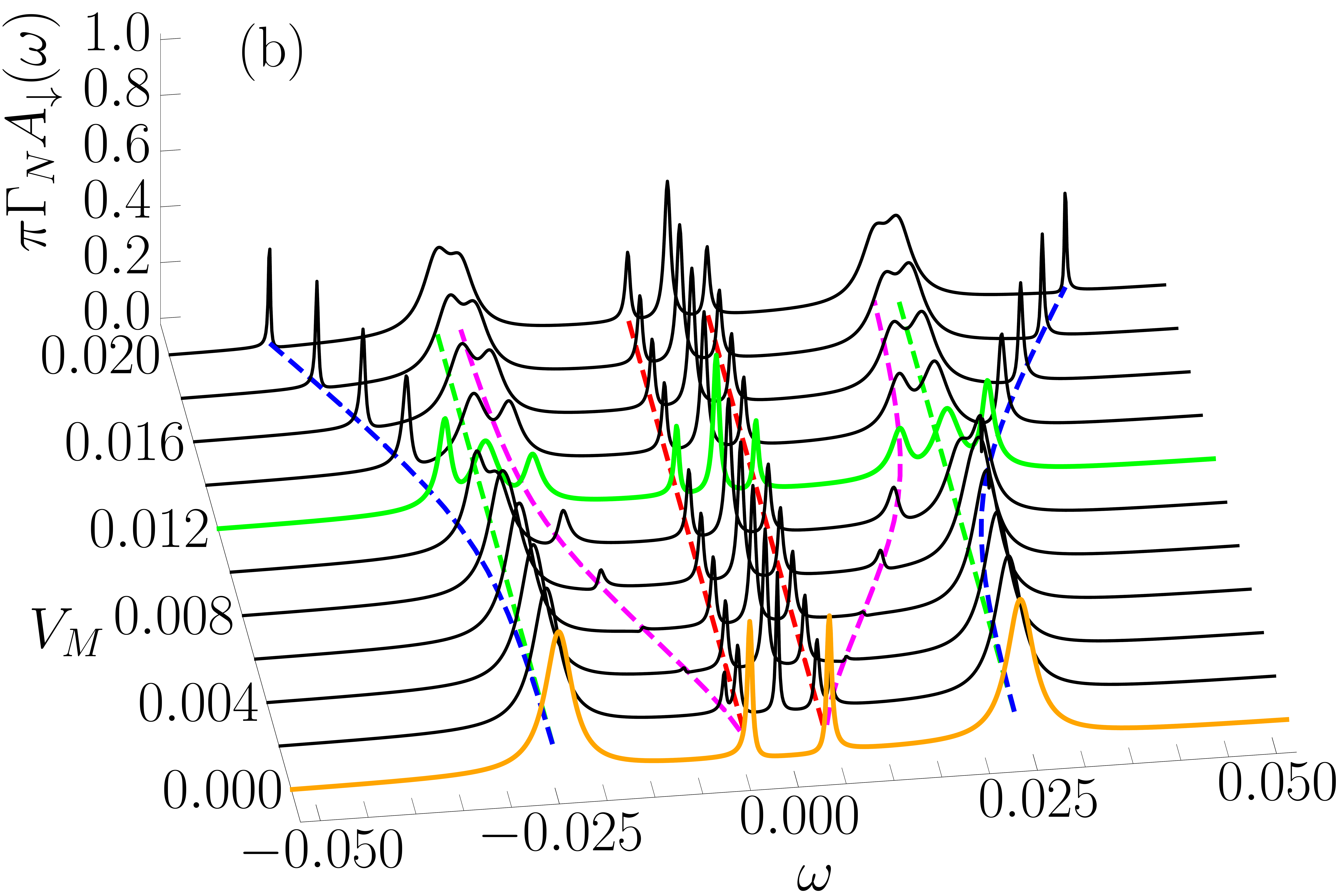}
\caption{
The normalized spectral function $\pi\Gamma_N A_{\sigma}(\omega)$ of
the first quantum dot for (a) spin-$\uparrow$ and (b) spin-$\downarrow$ electrons obtained for $\varepsilon_{1}=\varepsilon_{2}=0$, $\Gamma_{S}=0.02$,  $\Gamma_N=0.002$ and $t=0.01$. The quasiparticle energies $\varepsilon_{ADi}^{\pm}$ and $\varepsilon_{MDi}^{\pm}$,
given by Eqs.~(\ref{EA0}) and (\ref{EAM}), are marked by the dashed lines ($\varepsilon_{AD1}^{\pm}$--green, $\varepsilon_{AD2}^{\pm}$--red, $\varepsilon_{MD1}^{\pm}$--blue, $\varepsilon_{MD2}^{\pm}$--magenta). All energies are expressed in units of the bandwidth ($D\equiv 1$).}
\label{figA1_ed20}
\end{figure}

\begin{figure} 
\centering
\includegraphics[width=1\linewidth]{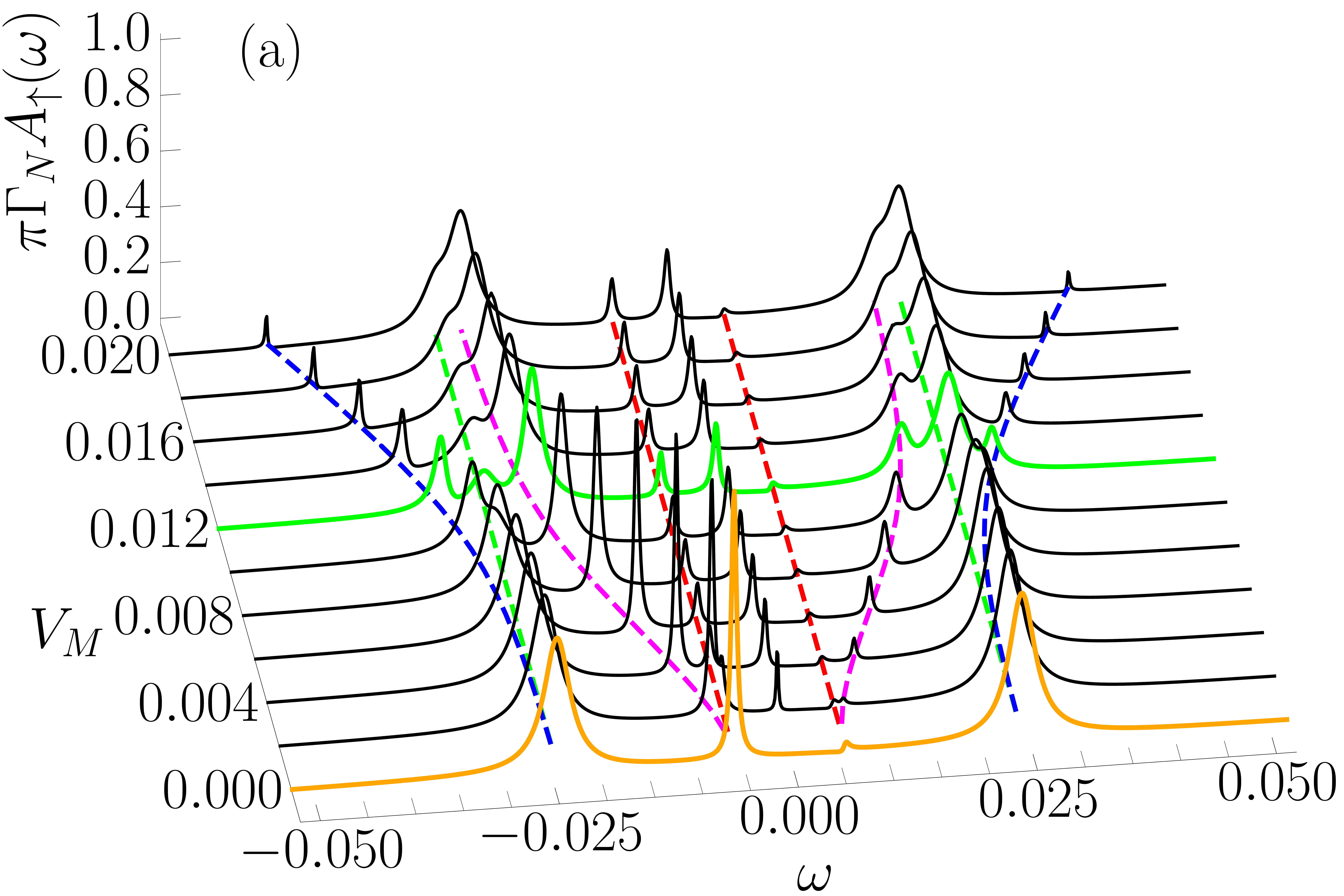}
\includegraphics[width=1\linewidth]{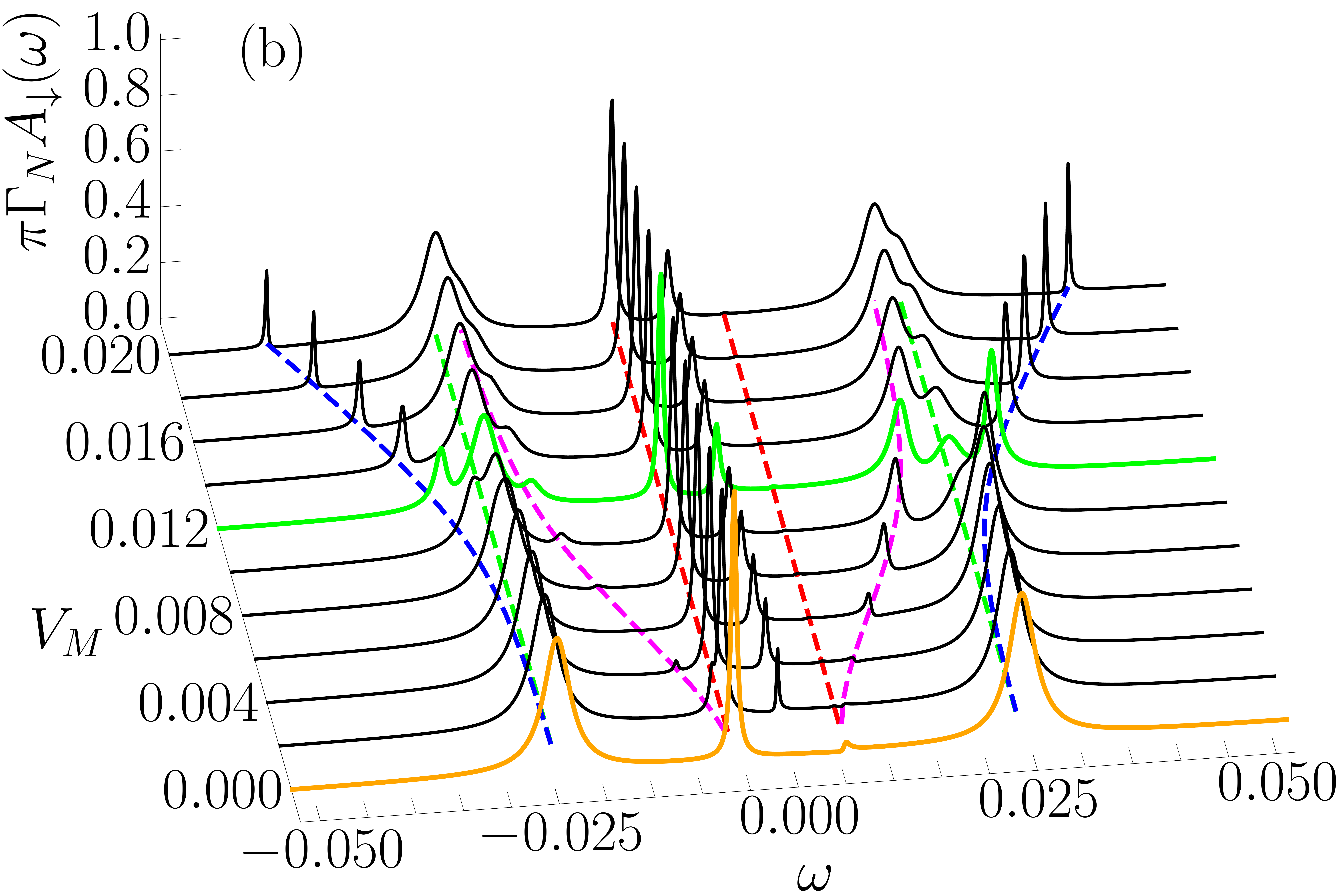}
\caption{(a) The spin-up and (b) spin-down normalized spectral function obtained for detuned energies of the quantum dots $\varepsilon_{1}=0$ and $\varepsilon_{2}=0.005$, using the same model  parameters as in Fig.~\ref{figA1_ed20}.}
\label{figA1_ed2}
\end{figure} 

To discuss the empirically measurable properties of our setup (Fig.\ \ref{scheme}), let us now examine the effective spectrum for a finite (yet small) coupling $\Gamma_{N}$. 
The continuous electronic spectrum of the normal lead broadens the subgap quasiparticle states, which acquire finite life-times.
Because the transport properties of the considered system
can be related to the spectral function of the first quantum dot,
in the following we shall focus on its behavior.
We have computed this spectral function
\begin{eqnarray}
A_{\sigma}(\omega)= -\frac{1}{\pi} \mbox{\rm Im}  \langle \langle  d_{1\sigma} ;d_{1\sigma}^{\dag} 
 \rangle \rangle  _{\omega+i0^{+}},
\label{spectral_fun}
\end{eqnarray}
choosing such model parameters that allow for clear identification of  the role played by the Majorana mode. For specific numerical calculations we use: $\Gamma_{N}=0.002$, $\Gamma_{S}=0.02$, $t=0.01$, and $\varepsilon_{1}=0$ in units of bandwidth $D\equiv 1$. Figure~\ref{figA1_ed20} refers to the case of identical quantum dot energies $\varepsilon_{1}=\varepsilon_{2}$, while Fig.~\ref{figA1_ed2} presents the case of detuned energies $\varepsilon_{1} \neq \varepsilon_{2}$, respectively.

In the absence of topological nanowire (see the orange curves in Figs. 
\ref{figA1_ed20} and \ref{figA1_ed2}), we notice two pairs of the Andreev peaks 
centered at $\varepsilon_{AD1}^\pm$ and $\varepsilon_{AD2}^\pm$ accompanied by 
the interferometric (Fano-type) structure at $\omega=\varepsilon_2$. The direct 
coupling of QD$_{2}$ to the topological nanowire indirectly affects the 
electronic states of QD$_{1}$. Such influence is transmitted via 
spin-$\downarrow$ electrons of the second quantum dot. In consequence, the 
leakage of Majorana mode shows up at zero energy of the spin-down spectral 
function $A_{\downarrow}(\omega)$. For $\varepsilon_{1}=\varepsilon_{2}$ and 
arbitrary couplings $V_M\neq 0$, we obtain the universal values 
$A_{\downarrow}(0)=1/2\pi\Gamma_N$ and $A_{\uparrow}(0)=0$, see 
Fig.~\ref{figA1_ed20},
in analogy to the previously reported behavior of single dot configurations \cite{Gorski-2018}.

\begin{figure}[t!]
\includegraphics[width=0.95\linewidth]{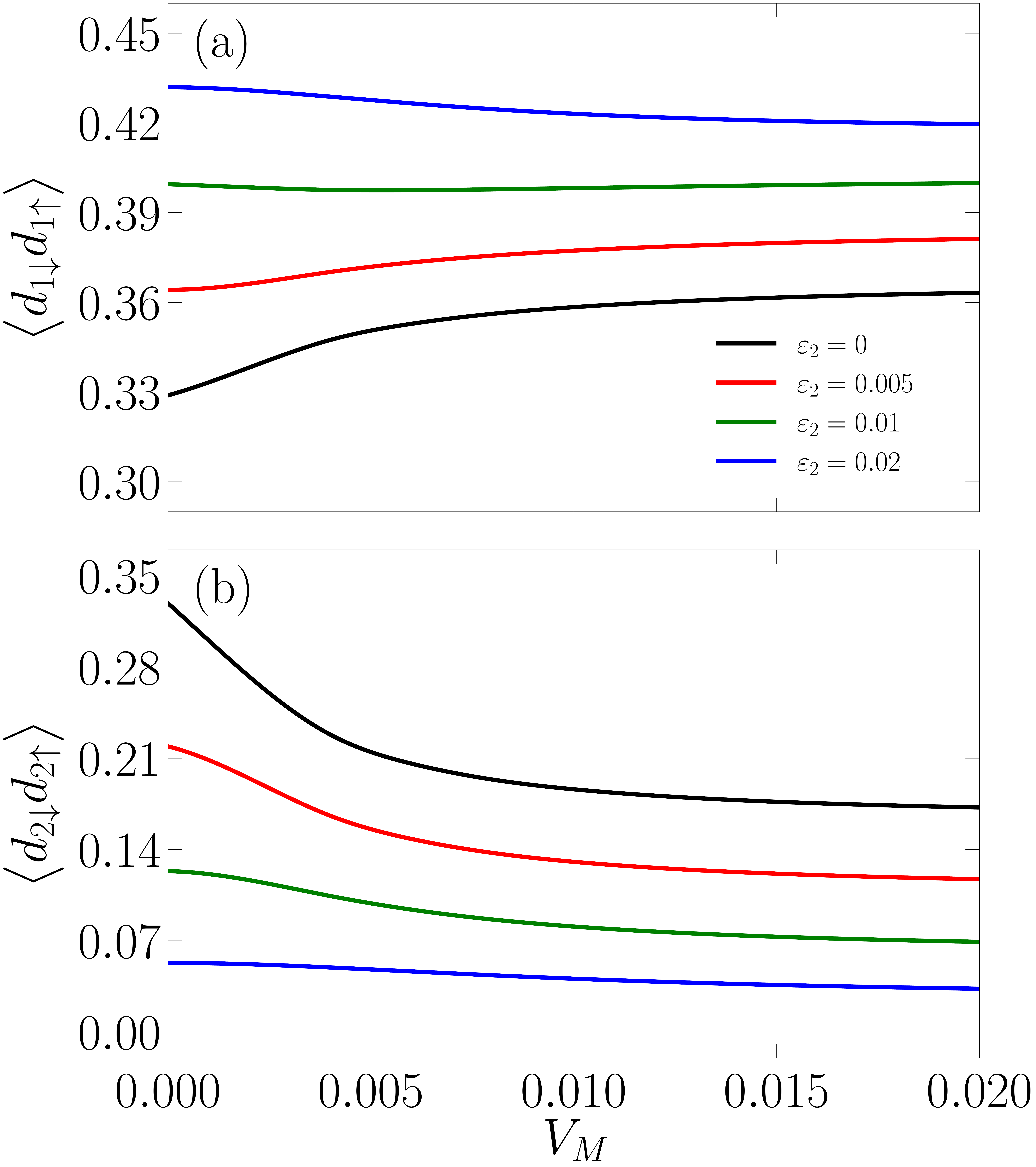}
\caption{The proximity induced pairing on (a) the first (QD$_{1}$) and (b) the second (QD$_{2}$) quantum dot as a function of the coupling $V_{M}$ to the Majorana wire obtained for $\varepsilon_{1}=0$ and several values of $\varepsilon_{2}$, as indicated. The other parameters are the same as in Fig.~\ref{figA1_ed20}.}
\label{fig_pair}
\end{figure}
 
The coupling $V_M$ is responsible for additional features, appearing at 
$\varepsilon_{MD1}^\pm$ and $\varepsilon_{MD2}^\pm$, whereas the quasiparticle 
states $\varepsilon_{AD1}^\pm$ and $\varepsilon_{AD2}^\pm$ practically do not 
change their energies (see green and red dashed lines), except for their 
spectral weights. 
For the weak coupling $V_M$, the quasiparticle states $\varepsilon_{MD1}^{\pm}$ 
coincide with $\varepsilon_{AD1}^{\pm}$, while the other states 
$\varepsilon_{MD2}^{\pm}$ are pushed to higher energies. On the other hand, for 
stronger couplings $V_{M}$, we observe the development of a molecular 
structure, in which $\varepsilon_{MD2}^{\pm}$ are mixed with 
$\varepsilon_{AD1}^{\pm}$. Figures \ref{figA1_ed20} and \ref{figA1_ed2} show 
that the quasiparticles $\varepsilon_{MD1}^{\pm}$ have the dominant spectral 
weights in spin-$\downarrow$ sector, whereas the states 
$\varepsilon_{MD2}^{\pm}$ gain their spectral weights mainly in the 
spin-$\uparrow$ sector.

Let us now inspect the role of QD$_{2}$ detuning from the particle-hole 
symmetry point ($\varepsilon_{2}\neq 0$). Under such conditions, the Majorana 
peak is observable in both spin components of the spectral function 
$A_{\sigma}(\omega)$, see Fig.~\ref{figA1_ed2}. The width of the zero-energy 
peak depends on the coupling $V_{M}$, while the height ratio 
$A_{\downarrow}(0)/A_{\uparrow}(0)$ is controlled by the energy level 
$\varepsilon_{2}$, keeping the total spectral function 
$A_{\uparrow}(0)+A_{\downarrow}(0) \approx 1/2\pi\Gamma_N$.
For the case of strong coupling $V_{M}$, we obtain 
$A_{\uparrow}(0)>A_{\downarrow}(0)$, and additionally there emerges a Fano-type 
dip in $A_{\uparrow}(\omega)$ at $\omega=\varepsilon_{2}$. Such interferometric 
feature is absent in $A_{\downarrow}(\omega)$ because of the hybridization of 
spin-$\downarrow$ electrons of the second quantum dot with the Majorana nanowire.

The second quantum dot (QD$_{2}$) is indirectly affected by the superconducting reservoir, absorbing the electron pairing  \cite{Tanaka-2008, Baranski-2011,Baranski-2012,Baranski-2020,Gorski-2021}. 
Figure~\ref{fig_pair} shows the variation of the pairing correlations 
$\expect{d_{i\down} d_{i\up}}$ induced on the first ($i=1$)
and the second ($i=2$) quantum dot with respect to the coupling $V_{M}$ for several values of energy level of QD$_{2}$.
For $\varepsilon_{2}=0$ and $V_{M}=0$, we reproduce the standard result $\left< d_{1\downarrow}d_{1\uparrow}\right> = \left< d_{2\downarrow}d_{2\uparrow}\right>$ reported in Ref.\ \cite{Tanaka-2008}. Upon detuning the quantum dot energy levels, we observe that pairing correlations of QD$_{2}$ are gradually suppressed, at an expense of enhancing the effective pairing on the central quantum dot (QD$_{1}$). The coupling of QD$_{2}$ to Majorana nanowire  has rather minor influence on all pairing sectors, it merely weakens the pairing correlations $\expect{ d_{2\downarrow}d_{2\uparrow}}$ on the second quantum dot.
 
\subsection{Linear Andreev conductance}
\label{sec:conductance}

The Majorana mode features leaking into the quantum dots could be detected by the charge transport measurements 
\cite{Deng-2012,Mourik-2012,Das-2012,Churchill-2013,Deng-2016}. 
In our setup, the central quantum dot (QD$_{1}$) is embedded between the normal and superconducting electrodes, so its subgap states should be observable in the Andreev current measurements \cite{Deacon-2010,Deacon-2010B, Hubler2012, Chang-2013}.
From theoretical side, the Andreev current can be calculated from
\begin{eqnarray} 
I_{A}(V) = \frac{e}{h} \int \!\!  d\omega \; T_{A}(\omega)
\left[ f(\omega\!-\!eV)\!-\!f(\omega\!+\!eV)\right],
\label{I_A}
\end{eqnarray} 
where $f(x)=\left[ 1 + \mbox{\rm exp}(x/k_{B}T) \right]^{-1}$ denotes the Fermi-Dirac distribution function and the energy-dependent Andreev transmittance is given by
\begin{equation}
T_{A}(\omega)= \Gamma_{N}^{2} \sum_\sigma \left| \langle \langle d_{1\sigma} ;d_{1\bar{\sigma}} \rangle\rangle_{\omega+i0^{+}} \right|^2.
\end{equation}
For $t=0$, the optimal value of the low-temperature Andreev conductance $G_{A}(V)=dI_{A}(V)/dV$ is $4e^2/h$ \cite{Buitelaar-2002,Deacon-2010,Koerting-2010,Baranski-2013}. 
Finite hopping between the dots, however, lowers this maximum
value. Moreover, the side-attached Majorana nanowire
can result in a further reduction of the Andreev conductance down to $ e^2/h$ \cite{Baranski-2017,Baranski-2019,Gorski-2018}.

\begin{figure}[t!] 
\centering
\includegraphics[width=0.95\linewidth]{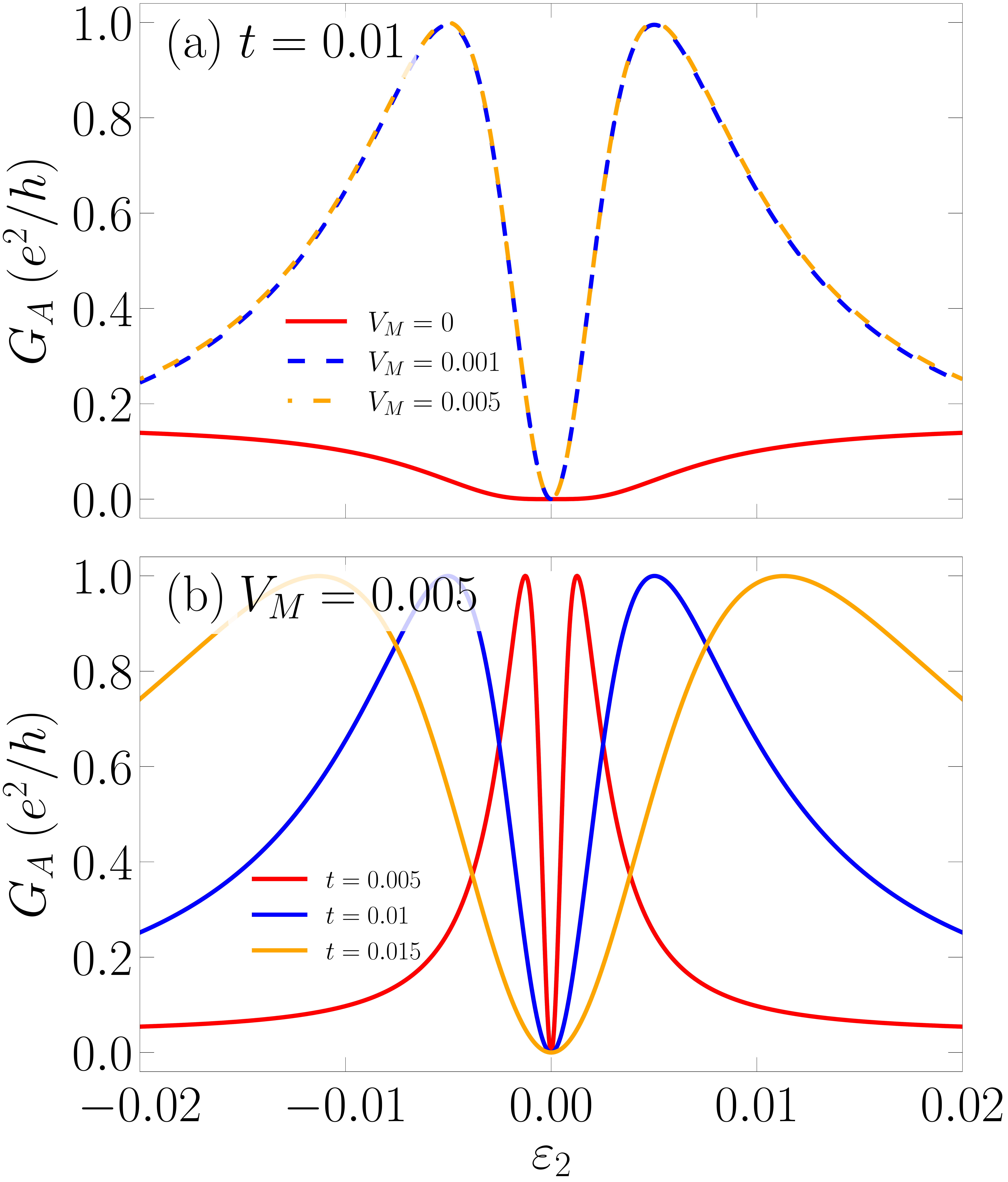}
\caption{
The linear (zero-bias) Andreev conductance $G_A$ as a function of the second quantum dot energy level $\varepsilon_2$ for (a) several values of the coupling $V_M$, assuming $t=0.01$, and (b) for different interdot couplings $t$, assuming $V_M=0.005$.
The other parameters are the same as in Fig.~\ref{figA1_ed20}.
}
\label{figGA}
\end{figure}

In what follows, we briefly investigate the influence of the Majorana mode on the Andreev conductance measured in the N-QD$_1$-S circuit. Figure~\ref{figGA} presents the linear (zero-bias) Andreev conductance $G_A$ as a function of $\varepsilon_2$ calculated for a fixed value of the first dot level position $\varepsilon_1=0$.
For nonzero $V_M$ the optimal Andreev conductance is equal to $e^2/h$, reflecting the fractional nature of the Majorana mode leaking to DQD. Exactly for $\varepsilon_2=0$, the conductance $G_A$ vanishes for all values of $t$ and $V_M$ because of the destructive quantum interference associated with the Fano effect. For a given interdot coupling $t$, the linear Andreev conductance rather weakly depends on the coupling $V_M$ [see Fig.~\ref{figGA}(a)]. However, for a fixed value of coupling $V_M$, $G_A$ is very sensitive to interdot hybridization $t$ [see Fig.~\ref{figGA}(b)].
In the weak inter-dot hopping limit, the conductance $G_A$ achieves its optimal value at small values of the second quantum dot energy level $\varepsilon_2$. On the other hand, when the inter-dot coupling becomes stronger, the position of the energy level $\varepsilon_2$, at which the Andreev conductance is optimal, substantially increases.

\begin{figure}
\centering
\includegraphics[width=0.95\linewidth]{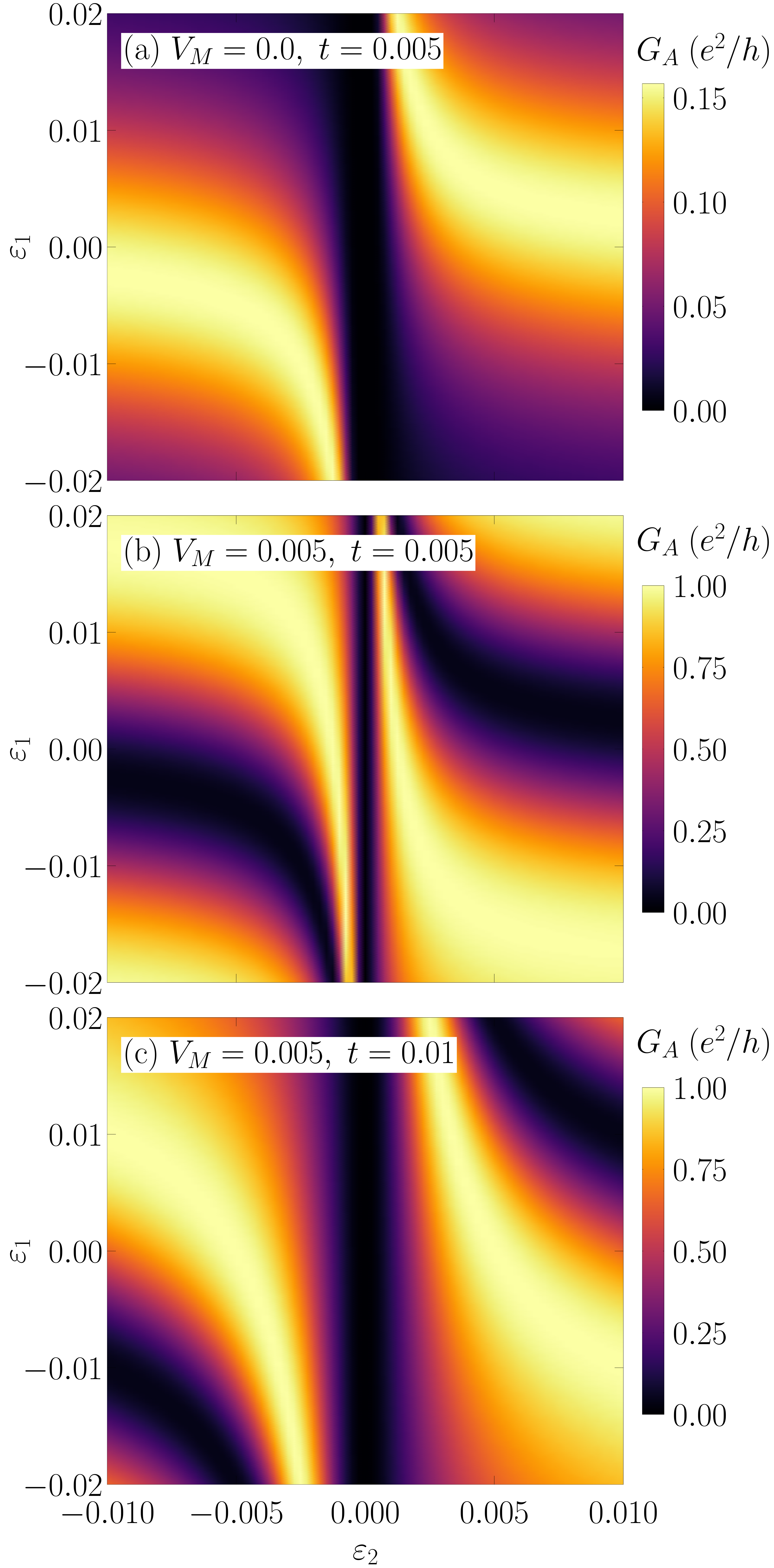}
\caption{
The linear Andreev conductance $G_A$ as a function of the first and second quantum dot energy levels calculated for (a) $V_M = 0, \; t = 0.005$, (b) $V_M = 0.005, \; t = 0.005$ and (c) $V_M = 0.005, \; t = 0.01$.}
\label{figGAed1}
\end{figure}

The linear Andreev conductance is also dependent on the first quantum dot energy level. This situation is shown in Fig.~\ref{figGAed1},
which displays the Andreev conductance as a function of both quantum dot energy levels for three selected values of $t$ and $V_M$.
As can be seen, no matter what the value of $\varepsilon_{1}$ is, $G_A$ always vanishes for $\varepsilon_{2}=0$. This characteristic feature is caused by the destructive quantum interference. Figure~\ref{figGAed1} displays the typical Fano line-shape of such entirely suppressed linear conductance of the considered setup.

\section{The case of the Kondo regime}
\label{sec:correlated}

We now address the situation when the Coulomb correlations in the quantum dots are relevant, $U\neq 0$. In this case new effects emerge that are exclusively due to electron correlations. One of such effects is the Kondo phenomenon \cite{Kondo-1964}, in which a correlated state is formed between the quantum dots and the conduction band electrons of the normal lead \cite{Hewson1993Jan}.
The main goal of this section is to examine the system's transport properties in the Kondo regime, and to shed light on the interplay of Kondo correlations, electron pairing and Majorana modes.

In the large pairing gap limit of the superconductor
the Kondo effect can be induced by the spin-exchange
coupling between QD$_{1}$ and the metallic lead \cite{Bauer-2007}.
Efficiency of such interaction is sensitive to the competition
between the on-dot pairing and the Coulomb repulsion.
The Kondo state would arise if QD$_{1}$ is in the singly occupied configuration
$\left| \sigma \right>$, which takes place when the Coulomb repulsion
dominates over the superconducting proximity effect, $\Gamma_S<U/2$.
Furthermore, the magnitude of the exchange interaction
is substantially enhanced near a transition between the spinful
$\left| \sigma \right>$ and the spinless (BCS-type) $u \left| 0 \right> -v \left| \uparrow\downarrow \right>$ configurations \cite{Zitko-2015b,Domanski-2016}.
Since this parity change of QD$_{1}$  is manifested by a crossing of the in-gap bound states,
therefore the influence of the zero-energy Majorana mode on this crossing would be very important. 

The relationship between the subgap Kondo effect
and the Majorana physics has been previously investigated in heterostructures
with a single correlated quantum dot  \cite{Gorski-2018,Baranski-2019}.
Here, we extend these studies to the double quantum dot system (Fig.~\ref{scheme}),
where the Majorana mode can affect QD$_{1}$ only indirectly, via the second quantum dot.
For efficiency of this two-stage Majorana leakage, one should also take into account
the Kondo effect developed in QD$_{2}$ \cite{Wojcik-2019,Weymann-2020,Majek-2021}.
With lowering the energy scale below the Kondo temperature $T_K$,
the Kondo effect first develops in the first quantum dot, giving rise to enhanced conductance \cite{Goldhaber-Gordon1998Jan}.
However,  for even lower energies, the spin on the second
quantum dot  becomes screened and, because this dot
is not directly coupled to the superconductor-normal lead circuit,
cf. Fig.~\ref{scheme},
the conductance becomes in turn suppressed.
This is known as the two-stage Kondo effect, which
is characterized by a dip in the spectral function of width
proportional to the second-stage Kondo temperature $T^*$
\cite{Pustilnik2001Nov,Cornaglia2005Feb,Chung2008Jan,Wojcik2015Apr}.
In what follows, we analyze a subtle interplay of these effects
and discuss their signatures observable in the charge transport properties of N-QD$_{1}$-S circuit.

\begin{figure}[t]
	\includegraphics[width=0.85\columnwidth]{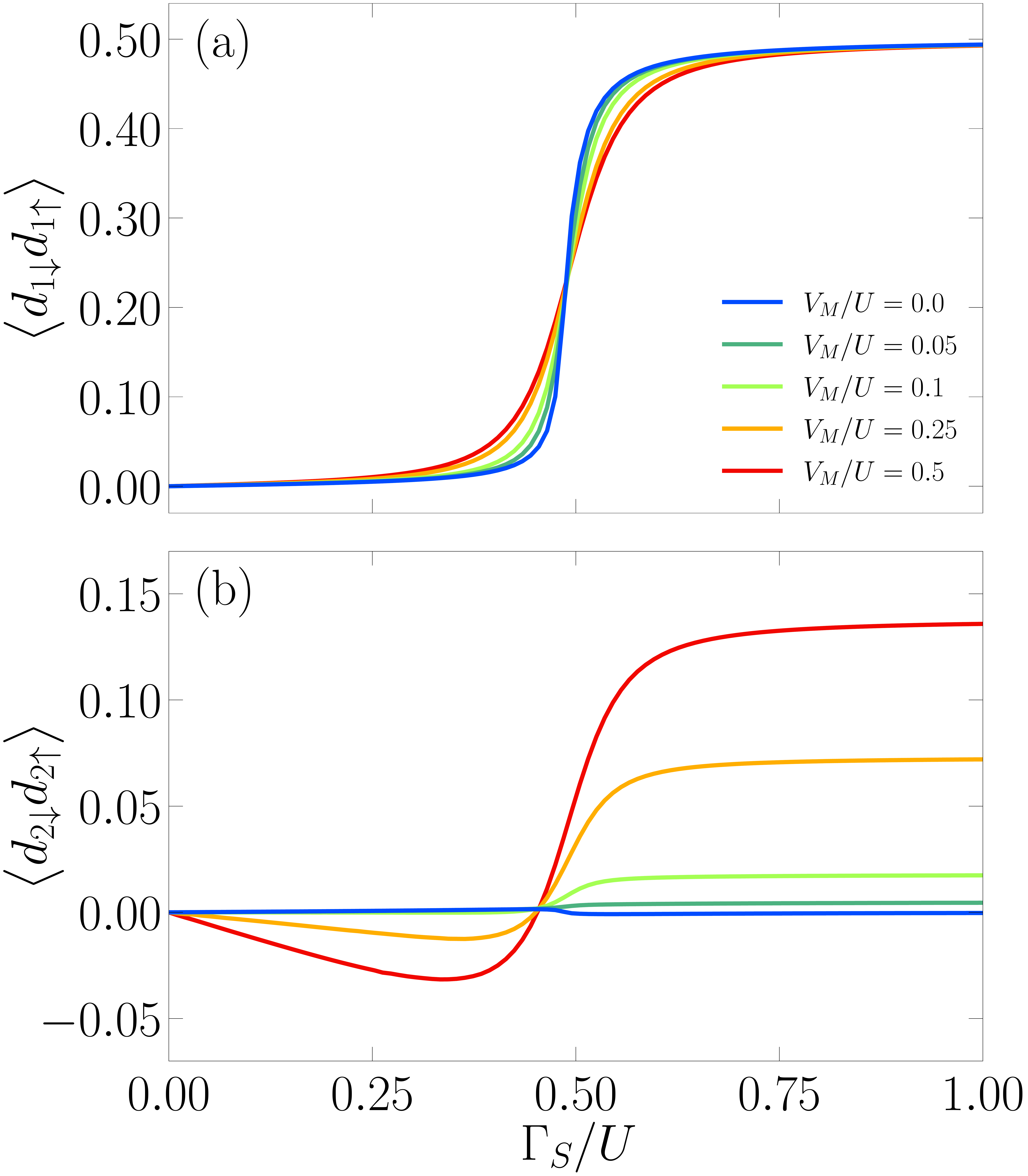}
	\caption{Superconducting pairing correlations
		of (a) the first and (b) the second quantum dot
		calculated as a function of the coupling to superconductor $\Gamma_S$
		for different values of $V_M$, as indicated. 
		The other parameters are: $t/U = 0.025$, $\Gamma_N/U=0.1$,
		$\varepsilon_1=\varepsilon_2=-U/2$ and 
		$U=0.1$ in units of bandwidth.}
	\label{SCcorrelations}
\end{figure}

To reliably treat the low and high energy features originating from the 
correlation effects and the proximity-induced pairing we perform the 
density-matrix numerical renormalization group calculations 
\cite{Wilson1975Oct,Andreas_broadening2007,Bulla2008Apr,NRG_code,Toth2008}.
We impose the band discretization parameter $\Lambda = 2 - 2.2$, keeping at least 
$2000$ states during the iterative diagonalization. The microscopic model 
(\ref{eq:H}) satisfies the $Z(2)$ parity, which we implement to
facilitate the numerical calculations. We perform the computations for the 
particle-hole symmetric case, $\varepsilon_{1}=\varepsilon_{2} =-U/2$,
assuming $U/D=0.1$, $\Gamma_{N}/D=0.01$.

\subsection{Superconducting pairing correlations}
\label{sec:pairing}

Before analyzing the spectral and transport properties of the system,
let us examine the behavior of the superconducting pairing correlations
induced in the quantum dots. These correlations are presented
in Fig.~\ref{SCcorrelations} as a function of the coupling to superconductor
$\Gamma_S$ for selected values of the coupling to Majorana wire $V_M$.
First of all, we note that the electron correlations ($U>0$)
strongly modify the behavior of 
$\left< d_{i\downarrow}d_{i\uparrow}\right> $
as compared to the uncorrelated case, cf. Figs.~\ref{fig_pair} and \ref{SCcorrelations}.
In the absence of coupling to Majorana wire
the pairing in the first dot exhibits  
the behavior, which is typical for superconductor-proximized quantum dots \cite{Domanski-2016};
$\left< d_{1\downarrow}d_{1\uparrow}\right> $
is suppressed in the doublet state, $\Gamma_S<U/2$,
and approaches $\left< d_{1\downarrow}d_{1\uparrow}\right> =1/2$
on the singlet side of the transition, i.e. for $\Gamma_S>U/2$.
Moreover, we find that in the strongly correlated case
$\left< d_{1\downarrow}d_{1\uparrow}\right>$
very weakly depends on $V_M$, see Fig.~\ref{SCcorrelations}(a).
This is contrary to the behavior of  $\left< d_{2\downarrow}d_{2\uparrow}\right> $,
which becomes enhanced when the coupling to Majorana wire is turned on, see Fig.~\ref{SCcorrelations}(b).
Furthermore, the dependence of the pairing induced
on the second dot on $\Gamma_S$ and $V_M$ is more subtle.
For $\Gamma_S<U/2$, $\left< d_{2\downarrow}d_{2\uparrow}\right> <0 $,
while for $\Gamma_S>U/2$, $\left< d_{2\downarrow}d_{2\uparrow}\right> >0 $.
The negative value on the doublet side of the transition
can be associated with the quantum interference
and the polarization of the second quantum dot
that is induced by the Majorana mode.

\subsection{Majorana features in spectral functions}
\label{sec:Kondo_spectra}
 
 \begin{figure}[b]
 	\includegraphics[width=0.85\columnwidth]{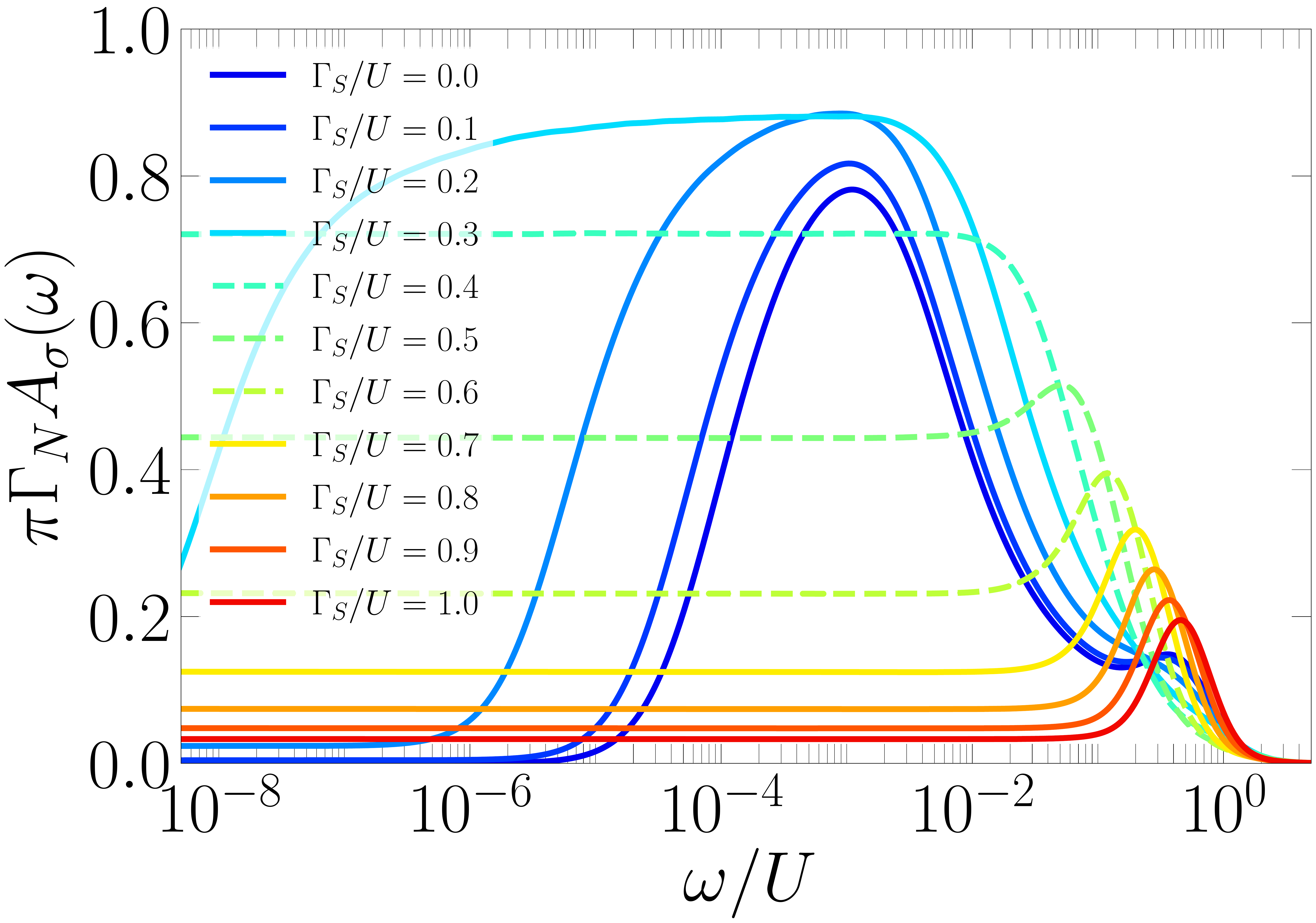}
 	\caption{The normalized spectral function $\pi\Gamma_N A_{\sigma}(\omega)$
 		of the half-filled central quantum dot obtained for
 		various couplings to superconductor $\Gamma_S$, as indicated, while $V_M=0$.
 		The other parameters are the same as in Fig.~\ref{SCcorrelations}.
 		The spectral  function is symmetric with respect to the Fermi energy,
 		therefore only positive energies are shown.
 		Note also the logarithmic energy scale.
 		The dashed curves correspond to the values of $\Gamma_S$ analyzed in
 		Figs.~\ref{Fig_spectr_up} and \ref{Fig_spectr_down}.}
 	\label{AGammaS}
 \end{figure}
 
 \begin{figure}[t]
 	\includegraphics[width=0.85\columnwidth]{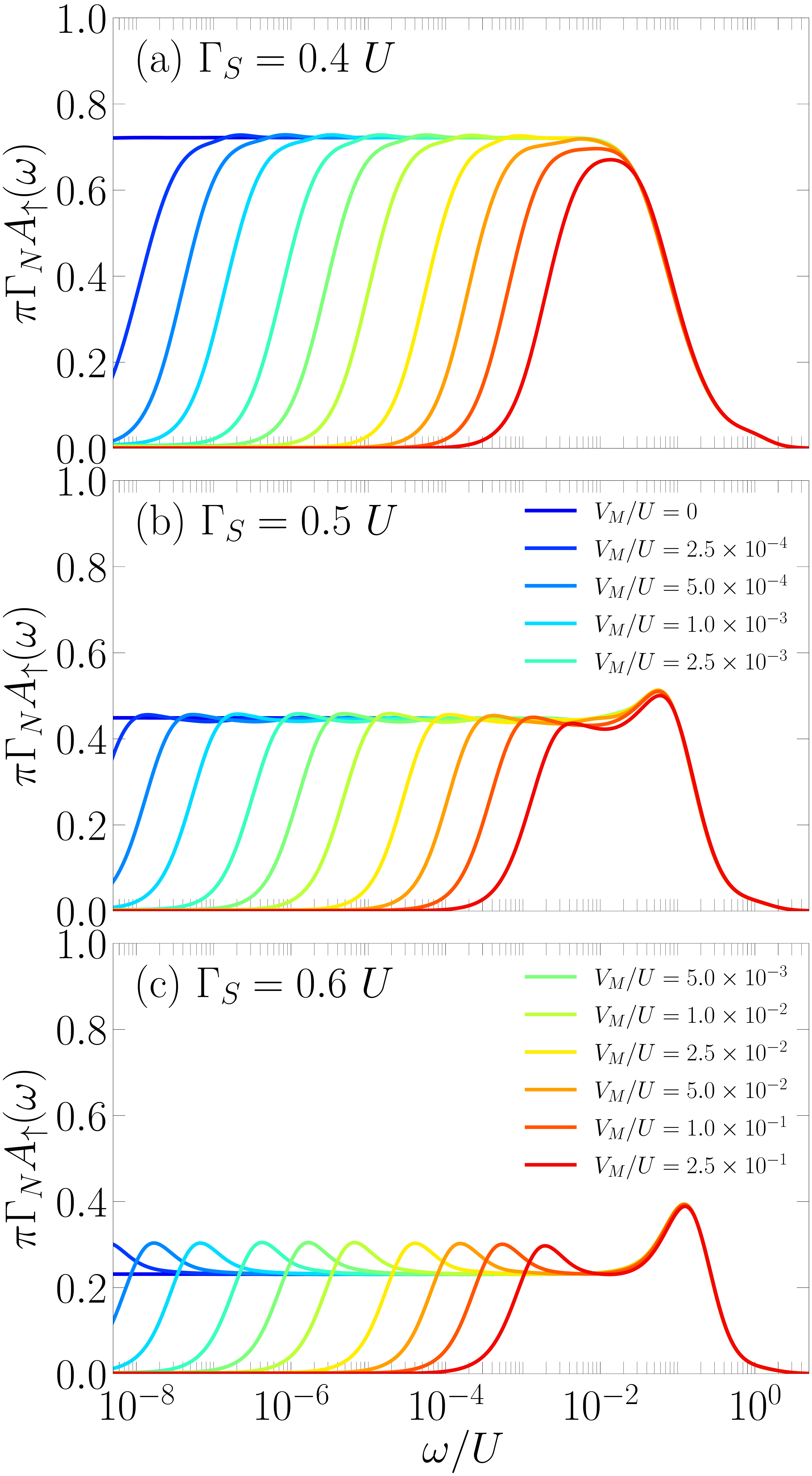}
 	\caption{The normalized spin-up spectral function $\pi\Gamma_N A_{\uparrow}(\omega)$
 		calculated for different couplings $V_M$ to the topological wire, as indicated.
 		The top, middle and bottom panels refer to $\Gamma_{S}/U=0.4$, $0.5$ and $0.6$, respectively.
 		The other parameters are the same as in Fig.~\ref{SCcorrelations}.}
 	\label{Fig_spectr_up}
 \end{figure}
 
 \begin{figure}[t]
 	\includegraphics[width=0.85\columnwidth]{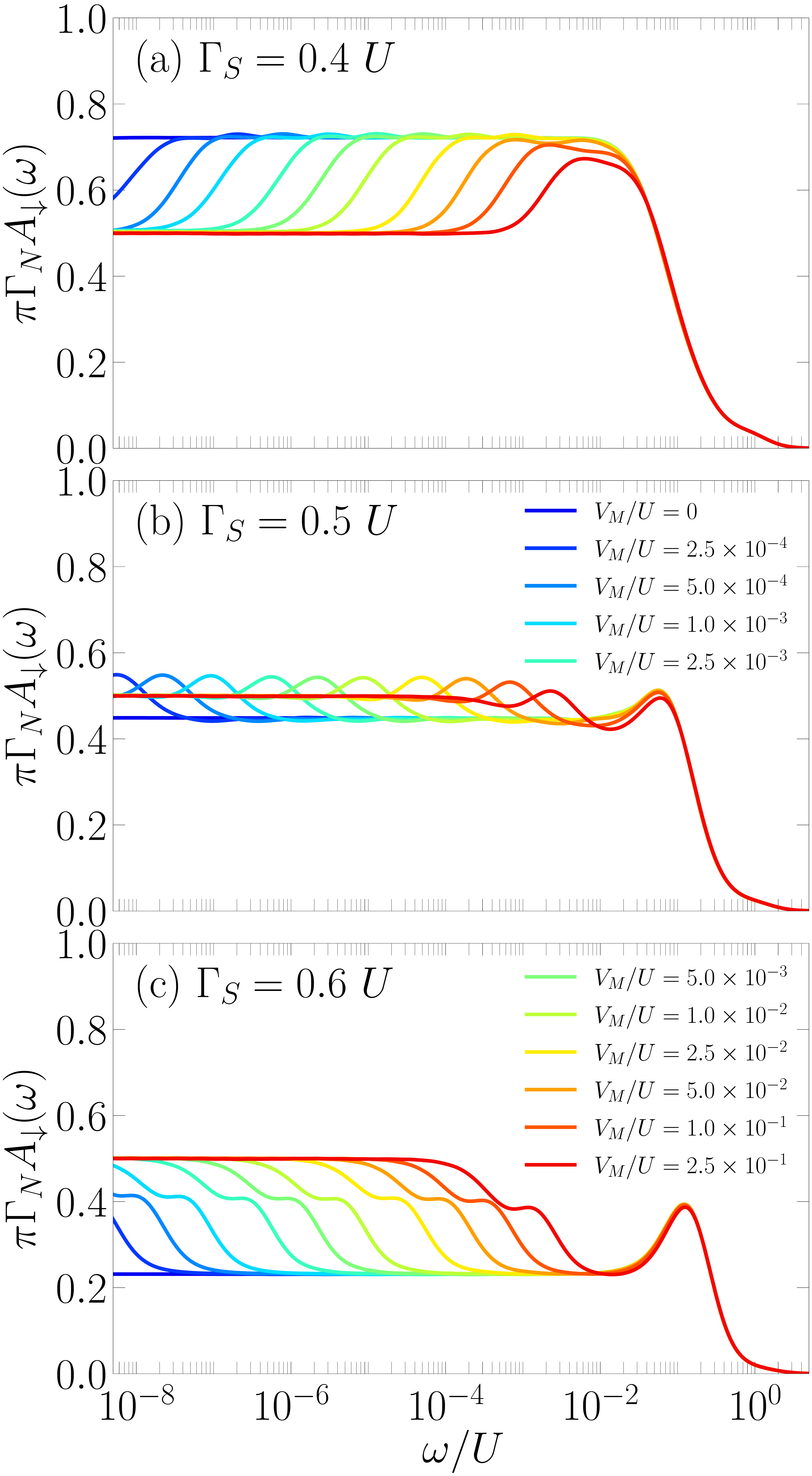}
 	\caption{The normalized spectral function for the spin-down electrons
 		$\pi\Gamma_N A_{\downarrow}(\omega)$
 		calculated for the same model parameters as in Fig.~\ref{Fig_spectr_up}.}
 	\label{Fig_spectr_down}
 \end{figure}
 
 \begin{figure*}
 	\includegraphics[width=1.6\columnwidth]{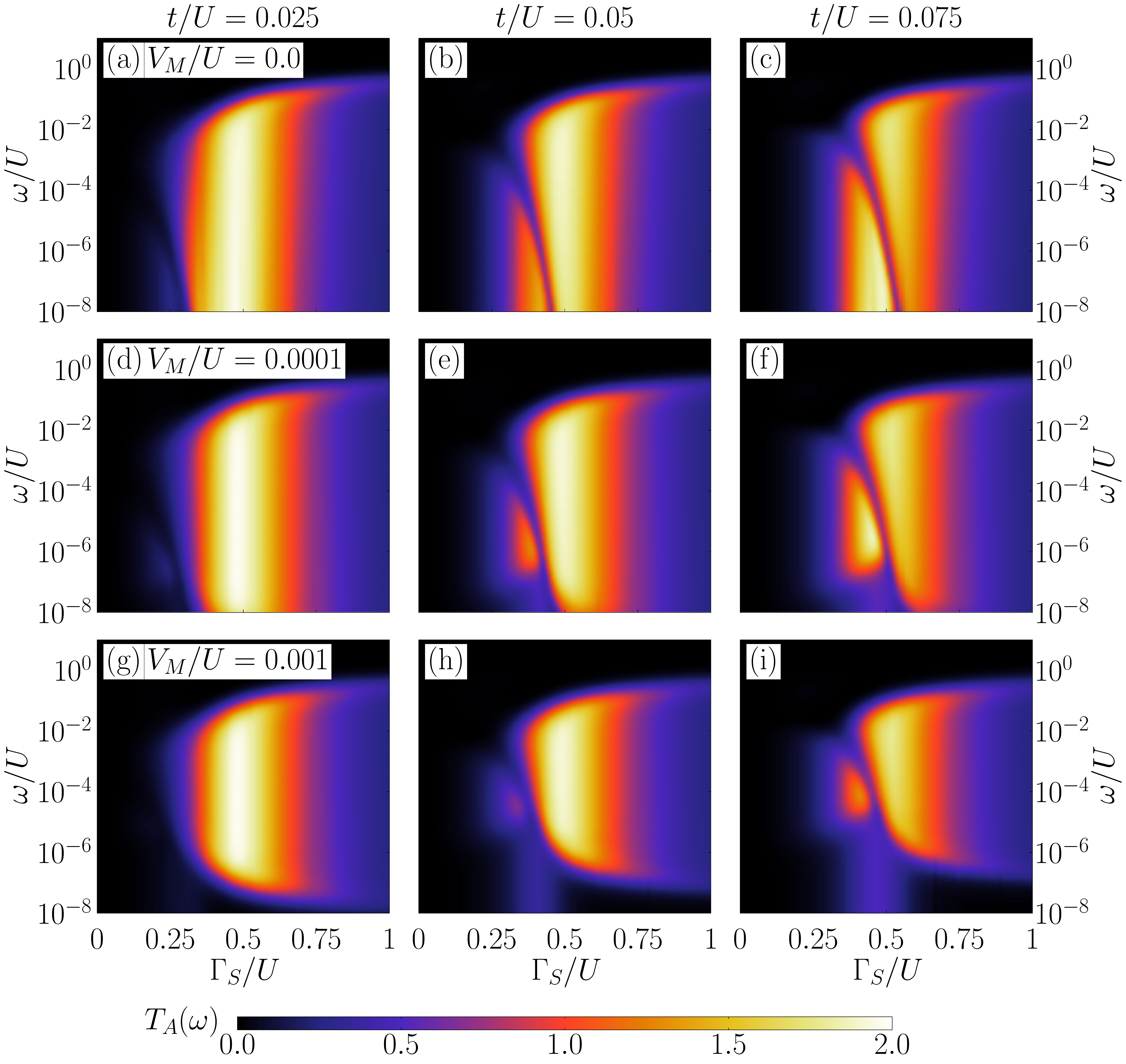}
 	\caption{The Andreev transmittance $T_A(\omega)$
 		as a function of $\Gamma_{S}/U$ and energy $\omega$
 		calculated for several values of the coupling to Majorana wire $V_{M}$
 		and interdot hopping $t$ (as indicated) obtained
 		by NRG. The other parameters are the same as in Fig.~\ref{SCcorrelations}.}
 	\label{Andreev_transmittance}
 \end{figure*}
 
To elucidate the role of the Majorana mode on the spectral density,
we first analyze the dependence of the spectral function
of the half-filled quantum dots on the strength of coupling  to the superconductor in the absence of
coupling to Majorana wire. The normalized spectral function
for spin $\sigma$ (note that for $V_M=0$ the spin degeneracy is restored)
is presented in Fig.~\ref{AGammaS}.
Let us recall that in the absence of the interdot coupling
($t=0$) the parity change of QD$_{1}$ from
$\left| \sigma \right>$ to ${u \left| 0 \right> -v \left| \uparrow\downarrow \right>}$
configuration occurs at $\Gamma_{S}=U/2$.
The subgap Kondo effect can thus be realized for $\Gamma_{S}< U/2$.
On the other hand, for finite $t$ but $\Gamma_S = 0$, one observes the behavior 
typical for the two-stage Kondo effect  \cite{Cornaglia2005Feb,Chung2008Jan,Wojcik2015Apr},
with suppressed spectral density for $\omega < T^*$.
In the case of finite $t$ and $\Gamma_S$,
the interplay of the on-dot pairing and Kondo correlations gives rise to 
an interesting behavior of the spectral function.
The proximity induced pairing affects $A_\sigma(\omega)$
by lifting the second-stage of screening and giving
rise to a finite value of $A_\sigma(0)$, which is the largest
around the transition between the doublet and BCS singlet state,
$\Gamma_S \approx U/2$, see Fig.~\ref{AGammaS}.

Let us now examine the behavior of the spectral functions 
in the case of finite coupling to topological nanowire,
focusing on values of $\Gamma_S$ around the singlet-doublet 
transition, cf. the dashed lines in Fig.~\ref{AGammaS}.
The corresponding spin-resolved spectral functions $A_\sigma (\omega)$
obtained for various couplings $V_M$ to the topological nanowire are presented in
Figs.~\ref{Fig_spectr_up} and \ref{Fig_spectr_down}.
First of all, we note that the numerical results
reveal qualitative differences between the spectra of the spin-up and spin-down electrons,
which should be attributed to the presence of Majorana mode.
Such influence is predominantly manifested in the low-energy sector,
therefore our plots are presented in a logarithmic scale.
The upper panels of the figures correspond 
to the situation when $\Gamma_{S}< U/2$
and the subgap Kondo effect can be realized,
therefore $A_\sigma(\omega)$ has a relatively large weight 
at the Fermi energy for $V_M=0$. On the other hand,
the lower panels display the case of  $\Gamma_{S}>U/2$,
where the weight is much reduced.
However, once $V_M$ is finite, these features may be completely changed.

Consider first the case of the spin-up spectrum of QD$_{1}$ shown in Fig. \ref{Fig_spectr_up}.
Each panel of the figure clearly displays a substantial depletion of 
the low-energy states driven by the Majorana mode. In the weak coupling limit, 
$V_{M}\ll U$, such destructive influence appears in the form of 
interferometric dip formed on top of either the subgap Kondo peak 
[in the strongly correlated limit $\Gamma_{S}/U=0.4$, see Fig.~\ref{Fig_spectr_up}(a)],
or on a flat background between the Andreev peaks [corresponding to the BCS-type configuration 
$\Gamma_{S}/U = 0.6$, see Fig.~\ref{Fig_spectr_up}(c)].
Upon increasing the coupling $V_{M}$, we observe 
the development of a molecular structure, where the Kondo effect gets suppressed. 
In other words, in the subgap Kondo regime, 
one can see a supporting influence of the coupling to the Majorana mode
on the second-stage Kondo effect,  i.e. there occurs a restoration of this effect for finite $V_M$.
Moreover, the stronger the coupling $V_M$ is, the higher the second-stage Kondo 
temperature $T^*$ becomes. This picture resembles itself through the 
singlet-doublet transition, lowering the maximum of the Kondo peak,
and finally suppressing the spectral function in the BCS regime,
where the Kondo effect does not develop.
Thus, the coupling to the topological nanowire has a rather destructive influence on the 
low-energy behavior of the spin-up spectral function.

In contrast to this tendency, the Majorana mode has a more complex impact
on the spin-down spectral function shown in Fig.\ \ref{Fig_spectr_down}.
In the subgap Kondo regime, finite coupling 
to the topological superconductor suppresses the low-energy spectral density by 
around $1/3$ to the universal value of $A_\downarrow (0) = 1/2 \pi \Gamma_N$.
However, the picture changes when the BCS configuration is formed
and the influence of $V_M$ becomes constructive.
For weak couplings, $V_{M}\ll U$, we notice a buildup of the narrow peak at zero-energy 
due to the leaking Majorana mode. Upon increasing $V_{M}$, the value of 
the spectral function $A_{\downarrow}(\omega=0)$ saturates, while the zero-energy 
peak gradually broadens. For larger values of $V_{M}$, this zero-energy 
quasiparticle state dominates over all other subgap features.
The  corresponding spectral function develops then its universal shape with the characteristic value 
$A_\downarrow (0) = 1/2 \pi \Gamma_N$, regardless of the Coulomb potential $U$.
Summarizing, the attachment of topological superconductor to the spin-$\downarrow$ 
electrons of QD$_{2}$ induces the zero-energy state in spin-$\downarrow$ sector 
of QD$_{1}$ with the fractional value of the low-energy spectral function.

\subsection{Subgap charge transport}
\label{sec:Kondo_transport}

Empirical detection of the quasiparticle spectra of QD$_{1}$
can be done with the use of the charge tunneling spectroscopy.
In the subgap region the only transport channel
is contributed by the particle-to-hole (Andreev) scattering mechanism,
which combines the both spin components.
The subgap tunneling spectroscopy would hence
provide an important information about the
convoluted  spin-up and spin-down spectra. 

At low temperatures the Andreev differential conductance
$G_{A}(V)=dI_{A}(V)/dV$ can be approximated by
the Andreev  transmittance $T_A(\omega)$ taken at $\omega=eV$,
${\rm lim}_{T\rightarrow 0}G_{A}(V)=\frac{2e^{2}}{h}T_{A}(\omega=eV)$.
We display $T_A(\omega)$ in  Fig.~\ref{Andreev_transmittance}
for several values of the inter-dot hopping $t$
(panels from left to right) and different couplings $V_{M}$ (panels from top to bottom).
In general, we notice, that the Andreev conductance
achieves its optimal value $4e^{2}/h$ near $\Gamma_{S} \approx U/2$ \cite{Baranski-2013}.
This situation corresponds to the ground state changeover of QD$_{1}$,
whose in-gap bound states tend to cross each other
and simultaneously the subgap Kondo peak is enhanced (on the doublet side).  

As regards the Majorana mode,
its influence shows up by suppression of the zero-bias conductance
(see the middle and bottom panels in Fig.~\ref{Andreev_transmittance}).
This effect comes merely from the destructive quantum interference of the Majorana mode
on the spin-$\uparrow$ sector of QD$_{1}$ spectrum.
Additionally, upon increasing the inter-dot hopping $t$,
we observe the signatures of emerging molecular bound states.
In particular, this is visible by a dark region splitting the optimal conductance.
We note that for a more precise and direct observability of all the spin-resolved
spectra of QD$_{1}$ one could use the spin-polarized Andreev spectroscopy
of  bound states \cite{Calzona-2021}.

\section{Summary}
\label{sec:summary}

We have studied the influence of the Majorana mode transmitted to 
the double quantum dot side-attached to a topological 
superconducting nanowire. This setup could be probed by tunneling 
spectroscopy through a circuit with the outer quantum dot (QD$_{1}$) 
sandwiched between the normal and superconducting leads. 
Proximity of DQD to the superconducting reservoir induces  
the in-gap bound states, whose complex structure depends on
the inter-dot coupling, as recently revealed in Ref.~\cite{Baumgartner-2021}.
Here, we have inspected the modification of these conventional bound  
states by the topological superconducting nanowire hosting 
the Majorana boundary mode. This issue might be important for 
designing the braiding protocols of Majorana quasiparticles.

In the absence of correlation effects we have derived analytical 
expressions for the resulting in-gap bound states, identifying  
the trivial Andreev quasiparticle branches and the additional 
structures induced by the Majorana mode. We have shown that 
these features manifest themselves differently in each spin sector. 
In particular, the zero-energy quasiparticle state induced at 
QD$_{1}$ appears in a form of destructive/constructive
interference pattern imprinted on the $\uparrow$/$\downarrow$ spin 
sectors. Since the subgap charge transport mixes both spins through 
the particle-to-hole scattering, the resulting tunneling 
characteristics are predominantly affected by these destructive signatures.
We have discussed them in detail, considering the linear subgap 
Andreev conductance.

We have also extended our considerations to the strongly correlated 
nanostructure, treating the competition of the superconducting 
proximity effect with the repulsive Coulomb interactions by 
the numerical renormalization group technique. Focusing on 
the half-filled quantum dots, we have studied both the Kondo regime,
when the Coulomb repulsion dominates over the superconducting 
proximity effect, as well as the opposite limit, where 
the superconducting proximity-induced pairing surpasses
the Kondo correlations. We have revealed qualitative differences
in these two regimes, evidenced in the spin-resolved spectral functions. 
We have predicted that the optimal Andreev conductance would occur  
near a crossover between the singly-occupied doublet to the  
BCS-type configurations of QD$_{1}$. Moreover, our numerical results obtained 
for the subgap spectroscopy indicate that the Majorona mode 
strongly suppresses the zero-bias conductance,
owing to its destructive influence on spin-$\uparrow$
sector of the outer quantum dot. 

\begin{acknowledgments}
This work was supported by the National Science Centre
in Poland through the Project No. 2018/29/B/ST3/00937.
The computing time at the Pozna\'n Supercomputing and Networking Center is acknowledged.
\end{acknowledgments}

\appendix
\section{Role of polarization}
\label{polarization}

\begin{figure}[t] 
\centering
\includegraphics[width=1\linewidth]{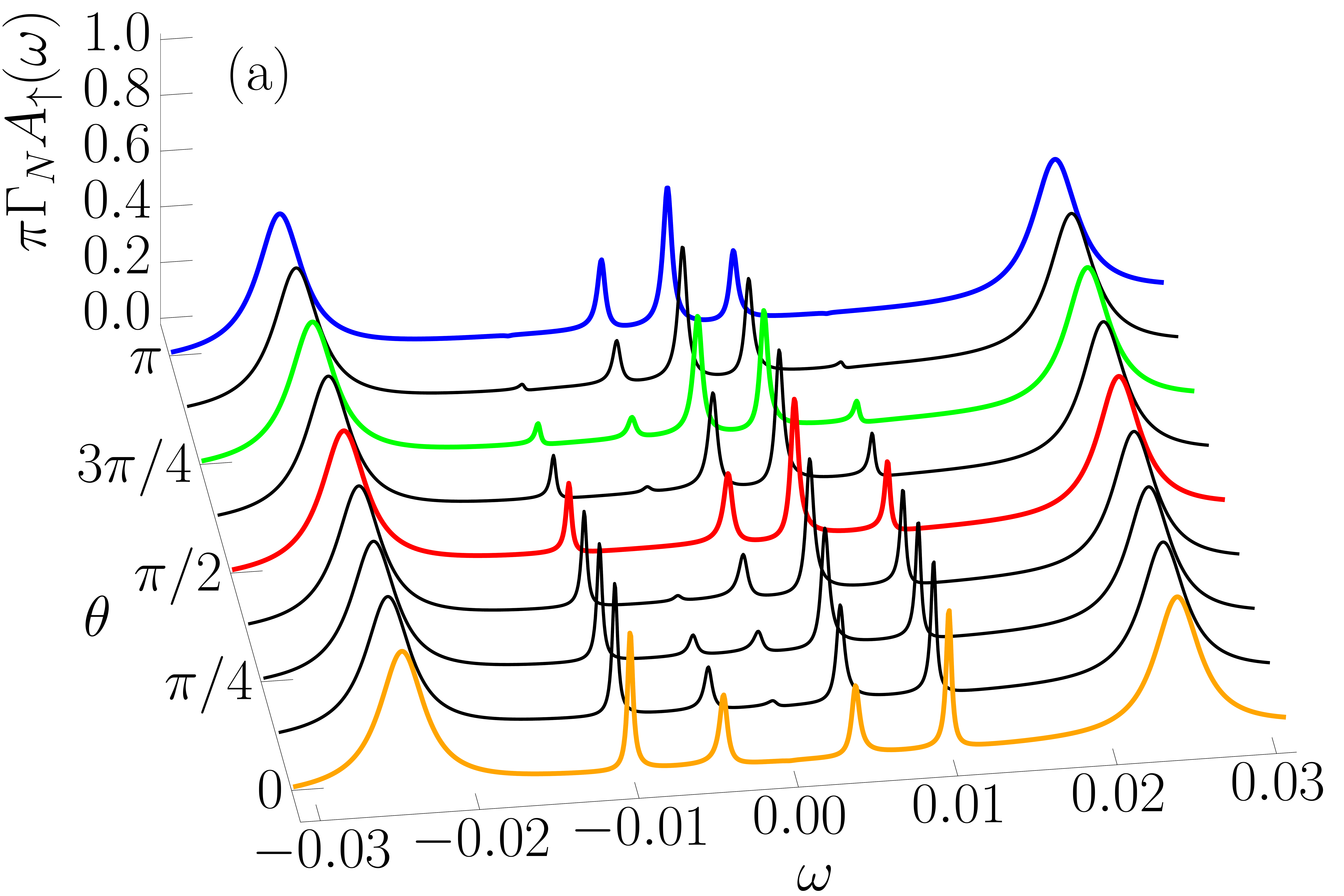}
\includegraphics[width=1\linewidth]{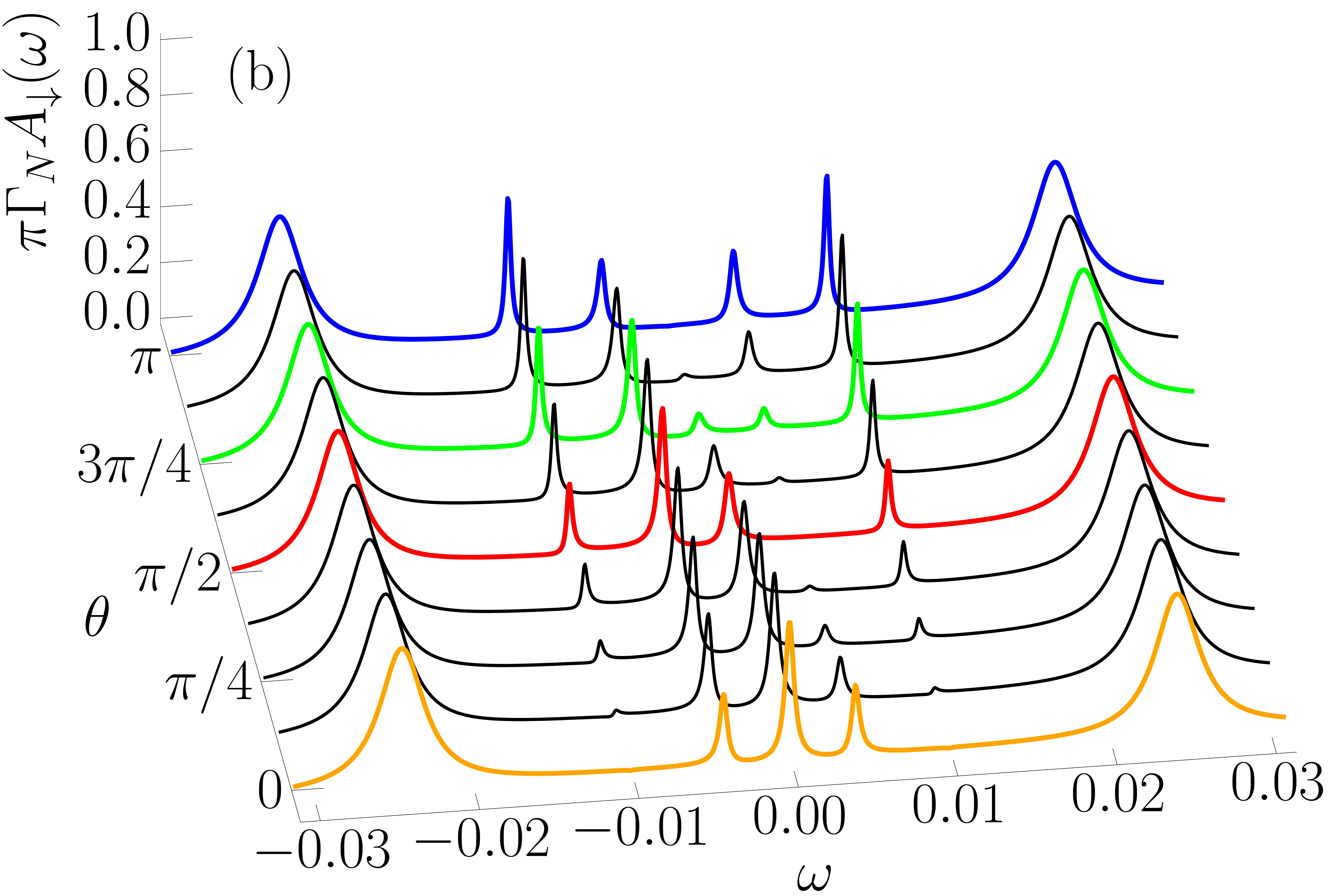}
\caption{The normalized spectral function $\pi\Gamma_N A_{\sigma}(\omega)$  for
	(a) the spin-up and (b) spin-down components as a function of the canting angle $\theta$.
	Results are obtained for the case of uncorrelated quantum dots
	using the following model parameters:
	$\varepsilon_{1}=\varepsilon_{2}=0$, 
	$V_M=0.005$, $t=0.01$, $\Gamma_{S}=0.02$,  $\Gamma_N=0.002$.}
\label{fig_polar}
\end{figure}

In this appendix we extend the discussion of 
the properties of our hybrid structure in the uncorrelated case
by assuming that both spins of QD$_{2}$ are coupled to the Majorana nanowire
though with different amplitudes $V_{M\sigma}$.
Hoffman {\it et al.} \cite{Klinovaja-2017} have shown that
the spin-dependent tunneling amplitudes between Majorana modes
and the quantum dot depend on the quantum dot distance to the topological section.
Moreover, such finite polarization of quantum dot attached to the topological nanowire
has been proposed as a suitable tool for probing the topology of Majorana
wave function \cite{Klinovaja-2017, Prada-2017,Deng-2018,Schuray-2018}.
To account for this effect, we use the following parametrization of 
the spin-dependent amplitudes
\begin{eqnarray}
V_{M \uparrow} & = & V_M \sin{\left( \frac{\theta}{2}\right)} \\
V_{M \downarrow} & = &V_M \cos{\left( \frac{\theta}{2}\right)}
\label{Vmsigma}
\end{eqnarray}
in terms of the canting angle $\theta$ \cite{Prada-2017},
corresponding to the rotation around the $y$-axis.

Figure~\ref{fig_polar} presents the variation of the spectral function $A_{\sigma}(\omega)$
of each spin sector with respect to the canting angle $\theta$.
One can clearly see the influence of $\theta$ on spectral weights
of the zero-energy peak, and also on the finite-energy quasiparticles.
In the fully polarized cases, $V_{M\uparrow}=0$ and $V_{M\downarrow}=V_M$
($V_{M\uparrow}=V_M$ and $V_{M\downarrow}=0$),
corresponding to $\theta=0$ ($\theta=\pi$),
we obtain the  Majorana peak of spin-$\downarrow$ (spin-$\uparrow$)
spectral function $A_{\downarrow}(0)=1/2 \pi \Gamma_N$ [$A_{\uparrow}(0)=1/2 \pi \Gamma_N$].
On the other hand, in the unpolarized case, i.e. $\theta=\frac{\pi}{2}$,
the zero-energy Majorana peak is identical in both spin sectors,
$A_{\uparrow}(0)=A_{\downarrow}(0)=1/4 \pi \Gamma_N$.
Additionally, we notice,that the canting angle neither affects the energies
of the Andreev bound states $\varepsilon_{ADi}^{\pm}$ [cf. Eq.~(\ref{EA0})]
nor the quasiparticle states $\varepsilon_{MDi}^{\pm}$ [cf. Eq.~(\ref{EAM})].
Its influence is visible merely in the spectral densities $A_{\sigma}(\varepsilon_{AD2}^{\pm})$ 
and $A_{\sigma}(\varepsilon_{MDi}^{\pm})$.

\begin{figure}[t]
	\centering
	\includegraphics[width=0.95\linewidth]{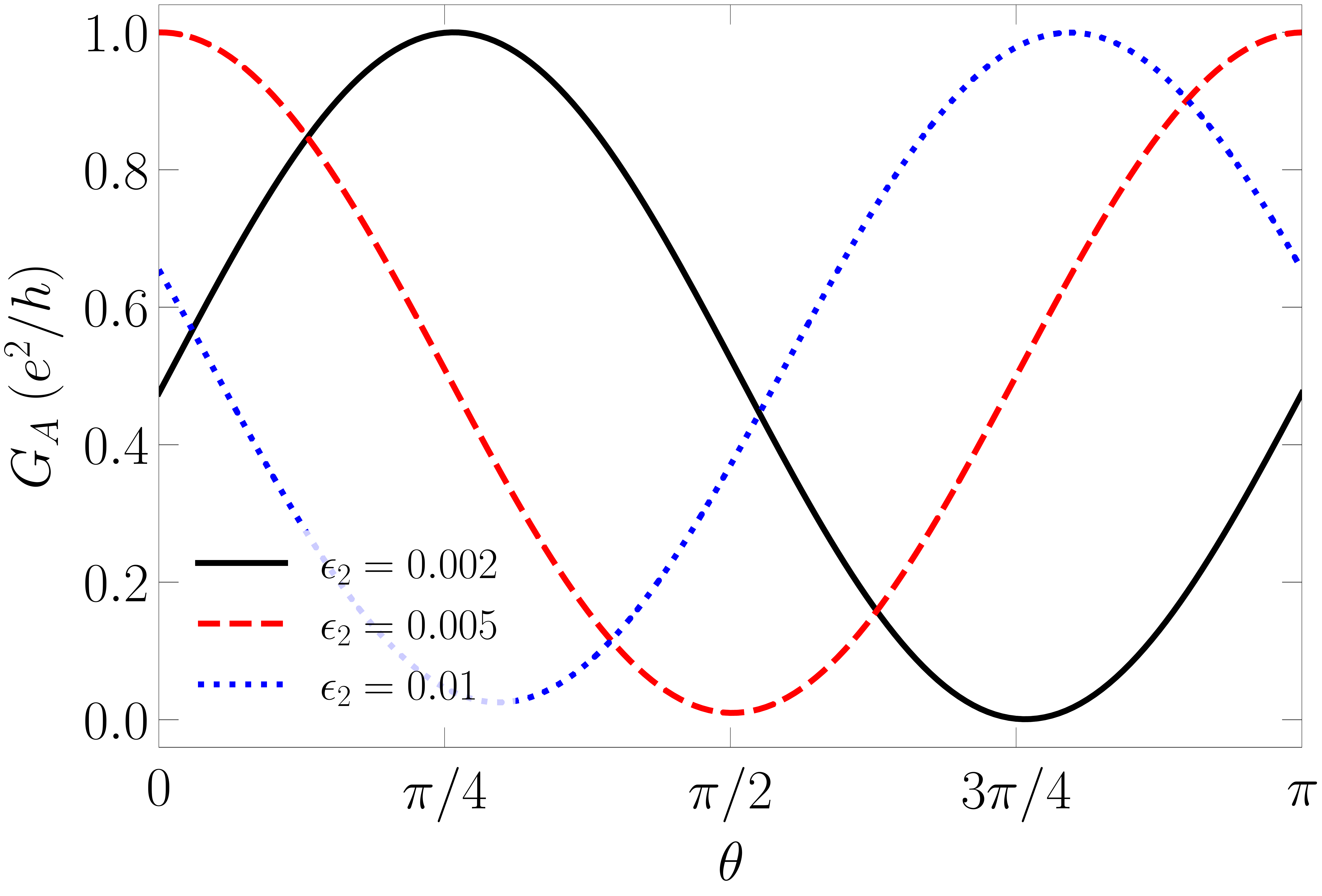}
	\caption{The linear Andreev conductance $G_A$  as a function of the spin 
		canting angle $\theta$ obtained for 
		$\varepsilon_{1}=0$, $V_M=0.005$, $t=0.01$, 
		$\Gamma_{S}=0.02$,  $\Gamma_N=0.002$ and several
		values of $\varepsilon_{2}$, as indicated.}
	\label{fig_GApolar}
\end{figure}

We have shown that $\theta$ affects the quasiparticle weights
of the diagonal spectral functions, but additionally it also strongly modifies
the off-diagonal spectral functions.
Such influence would be empirically observable in the subgap tunneling conductance.
Figure~\ref{fig_GApolar} presents the linear Andreev conductance $G_A$
as a function of the canting angle. We notice that $\theta$ leads to the variation of $G_A$ 
up to the maximal value  $G_{A}^{max}=e^2/h$.
For the full polarization of $V_{M\sigma}$  (i.e.\ $\theta=0$ or $\theta=\pi$),
we obtain identical values of the linear conductance,
because then the particle and hole degrees of freedom equally
participate in the Andreev scattering mechanism. 


\begin{thebibliography}{94}%
	\makeatletter
	\providecommand \@ifxundefined [1]{%
		\@ifx{#1\undefined}
	}%
	\providecommand \@ifnum [1]{%
		\ifnum #1\expandafter \@firstoftwo
		\else \expandafter \@secondoftwo
		\fi
	}%
	\providecommand \@ifx [1]{%
		\ifx #1\expandafter \@firstoftwo
		\else \expandafter \@secondoftwo
		\fi
	}%
	\providecommand \natexlab [1]{#1}%
	\providecommand \enquote  [1]{``#1''}%
	\providecommand \bibnamefont  [1]{#1}%
	\providecommand \bibfnamefont [1]{#1}%
	\providecommand \citenamefont [1]{#1}%
	\providecommand \href@noop [0]{\@secondoftwo}%
	\providecommand \href [0]{\begingroup \@sanitize@url \@href}%
	\providecommand \@href[1]{\@@startlink{#1}\@@href}%
	\providecommand \@@href[1]{\endgroup#1\@@endlink}%
	\providecommand \@sanitize@url [0]{\catcode `\\12\catcode `\$12\catcode
		`\&12\catcode `\#12\catcode `\^12\catcode `\_12\catcode `\%12\relax}%
	\providecommand \@@startlink[1]{}%
	\providecommand \@@endlink[0]{}%
	\providecommand \url  [0]{\begingroup\@sanitize@url \@url }%
	\providecommand \@url [1]{\endgroup\@href {#1}{\urlprefix }}%
	\providecommand \urlprefix  [0]{URL }%
	\providecommand \Eprint [0]{\href }%
	\providecommand \doibase [0]{http://dx.doi.org/}%
	\providecommand \selectlanguage [0]{\@gobble}%
	\providecommand \bibinfo  [0]{\@secondoftwo}%
	\providecommand \bibfield  [0]{\@secondoftwo}%
	\providecommand \translation [1]{[#1]}%
	\providecommand \BibitemOpen [0]{}%
	\providecommand \bibitemStop [0]{}%
	\providecommand \bibitemNoStop [0]{.\EOS\space}%
	\providecommand \EOS [0]{\spacefactor3000\relax}%
	\providecommand \BibitemShut  [1]{\csname bibitem#1\endcsname}%
	\let\auto@bib@innerbib\@empty
	\bibitem [{\citenamefont {van~der Wiel}\ \emph {et~al.}(2002)\citenamefont
		{van~der Wiel}, \citenamefont {De~Franceschi}, \citenamefont {Elzerman},
		\citenamefont {Fujisawa}, \citenamefont {Tarucha},\ and\ \citenamefont
		{Kouwenhoven}}]{vanderWiel-2002}%
	\BibitemOpen
	\bibfield  {author} {\bibinfo {author} {\bibfnamefont {W.~G.}\ \bibnamefont
			{van~der Wiel}}, \bibinfo {author} {\bibfnamefont {S.}~\bibnamefont
			{De~Franceschi}}, \bibinfo {author} {\bibfnamefont {J.~M.}\ 
			\bibnamefont
			{Elzerman}}, \bibinfo {author} {\bibfnamefont {T.}~\bibnamefont 
			{Fujisawa}},
		\bibinfo {author} {\bibfnamefont {S.}~\bibnamefont {Tarucha}}, \ and\
		\bibinfo {author} {\bibfnamefont {L.~P.}\ \bibnamefont {Kouwenhoven}},\
	}\bibfield  {title} {\enquote {\bibinfo {title} {Electron transport through
				double quantum dots},}\ }\href {\doibase 
				10.1103/RevModPhys.75.1} {\bibfield
		{journal} {\bibinfo  {journal} {Rev. Mod. Phys.}\ }\textbf {\bibinfo 
		{volume}
			{75}},\ \bibinfo {pages} {1--22} (\bibinfo {year} 
			{2002})}\BibitemShut
	{NoStop}%
	\bibitem [{\citenamefont {Hanson}\ \emph {et~al.}(2007)\citenamefont 
	{Hanson},
		\citenamefont {Kouwenhoven}, \citenamefont {Petta}, \citenamefont 
		{Tarucha},\
		and\ \citenamefont {Vandersypen}}]{Hanson-2007}%
	\BibitemOpen
	\bibfield  {author} {\bibinfo {author} {\bibfnamefont {R.}~\bibnamefont
			{Hanson}}, \bibinfo {author} {\bibfnamefont {L.~P.}\ \bibnamefont
			{Kouwenhoven}}, \bibinfo {author} {\bibfnamefont {J.~R.}\ 
			\bibnamefont
			{Petta}}, \bibinfo {author} {\bibfnamefont {S.}~\bibnamefont 
			{Tarucha}}, \
		and\ \bibinfo {author} {\bibfnamefont {L.~M.~K.}\ \bibnamefont
			{Vandersypen}},\ }\bibfield  {title} {\enquote {\bibinfo {title} 
			{Spins in
				few-electron quantum dots},}\ }\href {\doibase 
				10.1103/RevModPhys.79.1217}
	{\bibfield  {journal} {\bibinfo  {journal} {Rev. Mod. Phys.}\ }\textbf
		{\bibinfo {volume} {79}},\ \bibinfo {pages} {1217--1265} (\bibinfo 
		{year}
		{2007})}\BibitemShut {NoStop}%
	\bibitem [{\citenamefont {Nowack}\ \emph {et~al.}(2007)\citenamefont 
	{Nowack},
		\citenamefont {Koppens}, \citenamefont {Nazarov},\ and\ \citenamefont
		{Vandersypen}}]{Nowack-2007}%
	\BibitemOpen
	\bibfield  {author} {\bibinfo {author} {\bibfnamefont {K.C.}\ \bibnamefont
			{Nowack}}, \bibinfo {author} {\bibfnamefont {F.H.L.}\ \bibnamefont
			{Koppens}}, \bibinfo {author} {\bibfnamefont {Yu.V.}\ \bibnamefont
			{Nazarov}}, \ and\ \bibinfo {author} {\bibfnamefont {L.M.K.}\ 
			\bibnamefont
			{Vandersypen}},\ }\bibfield  {title} {\enquote {\bibinfo {title} 
			{Coherent
				control of a single electron spin with electric fields},}\ 
				}\href {\doibase
		10.1126/science.1148092} {\bibfield  {journal} {\bibinfo  {journal}
			{Science}\ }\textbf {\bibinfo {volume} {318}},\ \bibinfo {pages} 
			{1430}
		(\bibinfo {year} {2007})}\BibitemShut {NoStop}%
	\bibitem [{\citenamefont {Mart\'{i}n-Rodero}\ and\ \citenamefont
		{Levy~Yeyati}(2011)}]{Rodero-11}%
	\BibitemOpen
	\bibfield  {author} {\bibinfo {author} {\bibfnamefont {A.}~\bibnamefont
			{Mart\'{i}n-Rodero}}\ and\ \bibinfo {author} {\bibfnamefont 
			{A.}~\bibnamefont
			{Levy~Yeyati}},\ }\bibfield  {title} {\enquote {\bibinfo {title} 
			{Josephson
				and {A}ndreev transport through quantum dots},}\ }\href 
				{\doibase
		10.1080/00018732.2011.624266} {\bibfield  {journal} {\bibinfo  {journal}
			{Adv. Phys.}\ }\textbf {\bibinfo {volume} {60}},\ \bibinfo {pages} 
			{899}
		(\bibinfo {year} {2011})}\BibitemShut {NoStop}%
	\bibitem [{\citenamefont {Sherman}\ \emph {et~al.}(2017)\citenamefont
		{Sherman}, \citenamefont {Yodh}, \citenamefont {Albrecht}, \citenamefont
		{Nyg\aa{}rd}, \citenamefont {Krogstrup},\ and\ \citenamefont
		{Marcus}}]{Sherman.2017}%
	\BibitemOpen
	\bibfield  {author} {\bibinfo {author} {\bibfnamefont {D.}~\bibnamefont
			{Sherman}}, \bibinfo {author} {\bibfnamefont {J.S.}\ \bibnamefont 
			{Yodh}},
		\bibinfo {author} {\bibfnamefont {S.M.}\ \bibnamefont {Albrecht}}, 
		\bibinfo
		{author} {\bibfnamefont {J.}~\bibnamefont {Nyg\aa{}rd}}, \bibinfo 
		{author}
		{\bibfnamefont {P.}~\bibnamefont {Krogstrup}}, \ and\ \bibinfo {author}
		{\bibfnamefont {C.M.}\ \bibnamefont {Marcus}},\ }\bibfield  {title} 
		{\enquote
		{\bibinfo {title} {Normal, superconducting and topological regimes of 
		hybrid
				double quantum dots},}\ }\href {\doibase 
				10.1038/nnano.2016.227} {\bibfield
		{journal} {\bibinfo  {journal} {Nature Nanotechnol.}\ }\textbf {\bibinfo
			{volume} {12}},\ \bibinfo {pages} {212} (\bibinfo {year} 
			{2017})}\BibitemShut
	{NoStop}%
	\bibitem [{\citenamefont {Larsen}\ \emph {et~al.}(2015)\citenamefont 
	{Larsen},
		\citenamefont {Petersson}, \citenamefont {Kuemmeth}, \citenamefont
		{Jespersen}, \citenamefont {Krogstrup}, \citenamefont {Nyg\aa{}rd},\ 
		and\
		\citenamefont {Marcus}}]{Larsen-2015}%
	\BibitemOpen
	\bibfield  {author} {\bibinfo {author} {\bibfnamefont {T.~W.}\ \bibnamefont
			{Larsen}}, \bibinfo {author} {\bibfnamefont {K.~D.}\ \bibnamefont
			{Petersson}}, \bibinfo {author} {\bibfnamefont {F.}~\bibnamefont 
			{Kuemmeth}},
		\bibinfo {author} {\bibfnamefont {T.~S.}\ \bibnamefont {Jespersen}}, 
		\bibinfo
		{author} {\bibfnamefont {P.}~\bibnamefont {Krogstrup}}, \bibinfo 
		{author}
		{\bibfnamefont {J.}~\bibnamefont {Nyg\aa{}rd}}, \ and\ \bibinfo {author}
		{\bibfnamefont {C.~M.}\ \bibnamefont {Marcus}},\ }\bibfield  {title}
	{\enquote {\bibinfo {title} {Semiconductor-nanowire-based superconducting
				qubit},}\ }\href {\doibase 10.1103/PhysRevLett.115.127001} 
				{\bibfield
		{journal} {\bibinfo  {journal} {Phys. Rev. Lett.}\ }\textbf {\bibinfo
			{volume} {115}},\ \bibinfo {pages} {127001} (\bibinfo {year}
		{2015})}\BibitemShut {NoStop}%
	\bibitem [{\citenamefont {de~Lange}\ \emph {et~al.}(2015)\citenamefont
		{de~Lange}, \citenamefont {van Heck}, \citenamefont {Bruno}, 
		\citenamefont
		{van Woerkom}, \citenamefont {Geresdi}, \citenamefont {Plissard},
		\citenamefont {Bakkers}, \citenamefont {Akhmerov},\ and\ \citenamefont
		{DiCarlo}}]{deLange-2015}%
	\BibitemOpen
	\bibfield  {author} {\bibinfo {author} {\bibfnamefont {G.}~\bibnamefont
			{de~Lange}}, \bibinfo {author} {\bibfnamefont {B.}~\bibnamefont 
			{van Heck}},
		\bibinfo {author} {\bibfnamefont {A.}~\bibnamefont {Bruno}}, \bibinfo
		{author} {\bibfnamefont {D.~J.}\ \bibnamefont {van Woerkom}}, \bibinfo
		{author} {\bibfnamefont {A.}~\bibnamefont {Geresdi}}, \bibinfo {author}
		{\bibfnamefont {S.~R.}\ \bibnamefont {Plissard}}, \bibinfo {author}
		{\bibfnamefont {E.~P. A.~M.}\ \bibnamefont {Bakkers}}, \bibinfo {author}
		{\bibfnamefont {A.~R.}\ \bibnamefont {Akhmerov}}, \ and\ \bibinfo 
		{author}
		{\bibfnamefont {L.}~\bibnamefont {DiCarlo}},\ }\bibfield  {title} 
		{\enquote
		{\bibinfo {title} {Realization of microwave quantum circuits using 
		hybrid
				superconducting-semiconducting nanowire {J}osephson 
				elements},}\ }\href
	{\doibase 10.1103/PhysRevLett.115.127002} {\bibfield  {journal} {\bibinfo
			{journal} {Phys. Rev. Lett.}\ }\textbf {\bibinfo {volume} {115}},\ 
			\bibinfo
		{pages} {127002} (\bibinfo {year} {2015})}\BibitemShut {NoStop}%
	\bibitem [{\citenamefont {Luthi}\ \emph {et~al.}(2018)\citenamefont {Luthi},
		\citenamefont {Stavenga}, \citenamefont {Enzing}, \citenamefont {Bruno},
		\citenamefont {Dickel}, \citenamefont {Langford}, \citenamefont {Rol},
		\citenamefont {Jespersen}, \citenamefont {Nyg\aa{}rd}, \citenamefont
		{Krogstrup},\ and\ \citenamefont {DiCarlo}}]{Luthi-2018}%
	\BibitemOpen
	\bibfield  {author} {\bibinfo {author} {\bibfnamefont {F.}~\bibnamefont
			{Luthi}}, \bibinfo {author} {\bibfnamefont {T.}~\bibnamefont 
			{Stavenga}},
		\bibinfo {author} {\bibfnamefont {O.~W.}\ \bibnamefont {Enzing}}, 
		\bibinfo
		{author} {\bibfnamefont {A.}~\bibnamefont {Bruno}}, \bibinfo {author}
		{\bibfnamefont {C.}~\bibnamefont {Dickel}}, \bibinfo {author} 
		{\bibfnamefont
			{N.~K.}\ \bibnamefont {Langford}}, \bibinfo {author} {\bibfnamefont 
			{M.~A.}\
			\bibnamefont {Rol}}, \bibinfo {author} {\bibfnamefont {T.~S.}\ 
			\bibnamefont
			{Jespersen}}, \bibinfo {author} {\bibfnamefont {J.}~\bibnamefont
			{Nyg\aa{}rd}}, \bibinfo {author} {\bibfnamefont {P.}~\bibnamefont
			{Krogstrup}}, \ and\ \bibinfo {author} {\bibfnamefont 
			{L.}~\bibnamefont
			{DiCarlo}},\ }\bibfield  {title} {\enquote {\bibinfo {title} 
			{Evolution of
				nanowire transmon qubits and their coherence in a magnetic 
				field},}\ }\href
	{\doibase 10.1103/PhysRevLett.120.100502} {\bibfield  {journal} {\bibinfo
			{journal} {Phys. Rev. Lett.}\ }\textbf {\bibinfo {volume} {120}},\ 
			\bibinfo
		{pages} {100502} (\bibinfo {year} {2018})}\BibitemShut {NoStop}%
	\bibitem [{\citenamefont {Pita-Vidal}\ \emph {et~al.}(2020)\citenamefont
		{Pita-Vidal}, \citenamefont {Bargerbos}, \citenamefont {Yang}, 
		\citenamefont
		{van Woerkom}, \citenamefont {Pfaff}, \citenamefont {Haider}, 
		\citenamefont
		{Krogstrup}, \citenamefont {Kouwenhoven}, \citenamefont {de~Lange},\ 
		and\
		\citenamefont {Kou}}]{PitaVidal-2020}%
	\BibitemOpen
	\bibfield  {author} {\bibinfo {author} {\bibfnamefont {M.}~\bibnamefont
			{Pita-Vidal}}, \bibinfo {author} {\bibfnamefont {A.}~\bibnamefont
			{Bargerbos}}, \bibinfo {author} {\bibfnamefont {C.-K.}\ 
			\bibnamefont {Yang}},
		\bibinfo {author} {\bibfnamefont {D.~J.}\ \bibnamefont {van Woerkom}},
		\bibinfo {author} {\bibfnamefont {W.}~\bibnamefont {Pfaff}}, \bibinfo
		{author} {\bibfnamefont {N.}~\bibnamefont {Haider}}, \bibinfo {author}
		{\bibfnamefont {P.}~\bibnamefont {Krogstrup}}, \bibinfo {author}
		{\bibfnamefont {L.~P.}\ \bibnamefont {Kouwenhoven}}, \bibinfo {author}
		{\bibfnamefont {G.}~\bibnamefont {de~Lange}}, \ and\ \bibinfo {author}
		{\bibfnamefont {A.}~\bibnamefont {Kou}},\ }\bibfield  {title} {\enquote
		{\bibinfo {title} {Gate-tunable field-compatible fluxonium},}\ }\href
	{\doibase 10.1103/PhysRevApplied.14.064038} {\bibfield  {journal} {\bibinfo
			{journal} {Phys. Rev. Applied}\ }\textbf {\bibinfo {volume} {14}},\ 
			\bibinfo
		{pages} {064038} (\bibinfo {year} {2020})}\BibitemShut {NoStop}%
	\bibitem [{\citenamefont {Silva}\ and\ \citenamefont
		{Vernek}(2016)}]{Silva-2016}%
	\BibitemOpen
	\bibfield  {author} {\bibinfo {author} {\bibfnamefont {J.F.}\ \bibnamefont
			{Silva}}\ and\ \bibinfo {author} {\bibfnamefont {E.}~\bibnamefont 
			{Vernek}},\
	}\bibfield  {title} {\enquote {\bibinfo {title} {{A}ndreev and {M}ajorana
				bound states in single and double quantum dot structures},}\ 
				}\href {\doibase
		10.1088/0953-8984/28/43/435702} {\bibfield  {journal} {\bibinfo  
		{journal}
			{J. Phys.: Condens. Matter}\ }\textbf {\bibinfo {volume} {28}},\ 
			\bibinfo
		{pages} {435702} (\bibinfo {year} {2016})}\BibitemShut {NoStop}%
	\bibitem [{\citenamefont {Ivanov}(2017)}]{Ivanov-2017}%
	\BibitemOpen
	\bibfield  {author} {\bibinfo {author} {\bibfnamefont {T.~I.}\ \bibnamefont
			{Ivanov}},\ }\bibfield  {title} {\enquote {\bibinfo {title} 
			{Coherent
				tunneling through a double quantum dot coupled to {M}ajorana 
				bound states},}\
	}\href {\doibase 10.1103/PhysRevB.96.035417} {\bibfield  {journal} {\bibinfo
			{journal} {Phys. Rev. B}\ }\textbf {\bibinfo {volume} {96}},\ 
			\bibinfo
		{pages} {035417} (\bibinfo {year} {2017})}\BibitemShut {NoStop}%
	\bibitem [{\citenamefont {Su}\ \emph 
	{et~al.}(2017{\natexlab{a}})\citenamefont
		{Su}, \citenamefont {Tacla}, \citenamefont {Hocevar}, \citenamefont 
		{Car},
		\citenamefont {Plissard}, \citenamefont {Bakkers}, \citenamefont 
		{Daley},
		\citenamefont {Pekker},\ and\ \citenamefont {Frolov}}]{Su-2017}%
	\BibitemOpen
	\bibfield  {author} {\bibinfo {author} {\bibfnamefont {Z.}~\bibnamefont
			{Su}}, \bibinfo {author} {\bibfnamefont {A.B.}\ \bibnamefont 
			{Tacla}},
		\bibinfo {author} {\bibfnamefont {M.}~\bibnamefont {Hocevar}}, \bibinfo
		{author} {\bibfnamefont {D.}~\bibnamefont {Car}}, \bibinfo {author}
		{\bibfnamefont {S.R.}\ \bibnamefont {Plissard}}, \bibinfo {author}
		{\bibfnamefont {E.P.A.M.}\ \bibnamefont {Bakkers}}, \bibinfo {author}
		{\bibfnamefont {A.J.}\ \bibnamefont {Daley}}, \bibinfo {author}
		{\bibfnamefont {D.}~\bibnamefont {Pekker}}, \ and\ \bibinfo {author}
		{\bibfnamefont {S.M.}\ \bibnamefont {Frolov}},\ }\bibfield  {title} 
		{\enquote
		{\bibinfo {title} {Andreev molecules in semiconductor nanowire double 
		quantum
				dots},}\ }\href {\doibase 10.1038/s41467-017-00665-7} 
				{\bibfield  {journal}
		{\bibinfo  {journal} {Nature Communications}\ }\textbf {\bibinfo 
		{volume}
			{8}},\ \bibinfo {pages} {585} (\bibinfo {year}
		{2017}{\natexlab{a}})}\BibitemShut {NoStop}%
	\bibitem [{\citenamefont {Ran\v{c}i\'{c}}\ \emph {et~al.}(2019)\citenamefont
		{Ran\v{c}i\'{c}}, \citenamefont {Hoffman}, \citenamefont {Schrade},
		\citenamefont {Klinovaja},\ and\ \citenamefont {Loss}}]{Rancic-2019}%
	\BibitemOpen
	\bibfield  {author} {\bibinfo {author} {\bibfnamefont {M.J.}\ \bibnamefont
			{Ran\v{c}i\'{c}}}, \bibinfo {author} {\bibfnamefont 
			{S.}~\bibnamefont
			{Hoffman}}, \bibinfo {author} {\bibfnamefont {C.}~\bibnamefont 
			{Schrade}},
		\bibinfo {author} {\bibfnamefont {J.}~\bibnamefont {Klinovaja}}, \ and\
		\bibinfo {author} {\bibfnamefont {D.}~\bibnamefont {Loss}},\ }\bibfield
	{title} {\enquote {\bibinfo {title} {Entangling spins in double quantum dots
				and {M}ajorana bound states},}\ }\href {\doibase 
				10.1103/PhysRevB.99.165306}
	{\bibfield  {journal} {\bibinfo  {journal} {Phys. Rev. B}\ }\textbf 
	{\bibinfo
			{volume} {99}},\ \bibinfo {pages} {165306} (\bibinfo {year}
		{2019})}\BibitemShut {NoStop}%
	\bibitem [{\citenamefont {Cifuentes}\ and\ \citenamefont
		{da~Silva}(2019)}]{Cifuentes-2019}%
	\BibitemOpen
	\bibfield  {author} {\bibinfo {author} {\bibfnamefont {J.D.}\ \bibnamefont
			{Cifuentes}}\ and\ \bibinfo {author} {\bibfnamefont {Luis G. G. 
			V.~Dias}\
			\bibnamefont {da~Silva}},\ }\bibfield  {title} {\enquote {\bibinfo 
			{title}
			{Manipulating {M}ajorana zero modes in double quantum dots},}\ 
			}\href
	{\doibase 10.1103/PhysRevB.100.085429} {\bibfield  {journal} {\bibinfo
			{journal} {Phys. Rev. B}\ }\textbf {\bibinfo {volume} {100}},\ 
			\bibinfo
		{pages} {085429} (\bibinfo {year} {2019})}\BibitemShut {NoStop}%
	\bibitem [{\citenamefont {Weymann}\ \emph {et~al.}(2020)\citenamefont
		{Weymann}, \citenamefont {W\'ojcik},\ and\ \citenamefont
		{Majek}}]{Weymann-2020}%
	\BibitemOpen
	\bibfield  {author} {\bibinfo {author} {\bibfnamefont {I.}~\bibnamefont
			{Weymann}}, \bibinfo {author} {\bibfnamefont {K.~P.}\ \bibnamefont
			{W\'ojcik}}, \ and\ \bibinfo {author} {\bibfnamefont 
			{P.}~\bibnamefont
			{Majek}},\ }\bibfield  {title} {\enquote {\bibinfo {title}
			{{M}ajorana-{K}ondo interplay in {T}-shaped double quantum dots},}\ 
			}\href
	{\doibase 10.1103/PhysRevB.101.235404} {\bibfield  {journal} {\bibinfo
			{journal} {Phys. Rev. B}\ }\textbf {\bibinfo {volume} {101}},\ 
			\bibinfo
		{pages} {235404} (\bibinfo {year} {2020})}\BibitemShut {NoStop}%
	\bibitem [{\citenamefont {Prada}\ \emph
		{et~al.}(2020{\natexlab{a}})\citenamefont {Prada}, \citenamefont 
		{San-Jose},
		\citenamefont {de~Moor}, \citenamefont {Geresdi}, \citenamefont {Lee},
		\citenamefont {Klinovaja}, \citenamefont {Loss}, \citenamefont 
		{Nyg\aa{}rd},
		\citenamefont {Aguado},\ and\ \citenamefont
		{Kouwenhoven}}]{Kouwenhoven-2020}%
	\BibitemOpen
	\bibfield  {author} {\bibinfo {author} {\bibfnamefont {E.}~\bibnamefont
			{Prada}}, \bibinfo {author} {\bibfnamefont {P.}~\bibnamefont 
			{San-Jose}},
		\bibinfo {author} {\bibfnamefont {M.W.A.}\ \bibnamefont {de~Moor}}, 
		\bibinfo
		{author} {\bibfnamefont {A.}~\bibnamefont {Geresdi}}, \bibinfo {author}
		{\bibfnamefont {E.J.H.}\ \bibnamefont {Lee}}, \bibinfo {author}
		{\bibfnamefont {J.}~\bibnamefont {Klinovaja}}, \bibinfo {author}
		{\bibfnamefont {D.}~\bibnamefont {Loss}}, \bibinfo {author} 
		{\bibfnamefont
			{J.}~\bibnamefont {Nyg\aa{}rd}}, \bibinfo {author} {\bibfnamefont
			{R.}~\bibnamefont {Aguado}}, \ and\ \bibinfo {author} 
			{\bibfnamefont {L.P.}\
			\bibnamefont {Kouwenhoven}},\ }\bibfield  {title} {\enquote 
			{\bibinfo {title}
			{From {A}ndreev to {M}ajorana bound states in hybrid
				superconductor-semiconductor nanowires},}\ }\href {\doibase
		10.1038/s42254-020-0228-y} {\bibfield  {journal} {\bibinfo  {journal} 
		{Nat.
				Rev. Phys.}\ }\textbf {\bibinfo {volume} {2}},\ \bibinfo 
				{pages} {275}
		(\bibinfo {year} {2020}{\natexlab{a}})}\BibitemShut {NoStop}%
	\bibitem [{\citenamefont {Deng}\ \emph {et~al.}(2016)\citenamefont {Deng},
		\citenamefont {Vaitiekenas}, \citenamefont {Hansen}, \citenamefont 
		{Danon},
		\citenamefont {Leijnse}, \citenamefont {Flensberg}, \citenamefont 
		{Nyg{\r
				a}rd}, \citenamefont {Krogstrup},\ and\ \citenamefont 
				{Marcus}}]{Deng-2016}%
	\BibitemOpen
	\bibfield  {author} {\bibinfo {author} {\bibfnamefont {M.~T.}\ \bibnamefont
			{Deng}}, \bibinfo {author} {\bibfnamefont {S.}~\bibnamefont 
			{Vaitiekenas}},
		\bibinfo {author} {\bibfnamefont {E.~B.}\ \bibnamefont {Hansen}}, 
		\bibinfo
		{author} {\bibfnamefont {J.}~\bibnamefont {Danon}}, \bibinfo {author}
		{\bibfnamefont {M.}~\bibnamefont {Leijnse}}, \bibinfo {author} 
		{\bibfnamefont
			{K.}~\bibnamefont {Flensberg}}, \bibinfo {author} {\bibfnamefont
			{J.}~\bibnamefont {Nyg{\r a}rd}}, \bibinfo {author} {\bibfnamefont
			{P.}~\bibnamefont {Krogstrup}}, \ and\ \bibinfo {author} 
			{\bibfnamefont
			{C.~M.}\ \bibnamefont {Marcus}},\ }\bibfield  {title} {\enquote 
			{\bibinfo
			{title} {Majorana bound state in a coupled quantum-dot 
			hybrid-nanowire
				system},}\ }\href {\doibase 10.1126/science.aaf3961} 
				{\bibfield  {journal}
		{\bibinfo  {journal} {Science}\ }\textbf {\bibinfo {volume} {354}},\ 
		\bibinfo
		{pages} {1557} (\bibinfo {year} {2016})}\BibitemShut {NoStop}%
	\bibitem [{\citenamefont {Prada}\ \emph
		{et~al.}(2020{\natexlab{b}})\citenamefont {Prada}, \citenamefont 
		{San-Jose},
		\citenamefont {de~Moor}, \citenamefont {Geresdi}, \citenamefont {Lee},
		\citenamefont {Klinovaja}, \citenamefont {Loss}, \citenamefont 
		{Nyg\aa{}rd},
		\citenamefont {Aguado},\ and\ \citenamefont
		{Kouwenhoven}}]{Prada_review-2020}%
	\BibitemOpen
	\bibfield  {author} {\bibinfo {author} {\bibfnamefont {E.}~\bibnamefont
			{Prada}}, \bibinfo {author} {\bibfnamefont {P.}~\bibnamefont 
			{San-Jose}},
		\bibinfo {author} {\bibfnamefont {M.W.A.}\ \bibnamefont {de~Moor}}, 
		\bibinfo
		{author} {\bibfnamefont {A.}~\bibnamefont {Geresdi}}, \bibinfo {author}
		{\bibfnamefont {E.J.H.}\ \bibnamefont {Lee}}, \bibinfo {author}
		{\bibfnamefont {J.}~\bibnamefont {Klinovaja}}, \bibinfo {author}
		{\bibfnamefont {D.}~\bibnamefont {Loss}}, \bibinfo {author} 
		{\bibfnamefont
			{J.}~\bibnamefont {Nyg\aa{}rd}}, \bibinfo {author} {\bibfnamefont
			{R.}~\bibnamefont {Aguado}}, \ and\ \bibinfo {author} 
			{\bibfnamefont {L.P.}\
			\bibnamefont {Kouwenhoven}},\ }\bibfield  {title} {\enquote 
			{\bibinfo {title}
			{From {A}ndreev to {M}ajorana bound states in hybrid
				superconductor-semiconductor nanowires},}\ }\href {\doibase
		10.1038/s42254-020-0228-y} {\bibfield  {journal} {\bibinfo  {journal} 
		{Nat.
				Rev. Phys.}\ }\textbf {\bibinfo {volume} {2}},\ \bibinfo 
				{pages} {275}
		(\bibinfo {year} {2020}{\natexlab{b}})}\BibitemShut {NoStop}%
	\bibitem [{\citenamefont {Deng}\ \emph {et~al.}(2018)\citenamefont {Deng},
		\citenamefont {Vaitiek\ifmmode~\dot{e}\else \.{e}\fi{}nas}, 
		\citenamefont
		{Prada}, \citenamefont {San-Jose}, \citenamefont {Nyg\aa{}rd}, 
		\citenamefont
		{Krogstrup}, \citenamefont {Aguado},\ and\ \citenamefont
		{Marcus}}]{Deng-2018}%
	\BibitemOpen
	\bibfield  {author} {\bibinfo {author} {\bibfnamefont {M.-T.}\ \bibnamefont
			{Deng}}, \bibinfo {author} {\bibfnamefont {S.}~\bibnamefont
			{Vaitiek\ifmmode~\dot{e}\else \.{e}\fi{}nas}}, \bibinfo {author}
		{\bibfnamefont {E.}~\bibnamefont {Prada}}, \bibinfo {author} 
		{\bibfnamefont
			{P.}~\bibnamefont {San-Jose}}, \bibinfo {author} {\bibfnamefont
			{J.}~\bibnamefont {Nyg\aa{}rd}}, \bibinfo {author} {\bibfnamefont
			{P.}~\bibnamefont {Krogstrup}}, \bibinfo {author} {\bibfnamefont
			{R.}~\bibnamefont {Aguado}}, \ and\ \bibinfo {author} 
			{\bibfnamefont {C.~M.}\
			\bibnamefont {Marcus}},\ }\bibfield  {title} {\enquote {\bibinfo 
			{title}
			{Nonlocality of {M}ajorana modes in hybrid nanowires},}\ }\href 
			{\doibase
		10.1103/PhysRevB.98.085125} {\bibfield  {journal} {\bibinfo  {journal} 
		{Phys.
				Rev. B}\ }\textbf {\bibinfo {volume} {98}},\ \bibinfo {pages} 
				{085125}
		(\bibinfo {year} {2018})}\BibitemShut {NoStop}%
	\bibitem [{\citenamefont {G\'orski}\ \emph {et~al.}(2018)\citenamefont
		{G\'orski}, \citenamefont {Bara\'{n}ski}, \citenamefont {Weymann},\ and\
		\citenamefont {Doma\'{n}ski}}]{Gorski-2018}%
	\BibitemOpen
	\bibfield  {author} {\bibinfo {author} {\bibfnamefont {G.}~\bibnamefont
			{G\'orski}}, \bibinfo {author} {\bibfnamefont {J.}~\bibnamefont
			{Bara\'{n}ski}}, \bibinfo {author} {\bibfnamefont {I.}~\bibnamefont
			{Weymann}}, \ and\ \bibinfo {author} {\bibfnamefont 
			{T.}~\bibnamefont
			{Doma\'{n}ski}},\ }\bibfield  {title} {\enquote {\bibinfo {title} 
			{Interplay
				between correlations and {M}ajorana mode in proximitized 
				quantum dot},}\
	}\href {\doibase 10.1038/s41598-018-33529-1} {\bibfield  {journal} {\bibinfo
			{journal} {Sci. Rep.}\ }\textbf {\bibinfo {volume} {8}},\ \bibinfo 
			{pages}
		{15717} (\bibinfo {year} {2018})}\BibitemShut {NoStop}%
	\bibitem [{\citenamefont {Grove-Rasmussen}\ \emph 
	{et~al.}(2018)\citenamefont
		{Grove-Rasmussen}, \citenamefont {Steffensen}, \citenamefont 
		{Jellinggaard},
		\citenamefont {Madsen}, \citenamefont {\ifmmode~\check{Z}\else
			\v{Z}\fi{}itko}, \citenamefont {Paaske},\ and\ \citenamefont
		{Nyg\aa{}rd}}]{Grove_Rasmussen.2018}%
	\BibitemOpen
	\bibfield  {author} {\bibinfo {author} {\bibfnamefont {K.}~\bibnamefont
			{Grove-Rasmussen}}, \bibinfo {author} {\bibfnamefont 
			{G.}~\bibnamefont
			{Steffensen}}, \bibinfo {author} {\bibfnamefont {A.}~\bibnamefont
			{Jellinggaard}}, \bibinfo {author} {\bibfnamefont {M.H.}\ 
			\bibnamefont
			{Madsen}}, \bibinfo {author} {\bibfnamefont {R.}~\bibnamefont
			{\ifmmode~\check{Z}\else \v{Z}\fi{}itko}}, \bibinfo {author} 
			{\bibfnamefont
			{J.}~\bibnamefont {Paaske}}, \ and\ \bibinfo {author} {\bibfnamefont
			{J.}~\bibnamefont {Nyg\aa{}rd}},\ }\bibfield  {title} {\enquote 
			{\bibinfo
			{title} {{Y}u-{S}hiba-{R}usinov screening of spins in double 
			quantum dots},}\
	}\href {\doibase 10.1038/s41467-018-04683-x} {\bibfield  {journal} {\bibinfo
			{journal} {Nature Commun.}\ }\textbf {\bibinfo {volume} {9}},\ 
			\bibinfo
		{pages} {2376} (\bibinfo {year} {2018})}\BibitemShut {NoStop}%
	\bibitem [{\citenamefont {Estrada Salda\~na}\ \emph 
	{et~al.}(2018)\citenamefont
		{Estrada Salda\~na}, \citenamefont {Vekris}, \citenamefont {Steffensen},
		\citenamefont {\ifmmode~\check{Z}\else \v{Z}\fi{}itko}, \citenamefont
		{Krogstrup}, \citenamefont {Paaske}, \citenamefont {Grove-Rasmussen},\ 
		and\
		\citenamefont {Nyg\aa{}rd}}]{Estrada_Saldana.2018}%
	\BibitemOpen
	\bibfield  {author} {\bibinfo {author} {\bibfnamefont {J.~C.}\ \bibnamefont
			{Estrada Salda\~na}}, \bibinfo {author} {\bibfnamefont 
			{A.}~\bibnamefont
			{Vekris}}, \bibinfo {author} {\bibfnamefont {G.}~\bibnamefont 
			{Steffensen}},
		\bibinfo {author} {\bibfnamefont {R.}~\bibnamefont 
		{\ifmmode~\check{Z}\else
				\v{Z}\fi{}itko}}, \bibinfo {author} {\bibfnamefont 
				{P.}~\bibnamefont
			{Krogstrup}}, \bibinfo {author} {\bibfnamefont {J.}~\bibnamefont 
			{Paaske}},
		\bibinfo {author} {\bibfnamefont {K.}~\bibnamefont {Grove-Rasmussen}}, 
		\ and\
		\bibinfo {author} {\bibfnamefont {J.}~\bibnamefont {Nyg\aa{}rd}},\ 
		}\bibfield
	{title} {\enquote {\bibinfo {title} {Supercurrent in a double quantum
				dot},}\ }\href {\doibase 10.1103/PhysRevLett.121.257701} 
				{\bibfield
		{journal} {\bibinfo  {journal} {Phys. Rev. Lett.}\ }\textbf {\bibinfo
			{volume} {121}},\ \bibinfo {pages} {257701} (\bibinfo {year}
		{2018})}\BibitemShut {NoStop}%
	\bibitem [{\citenamefont {Bouman}\ \emph {et~al.}(2020)\citenamefont 
	{Bouman},
		\citenamefont {van Gulik}, \citenamefont {Steffensen}, \citenamefont
		{Pataki}, \citenamefont {Boross}, \citenamefont {Krogstrup}, 
		\citenamefont
		{Nyg\aa{}rd}, \citenamefont {Paaske}, \citenamefont {P\'alyi},\ and\
		\citenamefont {Geresdi}}]{Paaske-2020}%
	\BibitemOpen
	\bibfield  {author} {\bibinfo {author} {\bibfnamefont {D.}~\bibnamefont
			{Bouman}}, \bibinfo {author} {\bibfnamefont {R.~J.~J.}\ 
			\bibnamefont {van
				Gulik}}, \bibinfo {author} {\bibfnamefont {G.}~\bibnamefont 
				{Steffensen}},
		\bibinfo {author} {\bibfnamefont {D.}~\bibnamefont {Pataki}}, \bibinfo
		{author} {\bibfnamefont {P.r}\ \bibnamefont {Boross}}, \bibinfo {author}
		{\bibfnamefont {P.}~\bibnamefont {Krogstrup}}, \bibinfo {author}
		{\bibfnamefont {J.}~\bibnamefont {Nyg\aa{}rd}}, \bibinfo {author}
		{\bibfnamefont {J.}~\bibnamefont {Paaske}}, \bibinfo {author} 
		{\bibfnamefont
			{A.}~\bibnamefont {P\'alyi}}, \ and\ \bibinfo {author} 
			{\bibfnamefont
			{A.}~\bibnamefont {Geresdi}},\ }\bibfield  {title} {\enquote 
			{\bibinfo
			{title} {Triplet-blockaded {J}osephson supercurrent in double 
			quantum
				dots},}\ }\href {\doibase 10.1103/PhysRevB.102.220505} 
				{\bibfield  {journal}
		{\bibinfo  {journal} {Phys. Rev. B}\ }\textbf {\bibinfo {volume} 
		{102}},\
		\bibinfo {pages} {220505} (\bibinfo {year} {2020})}\BibitemShut 
		{NoStop}%
	\bibitem [{\citenamefont {Estrada Salda\~na}\ \emph 
	{et~al.}(2020)\citenamefont
		{Estrada Salda\~na}, \citenamefont {Vekris}, \citenamefont
		{\ifmmode~\check{Z}\else \v{Z}\fi{}itko}, \citenamefont {Steffensen},
		\citenamefont {Krogstrup}, \citenamefont {Paaske}, \citenamefont
		{Grove-Rasmussen},\ and\ \citenamefont {Nyg\aa{}rd}}]{Estrada-2020}%
	\BibitemOpen
	\bibfield  {author} {\bibinfo {author} {\bibfnamefont {J.~C.}\ \bibnamefont
			{Estrada Salda\~na}}, \bibinfo {author} {\bibfnamefont 
			{A.}~\bibnamefont
			{Vekris}}, \bibinfo {author} {\bibfnamefont {R.}~\bibnamefont
			{\ifmmode~\check{Z}\else \v{Z}\fi{}itko}}, \bibinfo {author} 
			{\bibfnamefont
			{G.}~\bibnamefont {Steffensen}}, \bibinfo {author} {\bibfnamefont
			{P.}~\bibnamefont {Krogstrup}}, \bibinfo {author} {\bibfnamefont
			{J.}~\bibnamefont {Paaske}}, \bibinfo {author} {\bibfnamefont
			{K.}~\bibnamefont {Grove-Rasmussen}}, \ and\ \bibinfo {author} 
			{\bibfnamefont
			{J.}~\bibnamefont {Nyg\aa{}rd}},\ }\bibfield  {title} {\enquote 
			{\bibinfo
			{title} {Two-impurity {Y}u-{S}hiba-{R}usinov states in coupled 
			quantum
				dots},}\ }\href {\doibase 10.1103/PhysRevB.102.195143} 
				{\bibfield  {journal}
		{\bibinfo  {journal} {Phys. Rev. B}\ }\textbf {\bibinfo {volume} 
		{102}},\
		\bibinfo {pages} {195143} (\bibinfo {year} {2020})}\BibitemShut 
		{NoStop}%
	\bibitem [{\citenamefont {Su}\ \emph 
	{et~al.}(2017{\natexlab{b}})\citenamefont
		{Su}, \citenamefont {Tacla}, \citenamefont {Hocevar}, \citenamefont 
		{Car},
		\citenamefont {Plissard}, \citenamefont {Bakkers}, \citenamefont 
		{Daley},
		\citenamefont {Pekker},\ and\ \citenamefont {Frolov}}]{Su.2017}%
	\BibitemOpen
	\bibfield  {author} {\bibinfo {author} {\bibfnamefont {Z.}~\bibnamefont
			{Su}}, \bibinfo {author} {\bibfnamefont {A.B.}\ \bibnamefont 
			{Tacla}},
		\bibinfo {author} {\bibfnamefont {M.}~\bibnamefont {Hocevar}}, \bibinfo
		{author} {\bibfnamefont {D.}~\bibnamefont {Car}}, \bibinfo {author}
		{\bibfnamefont {S.R.}\ \bibnamefont {Plissard}}, \bibinfo {author}
		{\bibfnamefont {E.P.A.M.}\ \bibnamefont {Bakkers}}, \bibinfo {author}
		{\bibfnamefont {A.J.}\ \bibnamefont {Daley}}, \bibinfo {author}
		{\bibfnamefont {D.}~\bibnamefont {Pekker}}, \ and\ \bibinfo {author}
		{\bibfnamefont {S.M.}\ \bibnamefont {Frolov}},\ }\bibfield  {title} 
		{\enquote
		{\bibinfo {title} {Andreev molecules in semiconductor nanowire double 
		quantum
				dots},}\ }\href {\doibase 10.1038/s41467-017-00665-7} 
				{\bibfield  {journal}
		{\bibinfo  {journal} {Nature Commun.}\ }\textbf {\bibinfo {volume} 
		{8}},\
		\bibinfo {pages} {585} (\bibinfo {year} 
		{2017}{\natexlab{b}})}\BibitemShut
	{NoStop}%
	\bibitem [{\citenamefont {Zarassi}\ \emph {et~al.}(2017)\citenamefont
		{Zarassi}, \citenamefont {Su}, \citenamefont {Danon}, \citenamefont
		{Schwenderling}, \citenamefont {Hocevar}, \citenamefont {Nguyen},
		\citenamefont {Yoo}, \citenamefont {Dayeh},\ and\ \citenamefont
		{Frolov}}]{Zarassi.2017}%
	\BibitemOpen
	\bibfield  {author} {\bibinfo {author} {\bibfnamefont {A.}~\bibnamefont
			{Zarassi}}, \bibinfo {author} {\bibfnamefont {Z.}~\bibnamefont 
			{Su}},
		\bibinfo {author} {\bibfnamefont {J.}~\bibnamefont {Danon}}, \bibinfo
		{author} {\bibfnamefont {J.}~\bibnamefont {Schwenderling}}, \bibinfo 
		{author}
		{\bibfnamefont {M.}~\bibnamefont {Hocevar}}, \bibinfo {author} 
		{\bibfnamefont
			{B.~M.}\ \bibnamefont {Nguyen}}, \bibinfo {author} {\bibfnamefont
			{J.}~\bibnamefont {Yoo}}, \bibinfo {author} {\bibfnamefont {S.~A.}\
			\bibnamefont {Dayeh}}, \ and\ \bibinfo {author} {\bibfnamefont 
			{S.~M.}\
			\bibnamefont {Frolov}},\ }\bibfield  {title} {\enquote {\bibinfo 
			{title}
			{Magnetic field evolution of spin blockade in {Ge/Si} nanowire 
			double quantum
				dots},}\ }\href {\doibase 10.1103/PhysRevB.95.155416} 
				{\bibfield  {journal}
		{\bibinfo  {journal} {Phys. Rev. B}\ }\textbf {\bibinfo {volume} {95}},\
		\bibinfo {pages} {155416} (\bibinfo {year} {2017})}\BibitemShut 
		{NoStop}%
	\bibitem [{\citenamefont {Cleuziou}\ \emph {et~al.}(2006)\citenamefont
		{Cleuziou}, \citenamefont {Wernsdorfer}, \citenamefont {Bouchiat},
		\citenamefont {Ondarcuhu},\ and\ \citenamefont 
		{Monthioux}}]{Cleuziou.2006}%
	\BibitemOpen
	\bibfield  {author} {\bibinfo {author} {\bibfnamefont {J.-P.}\ \bibnamefont
			{Cleuziou}}, \bibinfo {author} {\bibfnamefont {W.}~\bibnamefont
			{Wernsdorfer}}, \bibinfo {author} {\bibfnamefont {V.}~\bibnamefont
			{Bouchiat}}, \bibinfo {author} {\bibfnamefont {T.}~\bibnamefont 
			{Ondarcuhu}},
		\ and\ \bibinfo {author} {\bibfnamefont {M.}~\bibnamefont {Monthioux}},\
	}\bibfield  {title} {\enquote {\bibinfo {title} {Carbon nanotube
				superconducting quantum interference device},}\ }\href {\doibase
		10.1038/nnano.2006.54} {\bibfield  {journal} {\bibinfo  {journal} 
		{Nature
				Nanotechnol.}\ }\textbf {\bibinfo {volume} {1}},\ \bibinfo 
				{pages} {53}
		(\bibinfo {year} {2006})}\BibitemShut {NoStop}%
	\bibitem [{\citenamefont {Pillet}\ \emph {et~al.}(2013)\citenamefont 
	{Pillet},
		\citenamefont {Joyez}, \citenamefont {\ifmmode~\check{Z}\else
			\v{Z}\fi{}itko},\ and\ \citenamefont {Goffman}}]{Pillet.2013}%
	\BibitemOpen
	\bibfield  {author} {\bibinfo {author} {\bibfnamefont {J.-D.}\ \bibnamefont
			{Pillet}}, \bibinfo {author} {\bibfnamefont {P.}~\bibnamefont 
			{Joyez}},
		\bibinfo {author} {\bibfnamefont {R.}~\bibnamefont 
		{\ifmmode~\check{Z}\else
				\v{Z}\fi{}itko}}, \ and\ \bibinfo {author} {\bibfnamefont 
				{M.~F.}\
			\bibnamefont {Goffman}},\ }\bibfield  {title} {\enquote {\bibinfo 
			{title}
			{Tunneling spectroscopy of a single quantum dot coupled to a 
			superconductor:
				From {K}ondo ridge to {A}ndreev bound states},}\ }\href 
				{\doibase
		10.1103/PhysRevB.88.045101} {\bibfield  {journal} {\bibinfo  {journal} 
		{Phys.
				Rev. B}\ }\textbf {\bibinfo {volume} {88}},\ \bibinfo {pages} 
				{045101}
		(\bibinfo {year} {2013})}\BibitemShut {NoStop}%
	\bibitem [{\citenamefont {Ruby}\ \emph {et~al.}(2018)\citenamefont {Ruby},
		\citenamefont {Heinrich}, \citenamefont {Peng}, \citenamefont {von 
		Oppen},\
		and\ \citenamefont {Franke}}]{Ruby.2018}%
	\BibitemOpen
	\bibfield  {author} {\bibinfo {author} {\bibfnamefont {M.}~\bibnamefont
			{Ruby}}, \bibinfo {author} {\bibfnamefont {B.W.}\ \bibnamefont 
			{Heinrich}},
		\bibinfo {author} {\bibfnamefont {Y.}~\bibnamefont {Peng}}, \bibinfo 
		{author}
		{\bibfnamefont {F.}~\bibnamefont {von Oppen}}, \ and\ \bibinfo {author}
		{\bibfnamefont {K.J.}\ \bibnamefont {Franke}},\ }\bibfield  {title} 
		{\enquote
		{\bibinfo {title} {Wave-function hybridization in {Y}u-{S}hiba-{R}usinov
				dimers},}\ }\href {\doibase 10.1103/PhysRevLett.120.156803} 
				{\bibfield
		{journal} {\bibinfo  {journal} {Phys. Rev. Lett.}\ }\textbf {\bibinfo
			{volume} {120}},\ \bibinfo {pages} {156803} (\bibinfo {year}
		{2018})}\BibitemShut {NoStop}%
	\bibitem [{\citenamefont {Heinrich}\ \emph {et~al.}(2018)\citenamefont
		{Heinrich}, \citenamefont {Pascual},\ and\ \citenamefont
		{Franke}}]{Franke-2018}%
	\BibitemOpen
	\bibfield  {author} {\bibinfo {author} {\bibfnamefont {B.W.}\ \bibnamefont
			{Heinrich}}, \bibinfo {author} {\bibfnamefont {J.I.}\ \bibnamefont
			{Pascual}}, \ and\ \bibinfo {author} {\bibfnamefont {K.J.}\ 
			\bibnamefont
			{Franke}},\ }\bibfield  {title} {\enquote {\bibinfo {title} {Single 
			magnetic
				adsorbates on s-wave superconductors},}\ }\href {\doibase
		https://doi.org/10.1016/j.progsurf.2018.01.001} {\bibfield  {journal}
		{\bibinfo  {journal} {Prog. Surf. Science}\ }\textbf {\bibinfo {volume}
			{93}},\ \bibinfo {pages} {1} (\bibinfo {year} {2018})}\BibitemShut 
			{NoStop}%
	\bibitem [{\citenamefont {Choi}\ \emph {et~al.}(2018)\citenamefont {Choi},
		\citenamefont {Fern\'andez}, \citenamefont {Herrera}, \citenamefont
		{Rubio-Verd\'u}, \citenamefont {Ugeda}, \citenamefont {Guillam\'on},
		\citenamefont {Suderow}, \citenamefont {Pascual},\ and\ \citenamefont
		{Lorente}}]{Choi.2018}%
	\BibitemOpen
	\bibfield  {author} {\bibinfo {author} {\bibfnamefont {D.-J.}\ \bibnamefont
			{Choi}}, \bibinfo {author} {\bibfnamefont {C.G.}\ \bibnamefont
			{Fern\'andez}}, \bibinfo {author} {\bibfnamefont {E.}~\bibnamefont
			{Herrera}}, \bibinfo {author} {\bibfnamefont {C.}~\bibnamefont
			{Rubio-Verd\'u}}, \bibinfo {author} {\bibfnamefont {M.M.}\ 
			\bibnamefont
			{Ugeda}}, \bibinfo {author} {\bibfnamefont {I.}~\bibnamefont 
			{Guillam\'on}},
		\bibinfo {author} {\bibfnamefont {H.}~\bibnamefont {Suderow}}, \bibinfo
		{author} {\bibfnamefont {J.I.}\ \bibnamefont {Pascual}}, \ and\ \bibinfo
		{author} {\bibfnamefont {N.}~\bibnamefont {Lorente}},\ }\bibfield  
		{title}
	{\enquote {\bibinfo {title} {Influence of magnetic ordering between {C}r
				adatoms on the {Y}u-{S}hiba-{R}usinov states of the
				$\ensuremath{\beta}\text{\ensuremath{-}}${Bi}$_{2}${Pd} 
				superconductor},}\
	}\href {\doibase 10.1103/PhysRevLett.120.167001} {\bibfield  {journal}
		{\bibinfo  {journal} {Phys. Rev. Lett.}\ }\textbf {\bibinfo {volume} 
		{120}},\
		\bibinfo {pages} {167001} (\bibinfo {year} {2018})}\BibitemShut 
		{NoStop}%
	\bibitem [{\citenamefont {Kezilebieke}\ \emph {et~al.}(2019)\citenamefont
		{Kezilebieke}, \citenamefont {\v{Z}itko}, \citenamefont {Dvorak},\ and\
		\citenamefont {Liljeroth}}]{Kezilebieke.2019}%
	\BibitemOpen
	\bibfield  {author} {\bibinfo {author} {\bibfnamefont {S.}~\bibnamefont
			{Kezilebieke}}, \bibinfo {author} {\bibfnamefont {R.}~\bibnamefont
			{\v{Z}itko}}, \bibinfo {author} {\bibfnamefont {M.}~\bibnamefont 
			{Dvorak}}, \
		and\ \bibinfo {author} {\bibfnamefont {P.}~\bibnamefont {Liljeroth}},\
	}\bibfield  {title} {\enquote {\bibinfo {title} {Observation of coexistence
				of {Y}u-{S}hiba-{R}usinov states and spin-flip excitations},}\ 
				}\href
	{\doibase 10.1021/acs.nanolett.9b01583} {\bibfield  {journal} {\bibinfo
			{journal} {Nano Lett.}\ }\textbf {\bibinfo {volume} {19}},\ 
			\bibinfo {pages}
		{4614} (\bibinfo {year} {2019})}\BibitemShut {NoStop}%
	\bibitem [{\citenamefont {Choi}\ \emph {et~al.}(2000)\citenamefont {Choi},
		\citenamefont {Bruder},\ and\ \citenamefont {Loss}}]{Choi-2000}%
	\BibitemOpen
	\bibfield  {author} {\bibinfo {author} {\bibfnamefont {M.-S.}\ \bibnamefont
			{Choi}}, \bibinfo {author} {\bibfnamefont {C.}~\bibnamefont 
			{Bruder}}, \ and\
		\bibinfo {author} {\bibfnamefont {D.}~\bibnamefont {Loss}},\ }\bibfield
	{title} {\enquote {\bibinfo {title} {Spin-dependent {J}osephson current
				through double quantum dots and measurement of entangled 
				electron states},}\
	}\href {\doibase 10.1103/PhysRevB.62.13569} {\bibfield  {journal} {\bibinfo
			{journal} {Phys. Rev. B}\ }\textbf {\bibinfo {volume} {62}},\ 
			\bibinfo
		{pages} {13569} (\bibinfo {year} {2000})}\BibitemShut {NoStop}%
	\bibitem [{\citenamefont {Zhu}\ \emph {et~al.}(2002)\citenamefont {Zhu},
		\citenamefont {Sun},\ and\ \citenamefont {Lin}}]{Zhu-2002}%
	\BibitemOpen
	\bibfield  {author} {\bibinfo {author} {\bibfnamefont {Y.}~\bibnamefont
			{Zhu}}, \bibinfo {author} {\bibfnamefont {Q.-F.}\ \bibnamefont 
			{Sun}}, \ and\
		\bibinfo {author} {\bibfnamefont {T.-H.}\ \bibnamefont {Lin}},\ 
		}\bibfield
	{title} {\enquote {\bibinfo {title} {Probing spin states of coupled quantum
				dots by a dc {J}osephson current},}\ }\href {\doibase
		10.1103/PhysRevB.66.085306} {\bibfield  {journal} {\bibinfo  {journal} 
		{Phys.
				Rev. B}\ }\textbf {\bibinfo {volume} {66}},\ \bibinfo {pages} 
				{085306}
		(\bibinfo {year} {2002})}\BibitemShut {NoStop}%
	\bibitem [{\citenamefont {Tanaka}\ \emph {et~al.}(2010)\citenamefont 
	{Tanaka},
		\citenamefont {Kawakami},\ and\ \citenamefont {Oguri}}]{Tanaka.2010}%
	\BibitemOpen
	\bibfield  {author} {\bibinfo {author} {\bibfnamefont {Y.}~\bibnamefont
			{Tanaka}}, \bibinfo {author} {\bibfnamefont {N.}~\bibnamefont 
			{Kawakami}}, \
		and\ \bibinfo {author} {\bibfnamefont {A.}~\bibnamefont {Oguri}},\ 
		}\bibfield
	{title} {\enquote {\bibinfo {title} {Correlated electron transport through
				double quantum dots coupled to normal and superconducting 
				leads},}\ }\href
	{\doibase 10.1103/PhysRevB.81.075404} {\bibfield  {journal} {\bibinfo
			{journal} {Phys. Rev. B}\ }\textbf {\bibinfo {volume} {81}},\ 
			\bibinfo
		{pages} {075404} (\bibinfo {year} {2010})}\BibitemShut {NoStop}%
	\bibitem [{\citenamefont {\ifmmode~\check{Z}\else \v{Z}\fi{}itko}\ \emph
		{et~al.}(2010)\citenamefont {\ifmmode~\check{Z}\else \v{Z}\fi{}itko},
		\citenamefont {Lee}, \citenamefont {L\'opez}, \citenamefont {Aguado},\ 
		and\
		\citenamefont {Choi}}]{Zitko.2010}%
	\BibitemOpen
	\bibfield  {author} {\bibinfo {author} {\bibfnamefont {R.}~\bibnamefont
			{\ifmmode~\check{Z}\else \v{Z}\fi{}itko}}, \bibinfo {author} 
			{\bibfnamefont
			{M.}~\bibnamefont {Lee}}, \bibinfo {author} {\bibfnamefont 
			{R.}~\bibnamefont
			{L\'opez}}, \bibinfo {author} {\bibfnamefont {R.}~\bibnamefont 
			{Aguado}}, \
		and\ \bibinfo {author} {\bibfnamefont {M.-S.}\ \bibnamefont {Choi}},\
	}\bibfield  {title} {\enquote {\bibinfo {title} {Josephson current in
				strongly correlated double quantum dots},}\ }\href {\doibase
		10.1103/PhysRevLett.105.116803} {\bibfield  {journal} {\bibinfo  
		{journal}
			{Phys. Rev. Lett.}\ }\textbf {\bibinfo {volume} {105}},\ \bibinfo 
			{pages}
		{116803} (\bibinfo {year} {2010})}\BibitemShut {NoStop}%
	\bibitem [{\citenamefont {Eldridge}\ \emph {et~al.}(2010)\citenamefont
		{Eldridge}, \citenamefont {Pala}, \citenamefont {Governale},\ and\
		\citenamefont {K\"onig}}]{Konig.2010}%
	\BibitemOpen
	\bibfield  {author} {\bibinfo {author} {\bibfnamefont {J.}~\bibnamefont
			{Eldridge}}, \bibinfo {author} {\bibfnamefont {M.G.}\ \bibnamefont 
			{Pala}},
		\bibinfo {author} {\bibfnamefont {M.}~\bibnamefont {Governale}}, \ and\
		\bibinfo {author} {\bibfnamefont {J.}~\bibnamefont {K\"onig}},\ 
		}\bibfield
	{title} {\enquote {\bibinfo {title} {Superconducting proximity effect in
				interacting double-dot systems},}\ }\href {\doibase
		10.1103/PhysRevB.82.184507} {\bibfield  {journal} {\bibinfo  {journal} 
		{Phys.
				Rev. B}\ }\textbf {\bibinfo {volume} {82}},\ \bibinfo {pages} 
				{184507}
		(\bibinfo {year} {2010})}\BibitemShut {NoStop}%
	\bibitem [{\citenamefont {Droste}\ \emph {et~al.}(2012)\citenamefont 
	{Droste},
		\citenamefont {Andergassen},\ and\ \citenamefont
		{Splettstoesser}}]{Droste.2012}%
	\BibitemOpen
	\bibfield  {author} {\bibinfo {author} {\bibfnamefont {S.}~\bibnamefont
			{Droste}}, \bibinfo {author} {\bibfnamefont {S.}~\bibnamefont 
			{Andergassen}},
		\ and\ \bibinfo {author} {\bibfnamefont {J.}~\bibnamefont 
		{Splettstoesser}},\
	}\bibfield  {title} {\enquote {\bibinfo {title} {Josephson current through
				interacting double quantum dots with spin{\textendash}orbit 
				coupling},}\
	}\href {\doibase 10.1088/0953-8984/24/41/415301} {\bibfield  {journal}
		{\bibinfo  {journal} {J. Phys.: Condens. Matter}\ }\textbf {\bibinfo 
		{volume}
			{24}},\ \bibinfo {pages} {415301} (\bibinfo {year} 
			{2012})}\BibitemShut
	{NoStop}%
	\bibitem [{\citenamefont {Pfaller}\ \emph {et~al.}(2013)\citenamefont
		{Pfaller}, \citenamefont {Donarini},\ and\ \citenamefont
		{Grifoni}}]{Grifoni.2013}%
	\BibitemOpen
	\bibfield  {author} {\bibinfo {author} {\bibfnamefont {S.}~\bibnamefont
			{Pfaller}}, \bibinfo {author} {\bibfnamefont {A.}~\bibnamefont 
			{Donarini}}, \
		and\ \bibinfo {author} {\bibfnamefont {M.}~\bibnamefont {Grifoni}},\
	}\bibfield  {title} {\enquote {\bibinfo {title} {Subgap features due to
				quasiparticle tunneling in quantum dots coupled to 
				superconducting leads},}\
	}\href {\doibase 10.1103/PhysRevB.87.155439} {\bibfield  {journal} {\bibinfo
			{journal} {Phys. Rev. B}\ }\textbf {\bibinfo {volume} {87}},\ 
			\bibinfo
		{pages} {155439} (\bibinfo {year} {2013})}\BibitemShut {NoStop}%
	\bibitem [{\citenamefont {Sothmann}\ \emph {et~al.}(2014)\citenamefont
		{Sothmann}, \citenamefont {Weiss}, \citenamefont {Governale},\ and\
		\citenamefont {K\"onig}}]{Sothmann-2014}%
	\BibitemOpen
	\bibfield  {author} {\bibinfo {author} {\bibfnamefont {B.}~\bibnamefont
			{Sothmann}}, \bibinfo {author} {\bibfnamefont {S.}~\bibnamefont 
			{Weiss}},
		\bibinfo {author} {\bibfnamefont {M.}~\bibnamefont {Governale}}, \ and\
		\bibinfo {author} {\bibfnamefont {J.}~\bibnamefont {K\"onig}},\ 
		}\bibfield
	{title} {\enquote {\bibinfo {title} {Unconventional superconductivity in
				double quantum dots},}\ }\href {\doibase 
				10.1103/PhysRevB.90.220501}
	{\bibfield  {journal} {\bibinfo  {journal} {Phys. Rev. B}\ }\textbf 
	{\bibinfo
			{volume} {90}},\ \bibinfo {pages} {220501} (\bibinfo {year}
		{2014})}\BibitemShut {NoStop}%
	\bibitem [{\citenamefont {Yao}\ \emph {et~al.}(2014)\citenamefont {Yao},
		\citenamefont {Moca}, \citenamefont {Weymann}, \citenamefont {Sau},
		\citenamefont {Lukin}, \citenamefont {Demler},\ and\ \citenamefont
		{Zar{\ifmmode\acute{a}\else\'{a}\fi}nd}}]{Yao2014Dec}%
	\BibitemOpen
	\bibfield  {author} {\bibinfo {author} {\bibfnamefont {N.~Y.}\ \bibnamefont
			{Yao}}, \bibinfo {author} {\bibfnamefont {C.~P.}\ \bibnamefont 
			{Moca}},
		\bibinfo {author} {\bibfnamefont {I.}~\bibnamefont {Weymann}}, \bibinfo
		{author} {\bibfnamefont {J.~D.}\ \bibnamefont {Sau}}, \bibinfo {author}
		{\bibfnamefont {M.~D.}\ \bibnamefont {Lukin}}, \bibinfo {author}
		{\bibfnamefont {E.~A.}\ \bibnamefont {Demler}}, \ and\ \bibinfo {author}
		{\bibfnamefont {G.}~\bibnamefont 
		{Zar{\ifmmode\acute{a}\else\'{a}\fi}nd}},\
	}\bibfield  {title} {\enquote {\bibinfo {title} {{Phase diagram and
					excitations of a Shiba molecule}},}\ }\href {\doibase
		10.1103/PhysRevB.90.241108} {\bibfield  {journal} {\bibinfo  {journal} 
		{Phys.
				Rev. B}\ }\textbf {\bibinfo {volume} {90}},\ \bibinfo {pages} 
				{241108}
		(\bibinfo {year} {2014})}\BibitemShut {NoStop}%
	\bibitem [{\citenamefont {Meng}\ \emph {et~al.}(2015)\citenamefont {Meng},
		\citenamefont {Klinovaja}, \citenamefont {Hoffman}, \citenamefont 
		{Simon},\
		and\ \citenamefont {Loss}}]{Meng.2015}%
	\BibitemOpen
	\bibfield  {author} {\bibinfo {author} {\bibfnamefont {T.}~\bibnamefont
			{Meng}}, \bibinfo {author} {\bibfnamefont {J.}~\bibnamefont 
			{Klinovaja}},
		\bibinfo {author} {\bibfnamefont {S.}~\bibnamefont {Hoffman}}, \bibinfo
		{author} {\bibfnamefont {P.}~\bibnamefont {Simon}}, \ and\ \bibinfo 
		{author}
		{\bibfnamefont {D.}~\bibnamefont {Loss}},\ }\bibfield  {title} {\enquote
		{\bibinfo {title} {Superconducting gap renormalization around two 
		magnetic
				impurities: From {S}hiba to {A}ndreev bound states},}\ }\href 
				{\doibase
		10.1103/PhysRevB.92.064503} {\bibfield  {journal} {\bibinfo  {journal} 
		{Phys.
				Rev. B}\ }\textbf {\bibinfo {volume} {92}},\ \bibinfo {pages} 
				{064503}
		(\bibinfo {year} {2015})}\BibitemShut {NoStop}%
	\bibitem [{\citenamefont {\ifmmode~\check{Z}\else
			\v{Z}\fi{}itko}(2015)}]{Zitko-2015}%
	\BibitemOpen
	\bibfield  {author} {\bibinfo {author} {\bibfnamefont {R.}~\bibnamefont
			{\ifmmode~\check{Z}\else \v{Z}\fi{}itko}},\ }\bibfield  {title} 
			{\enquote
		{\bibinfo {title} {Numerical subgap spectroscopy of double quantum dots
				coupled to superconductors},}\ }\href {\doibase 
				10.1103/PhysRevB.91.165116}
	{\bibfield  {journal} {\bibinfo  {journal} {Phys. Rev. B}\ }\textbf 
	{\bibinfo
			{volume} {91}},\ \bibinfo {pages} {165116} (\bibinfo {year}
		{2015})}\BibitemShut {NoStop}%
	\bibitem [{\citenamefont {Wrze\'{s}niewski}\ and\ \citenamefont
		{Weymann}(2017)}]{Wrzesniewski-2017}%
	\BibitemOpen
	\bibfield  {author} {\bibinfo {author} {\bibfnamefont {K.}~\bibnamefont
			{Wrze\'{s}niewski}}\ and\ \bibinfo {author} {\bibfnamefont 
			{I.}~\bibnamefont
			{Weymann}},\ }\bibfield  {title} {\enquote {\bibinfo {title} {Kondo 
			physics
				in double quantum dot based {C}ooper pair splitters},}\ }\href 
				{\doibase
		10.1103/PhysRevB.96.195409} {\bibfield  {journal} {\bibinfo  {journal} 
		{Phys.
				Rev. B}\ }\textbf {\bibinfo {volume} {96}},\ \bibinfo {pages} 
				{195409}
		(\bibinfo {year} {2017})}\BibitemShut {NoStop}%
	\bibitem [{\citenamefont {Ptok}\ \emph {et~al.}(2017)\citenamefont {Ptok},
		\citenamefont {G\l{}odzik},\ and\ \citenamefont 
		{Doma\ifmmode~\acute{n}\else
			\'{n}\fi{}ski}}]{Glodzik.2017}%
	\BibitemOpen
	\bibfield  {author} {\bibinfo {author} {\bibfnamefont {A.}~\bibnamefont
			{Ptok}}, \bibinfo {author} {\bibfnamefont {S.}~\bibnamefont 
			{G\l{}odzik}}, \
		and\ \bibinfo {author} {\bibfnamefont {T.}~\bibnamefont
			{Doma\ifmmode~\acute{n}\else \'{n}\fi{}ski}},\ }\bibfield  {title} 
			{\enquote
		{\bibinfo {title} {{Y}u-{S}hiba-{R}usinov states of impurities in a
				triangular lattice of {NbSe}$_{2}$ with spin-orbit coupling},}\ 
				}\href
	{\doibase 10.1103/PhysRevB.96.184425} {\bibfield  {journal} {\bibinfo
			{journal} {Phys. Rev. B}\ }\textbf {\bibinfo {volume} {96}},\ 
			\bibinfo
		{pages} {184425} (\bibinfo {year} {2017})}\BibitemShut {NoStop}%
	\bibitem [{\citenamefont {Pekker}\ \emph {et~al.}(2021)\citenamefont 
	{Pekker},
		\citenamefont {Zhang},\ and\ \citenamefont {Frolov}}]{Frolov-2018}%
	\BibitemOpen
	\bibfield  {author} {\bibinfo {author} {\bibfnamefont {D.}~\bibnamefont
			{Pekker}}, \bibinfo {author} {\bibfnamefont {P.}~\bibnamefont 
			{Zhang}}, \
		and\ \bibinfo {author} {\bibfnamefont {S.M.}\ \bibnamefont {Frolov}},\
	}\bibfield  {title} {\enquote {\bibinfo {title} {Theory of {A}ndreev 
	blockade
				in a double quantum dot with a superconducting lead},}\ }\href 
				{\doibase
		10.21468/SciPostPhys.11.4.081} {\bibfield  {journal} {\bibinfo  
		{journal}
			{SciPost Phys.}\ }\textbf {\bibinfo {volume} {11}},\ \bibinfo 
			{pages} {81}
		(\bibinfo {year} {2021})}\BibitemShut {NoStop}%
	\bibitem [{\citenamefont {Scher\"ubl}\ \emph {et~al.}(2019)\citenamefont
		{Scher\"ubl}, \citenamefont {P\'alyi},\ and\ \citenamefont
		{Csonka}}]{Scherubl_2019}%
	\BibitemOpen
	\bibfield  {author} {\bibinfo {author} {\bibfnamefont {Z.}~\bibnamefont
			{Scher\"ubl}}, \bibinfo {author} {\bibfnamefont {A.}~\bibnamefont 
			{P\'alyi}},
		\ and\ \bibinfo {author} {\bibfnamefont {S.}~\bibnamefont {Csonka}},\
	}\bibfield  {title} {\enquote {\bibinfo {title} {Transport signatures of an
				{A}ndreev molecule in a quantum dot-superconductor-quantum dot 
				setup},}\
	}\href {\doibase 10.3762/bjnano.10.36} {\bibfield  {journal} {\bibinfo
			{journal} {Beilstein J. Nanotechnol.}\ }\textbf {\bibinfo {volume} 
			{10}},\
		\bibinfo {pages} {363} (\bibinfo {year} {2019})}\BibitemShut {NoStop}%
	\bibitem [{\citenamefont {Pokorn\'y}\ \emph {et~al.}(2020)\citenamefont
		{Pokorn\'y}, \citenamefont {\v{Z}onda}, \citenamefont {Loukeris},\ and\
		\citenamefont {Novotn\'y}}]{Zonda.2019}%
	\BibitemOpen
	\bibfield  {author} {\bibinfo {author} {\bibfnamefont {V.}~\bibnamefont
			{Pokorn\'y}}, \bibinfo {author} {\bibfnamefont {M.}~\bibnamefont
			{\v{Z}onda}}, \bibinfo {author} {\bibfnamefont {G.}~\bibnamefont 
			{Loukeris}},
		\ and\ \bibinfo {author} {\bibfnamefont {T}~\bibnamefont {Novotn\'y}},\
	}\bibfield  {title} {\enquote {\bibinfo {title} {Second order perturbation
				theory for a superconducting double quantum dot},}\ }\href 
				{\doibase
		10.7566/JPSCP.30.011002} {\bibfield  {journal} {\bibinfo  {journal} {JPS
				Conf. Proc.}\ }\textbf {\bibinfo {volume} {30}},\ \bibinfo 
				{pages} {011002}
		(\bibinfo {year} {2020})}\BibitemShut {NoStop}%
	\bibitem [{\citenamefont {Wang}\ \emph {et~al.}(2019)\citenamefont {Wang},
		\citenamefont {Zhang}, \citenamefont {Han},\ and\ \citenamefont
		{Gong}}]{Wang-2019}%
	\BibitemOpen
	\bibfield  {author} {\bibinfo {author} {\bibfnamefont {X.-Q.}\ \bibnamefont
			{Wang}}, \bibinfo {author} {\bibfnamefont {S.-F.}\ \bibnamefont 
			{Zhang}},
		\bibinfo {author} {\bibfnamefont {Y.}~\bibnamefont {Han}}, \ and\ 
		\bibinfo
		{author} {\bibfnamefont {W.-J.}\ \bibnamefont {Gong}},\ }\bibfield  
		{title}
	{\enquote {\bibinfo {title} {{F}ano-{A}ndreev effect in a parallel double
				quantum dot structure},}\ }\href {\doibase 
				10.1103/PhysRevB.100.115405}
	{\bibfield  {journal} {\bibinfo  {journal} {Phys. Rev. B}\ }\textbf 
	{\bibinfo
			{volume} {100}},\ \bibinfo {pages} {115405} (\bibinfo {year}
		{2019})}\BibitemShut {NoStop}%
	\bibitem [{\citenamefont {Li}\ and\ \citenamefont
		{Leijnse}(2019)}]{Leijnse.2019}%
	\BibitemOpen
	\bibfield  {author} {\bibinfo {author} {\bibfnamefont {Z.-Z.}\ \bibnamefont
			{Li}}\ and\ \bibinfo {author} {\bibfnamefont {M.}~\bibnamefont 
			{Leijnse}},\
	}\bibfield  {title} {\enquote {\bibinfo {title} {Quantum interference in
				transport through almost symmetric double quantum dots},}\ 
				}\href {\doibase
		10.1103/PhysRevB.99.125406} {\bibfield  {journal} {\bibinfo  {journal} 
		{Phys.
				Rev. B}\ }\textbf {\bibinfo {volume} {99}},\ \bibinfo {pages} 
				{125406}
		(\bibinfo {year} {2019})}\BibitemShut {NoStop}%
	\bibitem [{\citenamefont {Morr}\ and\ \citenamefont
		{Stavropoulos}(2003)}]{Morr-2003}%
	\BibitemOpen
	\bibfield  {author} {\bibinfo {author} {\bibfnamefont {D.~K.}\ \bibnamefont
			{Morr}}\ and\ \bibinfo {author} {\bibfnamefont {N.~A.}\ \bibnamefont
			{Stavropoulos}},\ }\bibfield  {title} {\enquote {\bibinfo {title} 
			{Quantum
				interference between impurities: Creating novel many-body 
				states in s-wave
				superconductors},}\ }\href {\doibase 
				10.1103/PhysRevB.67.020502} {\bibfield
		{journal} {\bibinfo  {journal} {Phys. Rev. B}\ }\textbf {\bibinfo 
		{volume}
			{67}},\ \bibinfo {pages} {020502} (\bibinfo {year} 
			{2003})}\BibitemShut
	{NoStop}%
	\bibitem [{\citenamefont {Morr}\ and\ \citenamefont 
	{Yoon}(2006)}]{Morr-2006}%
	\BibitemOpen
	\bibfield  {author} {\bibinfo {author} {\bibfnamefont {D.~K.}\ \bibnamefont
			{Morr}}\ and\ \bibinfo {author} {\bibfnamefont {J.}~\bibnamefont 
			{Yoon}},\
	}\bibfield  {title} {\enquote {\bibinfo {title} {Impurities, quantum
				interference, and quantum phase transitions in $s$-wave 
				superconductors},}\
	}\href {\doibase 10.1103/PhysRevB.73.224511} {\bibfield  {journal} {\bibinfo
			{journal} {Phys. Rev. B}\ }\textbf {\bibinfo {volume} {73}},\ 
			\bibinfo
		{pages} {224511} (\bibinfo {year} {2006})}\BibitemShut {NoStop}%
	\bibitem [{\citenamefont {Brunetti}\ \emph {et~al.}(2013)\citenamefont
		{Brunetti}, \citenamefont {Zazunov}, \citenamefont {Kundu},\ and\
		\citenamefont {Egger}}]{Brunetti-2013}%
	\BibitemOpen
	\bibfield  {author} {\bibinfo {author} {\bibfnamefont {A.}~\bibnamefont
			{Brunetti}}, \bibinfo {author} {\bibfnamefont {A.}~\bibnamefont 
			{Zazunov}},
		\bibinfo {author} {\bibfnamefont {A.}~\bibnamefont {Kundu}}, \ and\ 
		\bibinfo
		{author} {\bibfnamefont {R.}~\bibnamefont {Egger}},\ }\bibfield  {title}
	{\enquote {\bibinfo {title} {Anomalous {J}osephson current, incipient
				time-reversal symmetry breaking, and {M}ajorana bound states in 
				interacting
				multilevel dots},}\ }\href {\doibase 
				10.1103/PhysRevB.88.144515} {\bibfield
		{journal} {\bibinfo  {journal} {Phys. Rev. B}\ }\textbf {\bibinfo 
		{volume}
			{88}},\ \bibinfo {pages} {144515} (\bibinfo {year} 
			{2013})}\BibitemShut
	{NoStop}%
	\bibitem [{\citenamefont {J\"unger}\ \emph {et~al.}(2021)\citenamefont
		{J\"unger}, \citenamefont {Lehmann}, \citenamefont {Dick}, \citenamefont
		{Thelander}, \citenamefont {Sch\"onenberger},\ and\ \citenamefont
		{Baumgartner}}]{Baumgartner-2021}%
	\BibitemOpen
	\bibfield  {author} {\bibinfo {author} {\bibfnamefont {C.}~\bibnamefont
			{J\"unger}}, \bibinfo {author} {\bibfnamefont {S.}~\bibnamefont 
			{Lehmann}},
		\bibinfo {author} {\bibfnamefont {K.~A.}\ \bibnamefont {Dick}}, \bibinfo
		{author} {\bibfnamefont {C.}~\bibnamefont {Thelander}}, \bibinfo 
		{author}
		{\bibfnamefont {C.}~\bibnamefont {Sch\"onenberger}}, \ and\ \bibinfo 
		{author}
		{\bibfnamefont {A.}~\bibnamefont {Baumgartner}},\ }\href@noop {} 
		{\enquote
		{\bibinfo {title} {Intermediate states in {A}ndreev bound state 
		fusion},}\ }
	(\bibinfo {year} {2021}),\ \Eprint {http://arxiv.org/abs/2111.00651}
	{arXiv:2111.00651} \BibitemShut {NoStop}%
	\bibitem [{\citenamefont {Flensberg}\ \emph {et~al.}(2021)\citenamefont
		{Flensberg}, \citenamefont {von Oppen},\ and\ \citenamefont
		{Stern}}]{Flensberg_review-2021}%
	\BibitemOpen
	\bibfield  {author} {\bibinfo {author} {\bibfnamefont {K.}~\bibnamefont
			{Flensberg}}, \bibinfo {author} {\bibfnamefont {F.}~\bibnamefont 
			{von
				Oppen}}, \ and\ \bibinfo {author} {\bibfnamefont 
				{A.}~\bibnamefont {Stern}},\
	}\href@noop {} {\enquote {\bibinfo {title} {Engineered platforms for
				topological superconductivity and {M}ajorana zero modes},}\ } 
				(\bibinfo
	{year} {2021}),\ \Eprint {http://arxiv.org/abs/2103.05548} 
	{arXiv:2103.05548}
	\BibitemShut {NoStop}%
	\bibitem [{\citenamefont {Wilson}(1975)}]{Wilson1975Oct}%
	\BibitemOpen
	\bibfield  {author} {\bibinfo {author} {\bibfnamefont {Kenneth~G.}\
			\bibnamefont {Wilson}},\ }\bibfield  {title} {\enquote {\bibinfo 
			{title}
			{{The renormalization group: Critical phenomena and the Kondo 
			problem}},}\
	}\href {\doibase 10.1103/RevModPhys.47.773} {\bibfield  {journal} {\bibinfo
			{journal} {Rev. Mod. Phys.}\ }\textbf {\bibinfo {volume} {47}},\ 
			\bibinfo
		{pages} {773--840} (\bibinfo {year} {1975})}\BibitemShut {NoStop}%
	\bibitem [{\citenamefont {Weichselbaum}\ and\ \citenamefont {von
			Delft}(2007)}]{Andreas_broadening2007}%
	\BibitemOpen
	\bibfield  {author} {\bibinfo {author} {\bibfnamefont {A.}~\bibnamefont
			{Weichselbaum}}\ and\ \bibinfo {author} {\bibfnamefont 
			{J.}~\bibnamefont {von
				Delft}},\ }\bibfield  {title} {\enquote {\bibinfo {title} 
				{Sum-rule
				conserving spectral functions from the numerical 
				renormalization group},}\
	}\href {\doibase 10.1103/PhysRevLett.99.076402} {\bibfield  {journal}
		{\bibinfo  {journal} {Phys. Rev. Lett.}\ }\textbf {\bibinfo {volume} 
		{99}},\
		\bibinfo {pages} {076402} (\bibinfo {year} {2007})}\BibitemShut 
		{NoStop}%
	\bibitem [{\citenamefont {Bulla}\ \emph {et~al.}(2008)\citenamefont {Bulla},
		\citenamefont {Costi},\ and\ \citenamefont {Pruschke}}]{Bulla2008Apr}%
	\BibitemOpen
	\bibfield  {author} {\bibinfo {author} {\bibfnamefont {R.}~\bibnamefont
			{Bulla}}, \bibinfo {author} {\bibfnamefont {T.~A.}\ \bibnamefont 
			{Costi}}, \
		and\ \bibinfo {author} {\bibfnamefont {T.}~\bibnamefont {Pruschke}},\
	}\bibfield  {title} {\enquote {\bibinfo {title} {{Numerical renormalization
					group method for quantum impurity systems}},}\ }\href 
					{\doibase
		10.1103/RevModPhys.80.395} {\bibfield  {journal} {\bibinfo  {journal} 
		{Rev.
				Mod. Phys.}\ }\textbf {\bibinfo {volume} {80}},\ \bibinfo 
				{pages} {395--450}
		(\bibinfo {year} {2008})}\BibitemShut {NoStop}%
	\bibitem [{\citenamefont {T\'oth}\ \emph {et~al.}(2008)\citenamefont 
	{T\'oth},
		\citenamefont {Moca}, \citenamefont {Legeza},\ and\ \citenamefont
		{Zar\'and}}]{Toth2008}%
	\BibitemOpen
	\bibfield  {author} {\bibinfo {author} {\bibfnamefont {A.~I.}\ \bibnamefont
			{T\'oth}}, \bibinfo {author} {\bibfnamefont {C.~P.}\ \bibnamefont 
			{Moca}},
		\bibinfo {author} {\bibfnamefont {\"O.}\ \bibnamefont {Legeza}}, \ and\
		\bibinfo {author} {\bibfnamefont {G.}~\bibnamefont {Zar\'and}},\ 
		}\bibfield
	{title} {\enquote {\bibinfo {title} {Density matrix numerical 
	renormalization
				group for non-abelian symmetries},}\ }\href {\doibase
		10.1103/PhysRevB.78.245109} {\bibfield  {journal} {\bibinfo  {journal} 
		{Phys.
				Rev. B}\ }\textbf {\bibinfo {volume} {78}},\ \bibinfo {pages} 
				{245109}
		(\bibinfo {year} {2008})}\BibitemShut {NoStop}%
	\bibitem [{\citenamefont {Legeza}\ \emph {et~al.}(2008)\citenamefont 
	{Legeza},
		\citenamefont {Moca}, \citenamefont {T\'{o}th}, \citenamefont 
		{Weymann},\
		and\ \citenamefont {Zar\'{a}nd}}]{NRG_code}%
	\BibitemOpen
	\bibfield  {author} {\bibinfo {author} {\bibfnamefont {\"{O}.}\ \bibnamefont
			{Legeza}}, \bibinfo {author} {\bibfnamefont {C.~P.}\ \bibnamefont 
			{Moca}},
		\bibinfo {author} {\bibfnamefont {A.~I.}\ \bibnamefont {T\'{o}th}}, 
		\bibinfo
		{author} {\bibfnamefont {I.}~\bibnamefont {Weymann}}, \ and\ \bibinfo
		{author} {\bibfnamefont {G.}~\bibnamefont {Zar\'{a}nd}},\ }\href
	{http://arxiv.org/abs/0809.3143} {\enquote {\bibinfo {title} {{Manual for 
	the
					Flexible DM-NRG code}},}\ }\bibinfo {howpublished} 
					{arXiv:0809.3143v1}
	(\bibinfo {year} {2008}),\ \bibinfo {note} {(the open access Flexible DM-NRG
		Budapest code is available at
		\href{http://www.phy.bme.hu/\~dmnrg/}{http:/\!/www.phy.bme.hu/\textasciitilde{}dmnrg/}}\BibitemShut
	{NoStop}%
	\bibitem [{\citenamefont {Hoffman}\ \emph {et~al.}(2017)\citenamefont
		{Hoffman}, \citenamefont {Chevallier}, \citenamefont {Loss},\ and\
		\citenamefont {Klinovaja}}]{Klinovaja-2017}%
	\BibitemOpen
	\bibfield  {author} {\bibinfo {author} {\bibfnamefont {S.}~\bibnamefont
			{Hoffman}}, \bibinfo {author} {\bibfnamefont {D.}~\bibnamefont 
			{Chevallier}},
		\bibinfo {author} {\bibfnamefont {D.}~\bibnamefont {Loss}}, \ and\ 
		\bibinfo
		{author} {\bibfnamefont {J.}~\bibnamefont {Klinovaja}},\ }\bibfield  
		{title}
	{\enquote {\bibinfo {title} {Spin-dependent coupling between quantum dots 
	and
				topological quantum wires},}\ }\href {\doibase 
				10.1103/PhysRevB.96.045440}
	{\bibfield  {journal} {\bibinfo  {journal} {Phys. Rev. B}\ }\textbf 
	{\bibinfo
			{volume} {96}},\ \bibinfo {pages} {045440} (\bibinfo {year}
		{2017})}\BibitemShut {NoStop}%
	\bibitem [{\citenamefont {Prada}\ \emph {et~al.}(2017)\citenamefont {Prada},
		\citenamefont {Aguado},\ and\ \citenamefont {San-Jose}}]{Prada-2017}%
	\BibitemOpen
	\bibfield  {author} {\bibinfo {author} {\bibfnamefont {E.}~\bibnamefont
			{Prada}}, \bibinfo {author} {\bibfnamefont {R.}~\bibnamefont 
			{Aguado}}, \
		and\ \bibinfo {author} {\bibfnamefont {P.}~\bibnamefont {San-Jose}},\
	}\bibfield  {title} {\enquote {\bibinfo {title} {Measuring {M}ajorana
				nonlocality and spin structure with a quantum dot},}\ }\href 
				{\doibase
		10.1103/PhysRevB.96.085418} {\bibfield  {journal} {\bibinfo  {journal} 
		{Phys.
				Rev. B}\ }\textbf {\bibinfo {volume} {96}},\ \bibinfo {pages} 
				{085418}
		(\bibinfo {year} {2017})}\BibitemShut {NoStop}%
	\bibitem [{\citenamefont {Schuray}\ \emph {et~al.}(2018)\citenamefont
		{Schuray}, \citenamefont {Yeyati},\ and\ \citenamefont
		{Recher}}]{Schuray-2018}%
	\BibitemOpen
	\bibfield  {author} {\bibinfo {author} {\bibfnamefont {A.}~\bibnamefont
			{Schuray}}, \bibinfo {author} {\bibfnamefont {A.~L.}\ \bibnamefont 
			{Yeyati}},
		\ and\ \bibinfo {author} {\bibfnamefont {P.}~\bibnamefont {Recher}},\
	}\bibfield  {title} {\enquote {\bibinfo {title} {Influence of the {M}ajorana
				nonlocality on the supercurrent},}\ }\href {\doibase
		10.1103/PhysRevB.98.235301} {\bibfield  {journal} {\bibinfo  {journal} 
		{Phys.
				Rev. B}\ }\textbf {\bibinfo {volume} {98}},\ \bibinfo {pages} 
				{235301}
		(\bibinfo {year} {2018})}\BibitemShut {NoStop}%
	\bibitem [{\citenamefont {G\'orski}\ and\ \citenamefont
		{Kucab}(2021)}]{Gorski-2021}%
	\BibitemOpen
	\bibfield  {author} {\bibinfo {author} {\bibfnamefont {G.}~\bibnamefont
			{G\'orski}}\ and\ \bibinfo {author} {\bibfnamefont {K.}~\bibnamefont
			{Kucab}},\ }\bibfield  {title} {\enquote {\bibinfo {title} 
			{Transport
				properties of proximitized double quantum dots},}\ }\href 
				{\doibase
		https://doi.org/10.1016/j.physe.2020.114459} {\bibfield  {journal} 
		{\bibinfo
			{journal} {Physica E}\ }\textbf {\bibinfo {volume} {126}},\ 
			\bibinfo {pages}
		{114459} (\bibinfo {year} {2021})}\BibitemShut {NoStop}%
	\bibitem [{\citenamefont {Tanaka}\ \emph {et~al.}(2008)\citenamefont 
	{Tanaka},
		\citenamefont {Kawakami},\ and\ \citenamefont {Oguri}}]{Tanaka-2008}%
	\BibitemOpen
	\bibfield  {author} {\bibinfo {author} {\bibfnamefont {Y.}~\bibnamefont
			{Tanaka}}, \bibinfo {author} {\bibfnamefont {N.}~\bibnamefont 
			{Kawakami}}, \
		and\ \bibinfo {author} {\bibfnamefont {A.}~\bibnamefont {Oguri}},\ 
		}\bibfield
	{title} {\enquote {\bibinfo {title} {Andreev transport through side-coupled
				double quantum dots},}\ }\href {\doibase 
				10.1103/PhysRevB.78.035444}
	{\bibfield  {journal} {\bibinfo  {journal} {Phys. Rev. B}\ }\textbf 
	{\bibinfo
			{volume} {78}},\ \bibinfo {pages} {035444} (\bibinfo {year}
		{2008})}\BibitemShut {NoStop}%
	\bibitem [{\citenamefont {Bara\'{n}ski}\ and\ \citenamefont
		{Doma\'{n}ski}(2011)}]{Baranski-2011}%
	\BibitemOpen
	\bibfield  {author} {\bibinfo {author} {\bibfnamefont {J.}~\bibnamefont
			{Bara\'{n}ski}}\ and\ \bibinfo {author} {\bibfnamefont 
			{T.}~\bibnamefont
			{Doma\'{n}ski}},\ }\bibfield  {title} {\enquote {\bibinfo {title} 
			{Fano-type
				interference in quantum dots coupled between metallic and 
				superconducting
				leads},}\ }\href {\doibase 10.1103/PhysRevB.84.195424} 
				{\bibfield  {journal}
		{\bibinfo  {journal} {Phys. Rev. B}\ }\textbf {\bibinfo {volume} {84}},\
		\bibinfo {pages} {195424} (\bibinfo {year} {2011})}\BibitemShut 
		{NoStop}%
	\bibitem [{\citenamefont {Bara\'nski}\ and\ \citenamefont
		{Doma\'nski}(2012)}]{Baranski-2012}%
	\BibitemOpen
	\bibfield  {author} {\bibinfo {author} {\bibfnamefont {J.}~\bibnamefont
			{Bara\'nski}}\ and\ \bibinfo {author} {\bibfnamefont 
			{T.}~\bibnamefont
			{Doma\'nski}},\ }\bibfield  {title} {\enquote {\bibinfo {title} 
			{Decoherence
				effect on {F}ano line shapes in double quantum dots coupled 
				between normal
				and superconducting leads},}\ }\href {\doibase 
				10.1103/PhysRevB.85.205451}
	{\bibfield  {journal} {\bibinfo  {journal} {Phys. Rev. B}\ }\textbf 
	{\bibinfo
			{volume} {85}},\ \bibinfo {pages} {205451} (\bibinfo {year}
		{2012})}\BibitemShut {NoStop}%
	\bibitem [{\citenamefont {Bara\'nski}\ \emph {et~al.}(2020)\citenamefont
		{Bara\'nski}, \citenamefont {Zienkiewicz}, \citenamefont {Bara\'nska},\ 
		and\
		\citenamefont {Kapcia}}]{Baranski-2020}%
	\BibitemOpen
	\bibfield  {author} {\bibinfo {author} {\bibfnamefont {J.}~\bibnamefont
			{Bara\'nski}}, \bibinfo {author} {\bibfnamefont {T.}~\bibnamefont
			{Zienkiewicz}}, \bibinfo {author} {\bibfnamefont {M.}~\bibnamefont
			{Bara\'nska}}, \ and\ \bibinfo {author} {\bibfnamefont {K.J.}\ 
			\bibnamefont
			{Kapcia}},\ }\bibfield  {title} {\enquote {\bibinfo {title} 
			{Anomalous {F}ano
				resonance in double quantum dot system coupled to 
				superconductor.}}\ }\href
	{\doibase 10.1038/s41598-020-59498-y} {\bibfield  {journal} {\bibinfo
			{journal} {Sci. Rep.}\ }\textbf {\bibinfo {volume} {10}},\ \bibinfo 
			{pages}
		{2881} (\bibinfo {year} {2020})}\BibitemShut {NoStop}%
	\bibitem [{\citenamefont {Deng}\ \emph {et~al.}(2012)\citenamefont {Deng},
		\citenamefont {Yu}, \citenamefont {Huang}, \citenamefont {Larsson},
		\citenamefont {Caroff},\ and\ \citenamefont {Xu}}]{Deng-2012}%
	\BibitemOpen
	\bibfield  {author} {\bibinfo {author} {\bibfnamefont {M.~T.}\ \bibnamefont
			{Deng}}, \bibinfo {author} {\bibfnamefont {C.~L.}\ \bibnamefont 
			{Yu}},
		\bibinfo {author} {\bibfnamefont {G.~Y.}\ \bibnamefont {Huang}}, 
		\bibinfo
		{author} {\bibfnamefont {M.}~\bibnamefont {Larsson}}, \bibinfo {author}
		{\bibfnamefont {P.}~\bibnamefont {Caroff}}, \ and\ \bibinfo {author}
		{\bibfnamefont {H.~Q.}\ \bibnamefont {Xu}},\ }\bibfield  {title} 
		{\enquote
		{\bibinfo {title} {Anomalous zero-bias conductance peak in a {Nb–InSb}
				nanowire–{Nb} hybrid device},}\ }\href {\doibase 
				10.1021/nl303758w}
	{\bibfield  {journal} {\bibinfo  {journal} {Nano Lett.}\ }\textbf {\bibinfo
			{volume} {12}},\ \bibinfo {pages} {6414} (\bibinfo {year}
		{2012})}\BibitemShut {NoStop}%
	\bibitem [{\citenamefont {Mourik}\ \emph {et~al.}(2012)\citenamefont 
	{Mourik},
		\citenamefont {Zuo}, \citenamefont {Frolov}, \citenamefont {Plissard},
		\citenamefont {Bakkers},\ and\ \citenamefont 
		{Kouwenhoven}}]{Mourik-2012}%
	\BibitemOpen
	\bibfield  {author} {\bibinfo {author} {\bibfnamefont {V.}~\bibnamefont
			{Mourik}}, \bibinfo {author} {\bibfnamefont {K.}~\bibnamefont 
			{Zuo}},
		\bibinfo {author} {\bibfnamefont {S.~M.}\ \bibnamefont {Frolov}}, 
		\bibinfo
		{author} {\bibfnamefont {S.~R.}\ \bibnamefont {Plissard}}, \bibinfo 
		{author}
		{\bibfnamefont {E.~P. A.~M.}\ \bibnamefont {Bakkers}}, \ and\ \bibinfo
		{author} {\bibfnamefont {L.~P.}\ \bibnamefont {Kouwenhoven}},\ 
		}\bibfield
	{title} {\enquote {\bibinfo {title} {Signatures of {Majorana} fermions in
				hybrid superconductor-semiconductor nanowire devices},}\ }\href 
				{\doibase
		10.1126/science.1222360} {\bibfield  {journal} {\bibinfo  {journal}
			{Science}\ }\textbf {\bibinfo {volume} {336}},\ \bibinfo {pages} 
			{1003}
		(\bibinfo {year} {2012})}\BibitemShut {NoStop}%
	\bibitem [{\citenamefont {Das}\ \emph {et~al.}(2012)\citenamefont {Das},
		\citenamefont {Ronen}, \citenamefont {Most}, \citenamefont {Oreg},
		\citenamefont {Heiblum},\ and\ \citenamefont {Shtrikman}}]{Das-2012}%
	\BibitemOpen
	\bibfield  {author} {\bibinfo {author} {\bibfnamefont {A.}~\bibnamefont
			{Das}}, \bibinfo {author} {\bibfnamefont {Y.}~\bibnamefont 
			{Ronen}}, \bibinfo
		{author} {\bibfnamefont {Y.}~\bibnamefont {Most}}, \bibinfo {author}
		{\bibfnamefont {Y.}~\bibnamefont {Oreg}}, \bibinfo {author} 
		{\bibfnamefont
			{M.}~\bibnamefont {Heiblum}}, \ and\ \bibinfo {author} 
			{\bibfnamefont
			{H.}~\bibnamefont {Shtrikman}},\ }\bibfield  {title} {\enquote 
			{\bibinfo
			{title} {Zero-bias peaks and splitting in an al–inas nanowire 
			topological
				superconductor as a signature of {M}ajorana fermions},}\ }\href 
				{\doibase
		10.1038/nphys2479} {\bibfield  {journal} {\bibinfo  {journal} {Nat. 
		Phys.}\
		}\textbf {\bibinfo {volume} {8}},\ \bibinfo {pages} {887} (\bibinfo 
		{year}
		{2012})}\BibitemShut {NoStop}%
	\bibitem [{\citenamefont {Churchill}\ \emph {et~al.}(2013)\citenamefont
		{Churchill}, \citenamefont {Fatemi}, \citenamefont {Grove-Rasmussen},
		\citenamefont {Deng}, \citenamefont {Caroff}, \citenamefont {Xu},\ and\
		\citenamefont {Marcus}}]{Churchill-2013}%
	\BibitemOpen
	\bibfield  {author} {\bibinfo {author} {\bibfnamefont {H.~O.~H.}\
			\bibnamefont {Churchill}}, \bibinfo {author} {\bibfnamefont 
			{V.}~\bibnamefont
			{Fatemi}}, \bibinfo {author} {\bibfnamefont {K.}~\bibnamefont
			{Grove-Rasmussen}}, \bibinfo {author} {\bibfnamefont {M.~T.}\ 
			\bibnamefont
			{Deng}}, \bibinfo {author} {\bibfnamefont {P.}~\bibnamefont 
			{Caroff}},
		\bibinfo {author} {\bibfnamefont {H.~Q.}\ \bibnamefont {Xu}}, \ and\ 
		\bibinfo
		{author} {\bibfnamefont {C.~M.}\ \bibnamefont {Marcus}},\ }\bibfield  
		{title}
	{\enquote {\bibinfo {title} {Superconductor-nanowire devices from tunneling
				to the multichannel regime: {Zero-bias} oscillations and 
				magnetoconductance
				crossover},}\ }\href {\doibase 10.1103/PhysRevB.87.241401} 
				{\bibfield
		{journal} {\bibinfo  {journal} {Phys. Rev. B}\ }\textbf {\bibinfo 
		{volume}
			{87}},\ \bibinfo {pages} {241401} (\bibinfo {year} 
			{2013})}\BibitemShut
	{NoStop}%
	\bibitem [{\citenamefont {Deacon}\ \emph
		{et~al.}(2010{\natexlab{a}})\citenamefont {Deacon}, \citenamefont 
		{Tanaka},
		\citenamefont {Oiwa}, \citenamefont {Sakano}, \citenamefont {Yoshida},
		\citenamefont {Shibata}, \citenamefont {Hirakawa},\ and\ \citenamefont
		{Tarucha}}]{Deacon-2010}%
	\BibitemOpen
	\bibfield  {author} {\bibinfo {author} {\bibfnamefont {R.~S.}\ \bibnamefont
			{Deacon}}, \bibinfo {author} {\bibfnamefont {Y.}~\bibnamefont 
			{Tanaka}},
		\bibinfo {author} {\bibfnamefont {A.}~\bibnamefont {Oiwa}}, \bibinfo 
		{author}
		{\bibfnamefont {R.}~\bibnamefont {Sakano}}, \bibinfo {author} 
		{\bibfnamefont
			{K.}~\bibnamefont {Yoshida}}, \bibinfo {author} {\bibfnamefont
			{K.}~\bibnamefont {Shibata}}, \bibinfo {author} {\bibfnamefont
			{K.}~\bibnamefont {Hirakawa}}, \ and\ \bibinfo {author} 
			{\bibfnamefont
			{S.}~\bibnamefont {Tarucha}},\ }\bibfield  {title} {\enquote 
			{\bibinfo
			{title} {Kondo-enhanced {Andreev} transport in single 
			self-assembled {InAs}
				quantum dots contacted with normal and superconducting 
				leads},}\ }\href
	{\doibase 10.1103/PhysRevB.81.121308} {\bibfield  {journal} {\bibinfo
			{journal} {Phys. Rev. B}\ }\textbf {\bibinfo {volume} {81}},\ 
			\bibinfo
		{pages} {121308} (\bibinfo {year} {2010}{\natexlab{a}})}\BibitemShut
	{NoStop}%
	\bibitem [{\citenamefont {Deacon}\ \emph
		{et~al.}(2010{\natexlab{b}})\citenamefont {Deacon}, \citenamefont 
		{Tanaka},
		\citenamefont {Oiwa}, \citenamefont {Sakano}, \citenamefont {Yoshida},
		\citenamefont {Shibata}, \citenamefont {Hirakawa},\ and\ \citenamefont
		{Tarucha}}]{Deacon-2010B}%
	\BibitemOpen
	\bibfield  {author} {\bibinfo {author} {\bibfnamefont {R.~S.}\ \bibnamefont
			{Deacon}}, \bibinfo {author} {\bibfnamefont {Y.}~\bibnamefont 
			{Tanaka}},
		\bibinfo {author} {\bibfnamefont {A.}~\bibnamefont {Oiwa}}, \bibinfo 
		{author}
		{\bibfnamefont {R.}~\bibnamefont {Sakano}}, \bibinfo {author} 
		{\bibfnamefont
			{K.}~\bibnamefont {Yoshida}}, \bibinfo {author} {\bibfnamefont
			{K.}~\bibnamefont {Shibata}}, \bibinfo {author} {\bibfnamefont
			{K.}~\bibnamefont {Hirakawa}}, \ and\ \bibinfo {author} 
			{\bibfnamefont
			{S.}~\bibnamefont {Tarucha}},\ }\bibfield  {title} {\enquote 
			{\bibinfo
			{title} {Tunneling spectroscopy of {A}ndreev energy levels in a 
			quantum dot
				coupled to a superconductor},}\ }\href {\doibase
		10.1103/PhysRevLett.104.076805} {\bibfield  {journal} {\bibinfo  
		{journal}
			{Phys. Rev. Lett.}\ }\textbf {\bibinfo {volume} {104}},\ \bibinfo 
			{pages}
		{076805} (\bibinfo {year} {2010}{\natexlab{b}})}\BibitemShut {NoStop}%
	\bibitem [{\citenamefont {H\"ubler}\ \emph {et~al.}(2012)\citenamefont
		{H\"ubler}, \citenamefont {Wolf}, \citenamefont {Scherer}, \citenamefont
		{Wang}, \citenamefont {Beckmann},\ and\ \citenamefont
		{v.~L\"ohneysen}}]{Hubler2012}%
	\BibitemOpen
	\bibfield  {author} {\bibinfo {author} {\bibfnamefont {F.}~\bibnamefont
			{H\"ubler}}, \bibinfo {author} {\bibfnamefont {M.~J.}\ \bibnamefont 
			{Wolf}},
		\bibinfo {author} {\bibfnamefont {T.}~\bibnamefont {Scherer}}, \bibinfo
		{author} {\bibfnamefont {D.}~\bibnamefont {Wang}}, \bibinfo {author}
		{\bibfnamefont {D.}~\bibnamefont {Beckmann}}, \ and\ \bibinfo {author}
		{\bibfnamefont {H.}~\bibnamefont {v.~L\"ohneysen}},\ }\bibfield  {title}
	{\enquote {\bibinfo {title} {Observation of {A}ndreev bound states at
				spin-active interfaces},}\ }\href {\doibase 
				10.1103/PhysRevLett.109.087004}
	{\bibfield  {journal} {\bibinfo  {journal} {Phys. Rev. Lett.}\ }\textbf
		{\bibinfo {volume} {109}},\ \bibinfo {pages} {087004} (\bibinfo {year}
		{2012})}\BibitemShut {NoStop}%
	\bibitem [{\citenamefont {Chang}\ \emph {et~al.}(2013)\citenamefont {Chang},
		\citenamefont {Manucharyan}, \citenamefont {Jespersen}, \citenamefont
		{Nyg\aa{}rd},\ and\ \citenamefont {Marcus}}]{Chang-2013}%
	\BibitemOpen
	\bibfield  {author} {\bibinfo {author} {\bibfnamefont {W.}~\bibnamefont
			{Chang}}, \bibinfo {author} {\bibfnamefont {V.~E.}\ \bibnamefont
			{Manucharyan}}, \bibinfo {author} {\bibfnamefont {T.~S.}\ 
			\bibnamefont
			{Jespersen}}, \bibinfo {author} {\bibfnamefont {J.}~\bibnamefont
			{Nyg\aa{}rd}}, \ and\ \bibinfo {author} {\bibfnamefont {C.~M.}\ 
			\bibnamefont
			{Marcus}},\ }\bibfield  {title} {\enquote {\bibinfo {title} 
			{Tunneling
				spectroscopy of quasiparticle bound states in a spinful 
				{J}osephson
				junction},}\ }\href {\doibase 10.1103/PhysRevLett.110.217005} 
				{\bibfield
		{journal} {\bibinfo  {journal} {Phys. Rev. Lett.}\ }\textbf {\bibinfo
			{volume} {110}},\ \bibinfo {pages} {217005} (\bibinfo {year}
		{2013})}\BibitemShut {NoStop}%
	\bibitem [{\citenamefont {Buitelaar}\ \emph {et~al.}(2002)\citenamefont
		{Buitelaar}, \citenamefont {Nussbaumer},\ and\ \citenamefont
		{Sch\"onenberger}}]{Buitelaar-2002}%
	\BibitemOpen
	\bibfield  {author} {\bibinfo {author} {\bibfnamefont {M.~R.}\ \bibnamefont
			{Buitelaar}}, \bibinfo {author} {\bibfnamefont {T.}~\bibnamefont
			{Nussbaumer}}, \ and\ \bibinfo {author} {\bibfnamefont 
			{C.}~\bibnamefont
			{Sch\"onenberger}},\ }\bibfield  {title} {\enquote {\bibinfo 
			{title} {Quantum
				dot in the {K}ondo regime coupled to superconductors},}\ }\href 
				{\doibase
		10.1103/PhysRevLett.89.256801} {\bibfield  {journal} {\bibinfo  
		{journal}
			{Phys. Rev. Lett.}\ }\textbf {\bibinfo {volume} {89}},\ \bibinfo 
			{pages}
		{256801} (\bibinfo {year} {2002})}\BibitemShut {NoStop}%
	\bibitem [{\citenamefont {Koerting}\ \emph {et~al.}(2010)\citenamefont
		{Koerting}, \citenamefont {Andersen}, \citenamefont {Flensberg},\ and\
		\citenamefont {Paaske}}]{Koerting-2010}%
	\BibitemOpen
	\bibfield  {author} {\bibinfo {author} {\bibfnamefont {V.}~\bibnamefont
			{Koerting}}, \bibinfo {author} {\bibfnamefont {B.~M.}\ \bibnamefont
			{Andersen}}, \bibinfo {author} {\bibfnamefont {K.}~\bibnamefont 
			{Flensberg}},
		\ and\ \bibinfo {author} {\bibfnamefont {J.}~\bibnamefont {Paaske}},\
	}\bibfield  {title} {\enquote {\bibinfo {title} {Nonequilibrium transport 
	via
				spin-induced subgap states in superconductor/quantum dot/normal 
				metal
				cotunnel junctions},}\ }\href {\doibase 
				10.1103/PhysRevB.82.245108}
	{\bibfield  {journal} {\bibinfo  {journal} {Phys. Rev. B}\ }\textbf 
	{\bibinfo
			{volume} {82}},\ \bibinfo {pages} {245108} (\bibinfo {year}
		{2010})}\BibitemShut {NoStop}%
	\bibitem [{\citenamefont {Bara\'nski}\ and\ \citenamefont
		{Doma\'nski}(2013)}]{Baranski-2013}%
	\BibitemOpen
	\bibfield  {author} {\bibinfo {author} {\bibfnamefont {J.}~\bibnamefont
			{Bara\'nski}}\ and\ \bibinfo {author} {\bibfnamefont 
			{T.}~\bibnamefont
			{Doma\'nski}},\ }\bibfield  {title} {\enquote {\bibinfo {title} 
			{In-gap
				states of a quantum dot coupled between a normal and a 
				superconducting
				lead},}\ }\href 
				{http://stacks.iop.org/0953-8984/25/i=43/a=435305} {\bibfield
		{journal} {\bibinfo  {journal} {J. Phys.: Condens. Matter}\ }\textbf
		{\bibinfo {volume} {25}},\ \bibinfo {pages} {435305} (\bibinfo {year}
		{2013})}\BibitemShut {NoStop}%
	\bibitem [{\citenamefont {Bara\'nski}\ \emph {et~al.}(2017)\citenamefont
		{Bara\'nski}, \citenamefont {Kobia\l{}ka},\ and\ \citenamefont
		{Doma\'nski}}]{Baranski-2017}%
	\BibitemOpen
	\bibfield  {author} {\bibinfo {author} {\bibfnamefont {J.}~\bibnamefont
			{Bara\'nski}}, \bibinfo {author} {\bibfnamefont {A.}~\bibnamefont
			{Kobia\l{}ka}}, \ and\ \bibinfo {author} {\bibfnamefont 
			{T.}~\bibnamefont
			{Doma\'nski}},\ }\bibfield  {title} {\enquote {\bibinfo {title}
			{Spin-sensitive interference due to {Majorana} state on the 
			interface between
				normal and superconducting leads},}\ }\href
	{http://stacks.iop.org/0953-8984/29/i=7/a=075603} {\bibfield  {journal}
		{\bibinfo  {journal} {J. Phys.: Condens. Matter}\ }\textbf {\bibinfo 
		{volume}
			{29}},\ \bibinfo {pages} {075603} (\bibinfo {year} 
			{2017})}\BibitemShut
	{NoStop}%
	\bibitem [{\citenamefont {Zienkiewicz}\ \emph {et~al.}(2020)\citenamefont
		{Zienkiewicz}, \citenamefont {Bara{\'{n}}ski}, \citenamefont 
		{G{\'{o}}rski},\
		and\ \citenamefont {Doma{\'{n}}ski}}]{Baranski-2019}%
	\BibitemOpen
	\bibfield  {author} {\bibinfo {author} {\bibfnamefont {T.}~\bibnamefont
			{Zienkiewicz}}, \bibinfo {author} {\bibfnamefont {J.}~\bibnamefont
			{Bara{\'{n}}ski}}, \bibinfo {author} {\bibfnamefont 
			{G.}~\bibnamefont
			{G{\'{o}}rski}}, \ and\ \bibinfo {author} {\bibfnamefont 
			{T.}~\bibnamefont
			{Doma{\'{n}}ski}},\ }\bibfield  {title} {\enquote {\bibinfo {title} 
			{Leakage
				of {M}ajorana mode into correlated quantum dot nearby its 
				singlet-doublet
				crossover},}\ }\href {\doibase 10.1088/1361-648x/ab46d9} 
				{\bibfield
		{journal} {\bibinfo  {journal} {J. Phys.: Condens. Matter}\ }\textbf
		{\bibinfo {volume} {32}},\ \bibinfo {pages} {025302} (\bibinfo {year}
		{2020})}\BibitemShut {NoStop}%
	\bibitem [{\citenamefont {Kondo}(1964)}]{Kondo-1964}%
	\BibitemOpen
	\bibfield  {author} {\bibinfo {author} {\bibfnamefont {J.}~\bibnamefont
			{Kondo}},\ }\bibfield  {title} {\enquote {\bibinfo {title} 
			{Resistance
				minimum in dilute magnetic alloys},}\ }\href {\doibase 
				10.1143/PTP.32.37}
	{\bibfield  {journal} {\bibinfo  {journal} {Prog. Theor. Phys.}\ }\textbf
		{\bibinfo {volume} {32}},\ \bibinfo {pages} {37} (\bibinfo {year}
		{1964})}\BibitemShut {NoStop}%
	\bibitem [{\citenamefont {Hewson}(1993)}]{Hewson1993Jan}%
	\BibitemOpen
	\bibfield  {author} {\bibinfo {author} {\bibfnamefont {A.~C.}\ \bibnamefont
			{Hewson}},\ }\href {\doibase 10.1017/CBO9780511470752} {\emph 
			{\bibinfo
			{title} {{The Kondo Problem to Heavy Fermions}}}}\ (\bibinfo  
			{publisher}
	{Cambridge Univ. Press},\ \bibinfo {address} {Cambridge, UK},\ \bibinfo
	{year} {1993})\BibitemShut {NoStop}%
	\bibitem [{\citenamefont {Bauer}\ \emph {et~al.}(2007)\citenamefont {Bauer},
		\citenamefont {Oguri},\ and\ \citenamefont {Hewson}}]{Bauer-2007}%
	\BibitemOpen
	\bibfield  {author} {\bibinfo {author} {\bibfnamefont {J.}~\bibnamefont
			{Bauer}}, \bibinfo {author} {\bibfnamefont {A.}~\bibnamefont 
			{Oguri}}, \ and\
		\bibinfo {author} {\bibfnamefont {A.~C.}\ \bibnamefont {Hewson}},\ 
		}\bibfield
	{title} {\enquote {\bibinfo {title} {Spectral properties of locally
				correlated electrons in a {Bardeen-Cooper-Schrieffer} 
				superconductor},}\
	}\href {\doibase 10.1088/0953-8984/19/48/486211} {\bibfield  {journal}
		{\bibinfo  {journal} {J. Phys.: Condens. Matter}\ }\textbf {\bibinfo 
		{volume}
			{19}},\ \bibinfo {pages} {486211} (\bibinfo {year} 
			{2007})}\BibitemShut
	{NoStop}%
	\bibitem [{\citenamefont {\ifmmode~\check{Z}\else \v{Z}\fi{}itko}\ \emph
		{et~al.}(2015)\citenamefont {\ifmmode~\check{Z}\else \v{Z}\fi{}itko},
		\citenamefont {Lim}, \citenamefont {L\'opez},\ and\ \citenamefont
		{Aguado}}]{Zitko-2015b}%
	\BibitemOpen
	\bibfield  {author} {\bibinfo {author} {\bibfnamefont {R.}~\bibnamefont
			{\ifmmode~\check{Z}\else \v{Z}\fi{}itko}}, \bibinfo {author} 
			{\bibfnamefont
			{J.~Soo}\ \bibnamefont {Lim}}, \bibinfo {author} {\bibfnamefont
			{R.}~\bibnamefont {L\'opez}}, \ and\ \bibinfo {author} 
			{\bibfnamefont
			{R.}~\bibnamefont {Aguado}},\ }\bibfield  {title} {\enquote 
			{\bibinfo {title}
			{Shiba states and zero-bias anomalies in the hybrid 
			normal-superconductor
				{Anderson} model},}\ }\href {\doibase 
				10.1103/PhysRevB.91.045441} {\bibfield
		{journal} {\bibinfo  {journal} {Phys. Rev. B}\ }\textbf {\bibinfo 
		{volume}
			{91}},\ \bibinfo {pages} {045441} (\bibinfo {year} 
			{2015})}\BibitemShut
	{NoStop}%
	\bibitem [{\citenamefont {Doma\'{n}ski}\ \emph {et~al.}(2016)\citenamefont
		{Doma\'{n}ski}, \citenamefont {Weymann}, \citenamefont {Bara\'{n}ska},\ 
		and\
		\citenamefont {G\'orski}}]{Domanski-2016}%
	\BibitemOpen
	\bibfield  {author} {\bibinfo {author} {\bibfnamefont {T.}~\bibnamefont
			{Doma\'{n}ski}}, \bibinfo {author} {\bibfnamefont {I.}~\bibnamefont
			{Weymann}}, \bibinfo {author} {\bibfnamefont {M.}~\bibnamefont
			{Bara\'{n}ska}}, \ and\ \bibinfo {author} {\bibfnamefont 
			{G.}~\bibnamefont
			{G\'orski}},\ }\bibfield  {title} {\enquote {\bibinfo {title} 
			{Constructive
				influence of the induced electron pairing on the {K}ondo 
				state},}\ }\href
	{\doibase 10.1038/srep23336} {\bibfield  {journal} {\bibinfo  {journal} 
	{Sci.
				Rep.}\ }\textbf {\bibinfo {volume} {6}},\ \bibinfo {pages} 
				{23336} (\bibinfo
		{year} {2016})}\BibitemShut {NoStop}%
	\bibitem [{\citenamefont {W\'ojcik}\ and\ \citenamefont
		{Weymann}(2019)}]{Wojcik-2019}%
	\BibitemOpen
	\bibfield  {author} {\bibinfo {author} {\bibfnamefont {K.~P.}\ \bibnamefont
			{W\'ojcik}}\ and\ \bibinfo {author} {\bibfnamefont {I.}~\bibnamefont
			{Weymann}},\ }\bibfield  {title} {\enquote {\bibinfo {title} 
			{Nonlocal
				pairing as a source of spin exchange and {K}ondo screening},}\ 
				}\href
	{\doibase 10.1103/PhysRevB.99.045120} {\bibfield  {journal} {\bibinfo
			{journal} {Phys. Rev. B}\ }\textbf {\bibinfo {volume} {99}},\ 
			\bibinfo
		{pages} {045120} (\bibinfo {year} {2019})}\BibitemShut {NoStop}%
	\bibitem [{\citenamefont {Majek}\ and\ \citenamefont
		{Weymann}(2021)}]{Majek-2021}%
	\BibitemOpen
	\bibfield  {author} {\bibinfo {author} {\bibfnamefont {P.}~\bibnamefont
			{Majek}}\ and\ \bibinfo {author} {\bibfnamefont {I.}~\bibnamefont
			{Weymann}},\ }\bibfield  {title} {\enquote {\bibinfo {title} 
			{Majorana mode
				leaking into a spin-charge entangled double quantum dot},}\ 
				}\href {\doibase
		10.1103/PhysRevB.104.085416} {\bibfield  {journal} {\bibinfo  {journal}
			{Phys. Rev. B}\ }\textbf {\bibinfo {volume} {104}},\ \bibinfo 
			{pages}
		{085416} (\bibinfo {year} {2021})}\BibitemShut {NoStop}%
	\bibitem [{\citenamefont {Goldhaber-Gordon}\ \emph 
	{et~al.}(1998)\citenamefont
		{Goldhaber-Gordon}, \citenamefont {Shtrikman}, \citenamefont {Mahalu},
		\citenamefont {Abusch-Magder}, \citenamefont {Meirav},\ and\ 
		\citenamefont
		{Kastner}}]{Goldhaber-Gordon1998Jan}%
	\BibitemOpen
	\bibfield  {author} {\bibinfo {author} {\bibfnamefont {D.}~\bibnamefont
			{Goldhaber-Gordon}}, \bibinfo {author} {\bibfnamefont {Hadas}\ 
			\bibnamefont
			{Shtrikman}}, \bibinfo {author} {\bibfnamefont {D.}~\bibnamefont 
			{Mahalu}},
		\bibinfo {author} {\bibfnamefont {David}\ \bibnamefont {Abusch-Magder}},
		\bibinfo {author} {\bibfnamefont {U.}~\bibnamefont {Meirav}}, \ and\ 
		\bibinfo
		{author} {\bibfnamefont {M.~A.}\ \bibnamefont {Kastner}},\ }\bibfield
	{title} {\enquote {\bibinfo {title} {{Kondo effect in a single-electron
					transistor}},}\ }\href {\doibase 10.1038/34373} {\bibfield  
					{journal}
		{\bibinfo  {journal} {Nature}\ }\textbf {\bibinfo {volume} {391}},\ 
		\bibinfo
		{pages} {156--159} (\bibinfo {year} {1998})}\BibitemShut {NoStop}%
	\bibitem [{\citenamefont {Pustilnik}\ and\ \citenamefont
		{Glazman}(2001)}]{Pustilnik2001Nov}%
	\BibitemOpen
	\bibfield  {author} {\bibinfo {author} {\bibfnamefont {M.}~\bibnamefont
			{Pustilnik}}\ and\ \bibinfo {author} {\bibfnamefont {L.~I.}\ 
			\bibnamefont
			{Glazman}},\ }\bibfield  {title} {\enquote {\bibinfo {title} 
			{{Kondo Effect
					in Real Quantum Dots}},}\ }\href {\doibase 
					10.1103/PhysRevLett.87.216601}
	{\bibfield  {journal} {\bibinfo  {journal} {Phys. Rev. Lett.}\ }\textbf
		{\bibinfo {volume} {87}},\ \bibinfo {pages} {216601} (\bibinfo {year}
		{2001})}\BibitemShut {NoStop}%
	\bibitem [{\citenamefont {Cornaglia}\ and\ \citenamefont
		{Grempel}(2005)}]{Cornaglia2005Feb}%
	\BibitemOpen
	\bibfield  {author} {\bibinfo {author} {\bibfnamefont {P.~S.}\ \bibnamefont
			{Cornaglia}}\ and\ \bibinfo {author} {\bibfnamefont {D.~R.}\ 
			\bibnamefont
			{Grempel}},\ }\bibfield  {title} {\enquote {\bibinfo {title} 
			{{Strongly
					correlated regimes in a double quantum dot device}},}\ 
					}\href {\doibase
		10.1103/PhysRevB.71.075305} {\bibfield  {journal} {\bibinfo  {journal} 
		{Phys.
				Rev. B}\ }\textbf {\bibinfo {volume} {71}},\ \bibinfo {pages} 
				{075305}
		(\bibinfo {year} {2005})}\BibitemShut {NoStop}%
	\bibitem [{\citenamefont {Chung}\ \emph {et~al.}(2008)\citenamefont {Chung},
		\citenamefont {Zarand},\ and\ \citenamefont
		{W{\ifmmode\ddot{o}\else\"{o}\fi}lfle}}]{Chung2008Jan}%
	\BibitemOpen
	\bibfield  {author} {\bibinfo {author} {\bibfnamefont {Chung-Hou}\
			\bibnamefont {Chung}}, \bibinfo {author} {\bibfnamefont {Gergely}\
			\bibnamefont {Zarand}}, \ and\ \bibinfo {author} {\bibfnamefont 
			{Peter}\
			\bibnamefont {W{\ifmmode\ddot{o}\else\"{o}\fi}lfle}},\ }\bibfield  
			{title}
	{\enquote {\bibinfo {title} {{Two-stage Kondo effect in side-coupled quantum
					dots: Renormalized perturbative scaling theory and 
					numerical renormalization
					group analysis}},}\ }\href {\doibase 
					10.1103/PhysRevB.77.035120} {\bibfield
		{journal} {\bibinfo  {journal} {Phys. Rev. B}\ }\textbf {\bibinfo 
		{volume}
			{77}},\ \bibinfo {pages} {035120} (\bibinfo {year} 
			{2008})}\BibitemShut
	{NoStop}%
	\bibitem [{\citenamefont {W{\ifmmode\acute{o}\else\'{o}\fi}jcik}\ and\
		\citenamefont {Weymann}(2015)}]{Wojcik2015Apr}%
	\BibitemOpen
	\bibfield  {author} {\bibinfo {author} {\bibfnamefont {Krzysztof~P.}\
			\bibnamefont {W{\ifmmode\acute{o}\else\'{o}\fi}jcik}}\ and\ 
			\bibinfo {author}
		{\bibfnamefont {Ireneusz}\ \bibnamefont {Weymann}},\ }\bibfield  {title}
	{\enquote {\bibinfo {title} {{Two-stage Kondo effect in T-shaped double
					quantum dots with ferromagnetic leads}},}\ }\href {\doibase
		10.1103/PhysRevB.91.134422} {\bibfield  {journal} {\bibinfo  {journal} 
		{Phys.
				Rev. B}\ }\textbf {\bibinfo {volume} {91}},\ \bibinfo {pages} 
				{134422}
		(\bibinfo {year} {2015})}\BibitemShut {NoStop}%
	\bibitem [{\citenamefont {Calzona}\ and\ \citenamefont
		{Trauzettel}(2021)}]{Calzona-2021}%
	\BibitemOpen
	\bibfield  {author} {\bibinfo {author} {\bibfnamefont {A.}~\bibnamefont
			{Calzona}}\ and\ \bibinfo {author} {\bibfnamefont {B.}~\bibnamefont
			{Trauzettel}},\ }\href@noop {} {\enquote {\bibinfo {title} 
			{Spin-resolved
				spectroscopy of helical {A}ndreev bound states},}\ } (\bibinfo 
				{year}
	{2021}),\ \Eprint {http://arxiv.org/abs/2111.07696} {arXiv:2111.07696}
	\BibitemShut {NoStop}%
\end{thebibliography}

%


\end{document}